\documentclass[aps,10pt,reprint,groupedaddress,superscriptaddressm, prx,nofootinbib]{revtex4-2}

\usepackage[usenames,dvipsnames]{xcolor}
\usepackage{amssymb}
\usepackage{graphicx}
\usepackage{amsmath}
\usepackage{bbm}
\usepackage[bookmarks=true,colorlinks,citecolor=blue,urlcolor=blue]{hyperref}
\usepackage{braket}
\usepackage{babel}
\usepackage{blindtext}
\usepackage[compat=1.1.0]{tikz-feynman}
\usepackage[normalem]{ulem}
\usepackage{physics}

\usepackage{mathtools}

\usepackage{dsfont}
\usepackage{ulem}

\definecolor{mypurp}{rgb}{0.35, 0, 0.7}

\usepackage{amsthm}
\usepackage{txfonts}

\theoremstyle{definition}

\DeclareMathOperator*{\Motimes}{\text{$\vcenter{\hbox{\scalebox{0.6}{$\bigotimes$}}}$}}

\begin{document}
\def\papertitle{{Quantum phase transition between symmetry enriched topological phases in tensor-network states}}
\newcommand{\TUM}{\affiliation{Technical University of Munich, TUM School of Natural Sciences, Physics Department, 85748 Garching, Germany}}
\newcommand{\MCQST}{\affiliation{Munich Center for Quantum Science and Technology (MCQST), Schellingstr. 4, 80799 M{\"u}nchen, Germany}}

\author{Lukas Haller$^\dagger$} \TUM \MCQST
\author{Wen-Tao Xu$^\dagger$} \TUM \MCQST
\def\thefootnote{$\dagger$}\footnotetext{These authors contributed equally.}
\author{Yu-Jie Liu} \TUM \MCQST
\author{Frank Pollmann}  \TUM \MCQST

\title{\papertitle}

\let\oldmaketitle\maketitle
\renewcommand{\maketitle}{\oldmaketitle\setcounter{footnote}{0}}
\renewcommand{\thefootnote}{\arabic{footnote}}

\begin{abstract}
Quantum phase transitions between different topologically ordered phases exhibit rich structures and are generically challenging to study in microscopic lattice models. 
In this work, we propose a tensor-network solvable model that allows us to tune between different symmetry enriched topological (SET) phases. 
Concretely, we consider a decorated two-dimensional toric code model for which the ground state can be expressed as a two-dimensional tensor-network state with bond dimension $D=3$ and two tunable parameters.
We find that the time-reversal (TR) symmetric system exhibits three distinct phases
(i) an SET toric code phase in which anyons transform non-trivially under TR, 
(ii) a toric code phase in which TR does not fractionalize, and
(iii) a topologically  trivial phase that is adiabatically connected to a product state. 
We characterize the different phases using the topological entanglement entropy and a membrane order parameter that distinguishes the two SET phases. 
Along the phase boundary between the SET toric code phase and the toric code phase, the model has an enhanced $U(1)$ symmetry and the ground state is a quantum critical loop gas wavefunction whose squared norm is equivalent to the partition function of the classical $O(2)$ model.
By duality transformations, this tensor-network solvable model can also be used to describe transitions between SET double-semion phases and between $\mathbb{Z}_2\times\mathbb{Z}_2^T$ symmetry protected topological phases in two dimensions.
\end{abstract}
\maketitle

\begin{figure}[t]
    \centering
    \includegraphics[width=8cm]{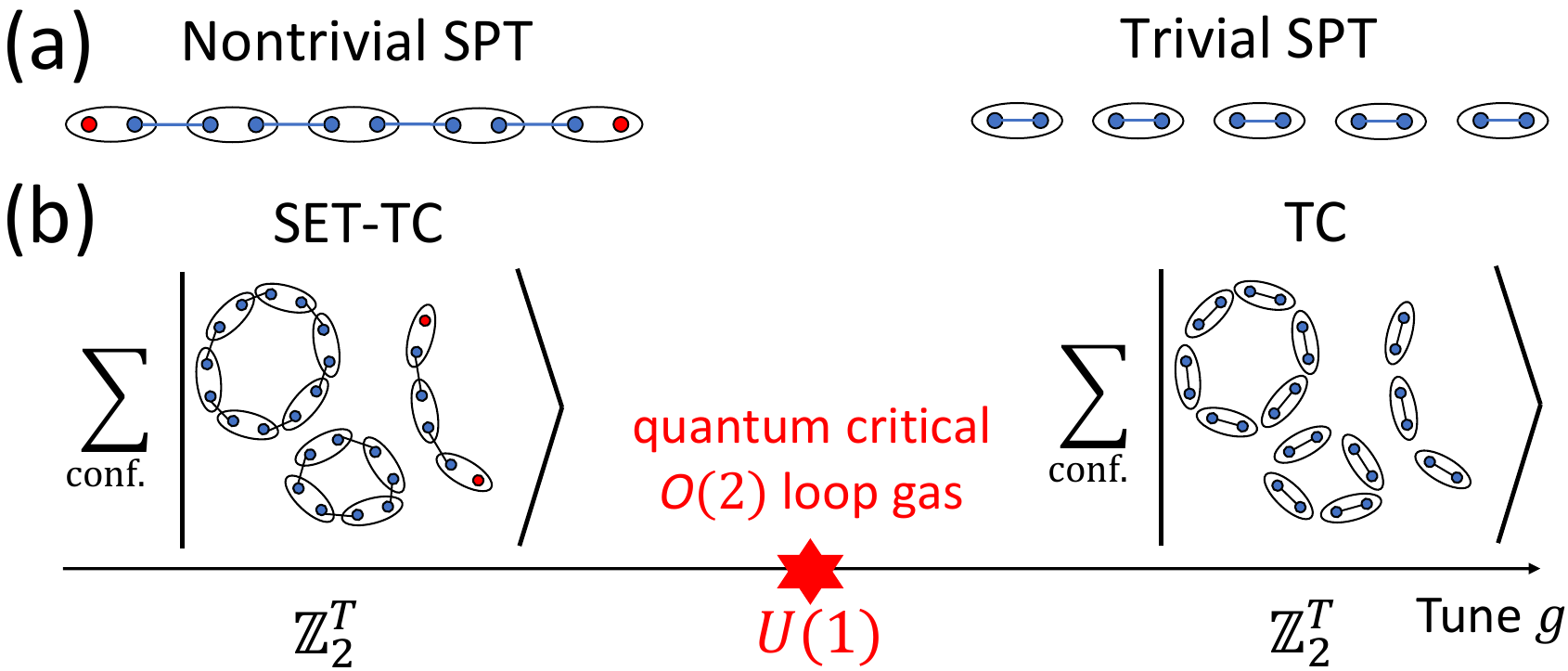}
    \caption{ Schematic illustration of distinct SET phases and the transition between them. (a) The symmetry fractionalizes over the edges in the 1D non-trivial SPT phase (red dots). By tuning a parameter $g$, the non-trivial SPT phase becomes trivial through a topological phase transition.  (b) A state in the SET toric code (SET-TC) phase or toric code (TC) phase with a pair of charge excitations at the ends of the broken loops. The sum runs over all the configurations with closed SPT loops and a broken SPT loop whose two ends are fixed. 
    The global symmetry fractionalizes over the charges. Notice that for the ground state on a closed manifold, the configurations only contain closed loops. Using the 1D SPT phase transition, we can construct a direct continuous phase transition from the SET-TC phase to the TC phase.} 
    \label{Figure_0}
\end{figure}

Over the past decades, significant progress has been made in understanding quantum phases of matter~\cite{tsui:1982,dijkgraaf:1990,WEN:1990}. In the absence of any symmetries, quantum systems in two or higher spatial dimensions can host distinct topologically ordered phases separated by quantum phase transitions (QPT)~\cite{Wen:2017}. 
When (intrinsic) topological order is absent, the presence of symmetries alone can lead to different symmetry protected topological (SPT) phases~\cite{gu:2009,Pollmann:2009, chen:2011b,Pschuch:2011,chen:2011a,Pollmann:2012a,chen:2013}. 
If both topological order and symmetries are present, distinct \emph{symmetry enriched topological} (SET) phases can emerge, which are characterized by how symmetry operations act on the anyonic quasiparticle excitations~\cite{Essin:2013,Mesaros:2013}. 
A remarkable experimental manifestation of SET order is the $\nu = 1/3$ Laughlin's fractional quantum Hall state~\cite{tsui:1982,laughlin:1983}, where the anyons carry fractional charges under the global $U(1)$ symmetry. 
The classification and characterization of bosonic and fermionic SET phases have been intensively investigated~\cite{Essin:2013,Mesaros:2013,Tarantino:2016,Lu:2016,Heinrich:2016,Meng:2017,Barkeshli:2019,aasen:2021,barkeshli:2022}. 
Certain phase transitions between different SET phases can be understood via anyon condensation~\cite{bais:2009,Garre-Rubio:2017,williamson:2017}, or as SPT phase transitions after gauging the global symmetries~\cite{levin:2012,Barkeshli:2019}. 
Simple toy models realizing different SET phases can be constructed in the following way: 
Starting from a $\mathbb{Z}_2$ topologically ordered system (for example, the toric code~\cite{Kitaev:2003}), different SET phases protected by a global symmetry $G$ can be constructed by decorating the loops in the topologically ordered state with one-dimensional (1D) SPT states protected by the symmetry $G$~\cite{Chen:2014,Li:2014,Huang:2014,Ben-Zion:2016}. 
As illustrated in Fig.~\ref{Figure_0}, the resulting state is a condensate of SPT loops and the symmetry will fractionalize between the anyons in a similar fashion as the symmetry fractionalizes at the boundaries of a 1D SPT chain with open boundary conditions \cite{Essin:2013,Chen:2014}.

In the present work, we follow this idea and construct a parameterized tensor-network solvable model that realizes a direct continuous transition between SET phases with an antiunitary time-reversal symmetry $\mathbb{Z}_2^T$.
In particular, we derive a tunable model for which the ground state is given by a tensor-network state (TNS)~\cite{maeshima:2001,verstraete2004}. This family of exact TNS corresponds to states of decorated loops with string tension and a tunable internal parameter, which are able to describe two distinct $\mathbb{Z}_2$ topologically ordered SET phases with different symmetry fractionalization patterns and a continuous phase transition between them (Fig.~\ref{Figure_0}). 
We numerically determine the phase diagram of the system by examining the correlation length, the topological entanglement entropy~\cite{Levin:2006,Kitaev_TEE:2006}, and a membrane order parameter~\cite{Huang:2014}. 
Along the phase boundary between the two SET phases, the amplitudes of the wavefunction can be exactly mapped to the partition function of the classical $O(2)$ loop model in the dense loop phase, described by the compactified free boson conformal field theory (CFT) with central charge $c=1$. 
The model exhibits an additional $U(1)$ symmetry at the $O(2)$ critical points. The additional $U(1)$ symmetry is an example of a pivot symmetry, which has recently been studied in the context of SPT phase transitions~\cite{Tantivasadakarn:2021}.
These transition points are, similar to the Rokhsar–Kivelson point on a square lattice~\cite{rokhsar:1988}, $(2+0)$D conformal critical points~\cite{Ardonne:2004,castelnovo:2010}, which have also appeared in several Abelian and non-Abelian topological phase transitions described by TNS~\cite{Verstraete:2006,xu:2018,zhu:2019,xu:2020,zhang:2020,Xu:2021,Xu:2022}. 
Finally, we discuss how the constructed example is dual to tensor-network solvable paths for the symmetry enriched double-semion model~\cite{Freedman:2004,levin:2005} and $(2+1)$D SPT states protected by $\mathbb{Z}_2\times \mathbb{Z}_2^T$.

The paper is organized as follows:
In Sec.~\ref{sec:tns}, we review the 1D and 2D examples of phase transitions in TNS which are used for the construction. In Sec.~\ref{sec:decorate} and Sec.~\ref{PEPS_rep}, we construct the decorated TNS for SET phase transitions. In Sec.~\ref{Sec_parent_hamitonian}, we show the parent Hamiltonian for the decorated TNS. In Sec.~\ref{sec:phase_diagram}, we show the numerical results of the phase diagram of the model and the order parameters. In Sec.~\ref{discussion}, we summarize the result and discuss several generalizations beyond the current example.

\section{Quantum Phase Transitions in Tensor-Network States}\label{sec:tns}

In this section, we review the two main ingredients for our construction. We begin by first reviewing the 1D SPT phase transition described by a family of 1D TNS, namely matrix-product states (MPS) that will be used for the decoration of the loops. We then recall the definition of the toric code model with a tunable string tension on a honeycomb lattice. In this paper, we use the standard notation $\{X,Z\}$ for Pauli matrices, and their eigenstates are denoted as $Z\ket{0}=\ket{0}$, $Z\ket{1}=-\ket{1}$, $X\ket{\pm}=\pm\ket{+}$, where $\ket{\pm}=(\ket{0}\pm\ket{1})/\sqrt{2}$. The Greenberger–Horne–Zeilinger (GHZ) state is defined as $(\ket{00\cdots 0}+\ket{11\cdots 1})/\sqrt{2}$.

\subsection{1D $\mathbb{Z}_2^T$-symmetric SPT phase transition in matrix product states}

 We consider the antiunitary $\mathbb{Z}_2^T$ time-reversal symmetry $K\prod_iX_i$, which is a combination of the global spin flip operator and complex conjugation $K$. A Hamiltonian describing a phase transition between two 1D SPT phases protected by the $\mathbb{Z}_2^T$ symmetry is~\cite{Wolf:2006-MPS_phase_transition}
\begin{equation}\label{SPT_Hamiltonian}
		H(g)=\sum_{i} \left[2(g^2 -1 ) Z_i Z_{i+1} - (1+g)^2 X_i + (1-g)^2 Z_i X_{i+1} Z_{i+2}\right],
	\end{equation}
where $g\in[-1,1]$ is the tuning parameter. When $g=1$, $H=-4\sum_iX_i$ and the ground state is a product state $|\psi(1)\rangle=\otimes_i|+\rangle_i$. When $g=-1$, $H$ reduces to the cluster model $H=4\sum_iZ_{i-1}X_iZ_{i+1}$ with the ground state $|\psi(-1)\rangle=\prod_{i}CZ_{i,i+1}\prod_{i}Z_i|\psi(1)\rangle$, where the control $Z$ gate $CZ_{i,i+1}$ acts on qubits $i$ and $i+1$, and $CZ_{i,i+1}=-1$ if both qubits are $1$ and $CZ_{i,i+1}=1$ otherwise.
The two limits $g = \pm 1$ exactly correspond to two fixed points of time-reversal symmetric SPT phases~\cite{Pschuch:2011,chen:2011a}. A phase transition occurs at $g = 0$, which is a multi-critical point characterized by a dynamical critical exponent $z=2$~\cite{Wolf:2006-MPS_phase_transition,Jones:2021}.

The ground states of this Hamiltonian are exactly described by a one-parameter family of MPS with bond dimension $\chi = 2$~\cite{Wolf:2006-MPS_phase_transition}   
\begin{equation}\label{MPS_with_g}
  |\psi(g)\rangle=\frac{1}{\sqrt{\mathcal{N}(g)}}\sum_{\{s_i\}}\text{Tr}(M^{[s_1]}M^{[s_2]}\cdots M^{[s_N]})|s_1,s_2,\cdots,s_N\rangle,
\end{equation}
where the MPS tensors are given by
\begin{equation}
		\label{eq:MPS(g)}
		M^{[0]}=
		\begin{pmatrix} 
			0 & 0 \\ 
			1 & 1  
		\end{pmatrix},
		\qquad M^{[1]}=
		\begin{pmatrix} 
			1 & g \\ 
			0 & 0  
		\end{pmatrix},
	\end{equation}
 and $\mathcal{N}(g)$ is the normalization coefficient (or simply squared norm) of the MPS. Notice that at the phase transition point $g=0$, the MPS becomes a GHZ state.

\subsection{2D toric code with string tension}
Let us now consider a honeycomb lattice with qubits on the edges, as shown in Fig.~\ref{Figure_1}. Each vertex $v$ is a set of three edges and each plaquette $p$ is a set of six edges. The toric code Hamiltonian is a sum of local and commuting projectors~\cite{Kitaev:2003}
\begin{equation}\label{eq:fp_h}
    H_{\text{TC}} = \sum_v A_v +\sum_p B_p ,
\end{equation}
where the star projector around each vertex $v$ is $A_v = \frac{1}{2}\left(1-\prod_{e\in v}Z_e\right)$. The plaquette projectors have the form $B_p = \frac{1}{2}\left(1- \prod_{e\in p}X_e\right)$. 
The Hamiltonian has a ground state energy of zero. As shown in Fig.~\ref{Figure_1}, an edge of state $\ket{1}$ is said to be occupied by a loop segment (or a string) and the state $\ket{0}$ is empty (vacuum). The ground state of the toric code is then an equal-weight superposition of closed-loop configurations on the edges of the lattice. The excitations in the toric code are denoted as electric $\pmb{e}$ with $\langle A_v\rangle = 1$ and magnetic $\pmb{m}$ with $\langle B_p\rangle = 1$. Their composite forms a fermion, which we denote by $\pmb{f}$. We further denote the trivial (null) excitation as $\pmb{1}$.

As we will discuss in Sec.~\ref{sec:decorate}, it turns out to be convenient to introduce a tunable string tension $\eta>0$ on the loops in the toric code~\cite{Castelnovo:2008,castelnovo:2009}.
The ground state is then modified to be a weighted superposition of closed-loop configurations
\begin{equation}
    \ket{\Psi(\eta)} \propto \sum_C \eta^{L(C)}\ket{C},\label{eq:TC_with_tension}
\end{equation}
where $C$ denotes the closed loop configurations on the honeycomb lattice and $L(C)$ is the total length of all loops in $C$. A parent Hamiltonian of the modified ground state is given in Sec.~\ref{Sec_parent_hamitonian}. For $\eta = 1$, we recover the toric code ground state $\ket{\Psi(1)}=\ket{\Psi_{\text{TC}}}$. At large string tension ($\eta \rightarrow 0$), the state becomes fully polarized.
The amplitude $\eta^{L(C)}$ can be mapped to the Boltzmann weight of the 2D classical Ising model and the critical string tension can be identified from the critical temperature of the Ising model as $\eta_c=3^{-1/4}$~\cite{castelnovo:2009}.
     
Moreover, the one-parameter family of wavefunctions in Eq.~\eqref{eq:TC_with_tension} can be expressed in terms of the ``single-line'' TNS
   \begin{equation}\label{single_line_TNS}
   |\Psi(\eta)\rangle=\frac{1}{\sqrt{\mathcal{N}(\eta)}}\sum_{\{s_e\}}\text{tTr}\left(\Motimes_v V\Motimes_e E^{[s_e]}(\eta)\right)|\cdots s_e\cdots\rangle
   \end{equation}
   with bond dimension $D=2$~\cite{Gu_PEPS_rep_2009},
   where the superscripts (subscripts) are the physical (virtual) indices which take $0$ or $1$, tTr denotes the tensor contraction over all virtual indices and
\begin{equation}\label{TC_tensors} V_{\alpha\beta\gamma}=\delta_{\text{mod}(\alpha+\beta+\gamma,2),0}, \quad E_{\alpha\beta}^{[s]}(\eta)=\eta^{s}\delta_{\alpha\beta}\delta_{\alpha p},
   \end{equation}
   are tensors placed at the vertices and edges of the honeycomb lattice, respectively. $\mathcal{N(\eta)}$ is the squared norm of the TNS. The tensor $V$ imposes the $\mathbb{Z}_2$ Gauss law on each vertex, and the tensor $E$ promotes the virtual degrees of freedom to the physical level and implements the string tension.

\section{Decorating the toric code}\label{sec:decorate}

\begin{figure}
    \centering
    \includegraphics[width=8cm]{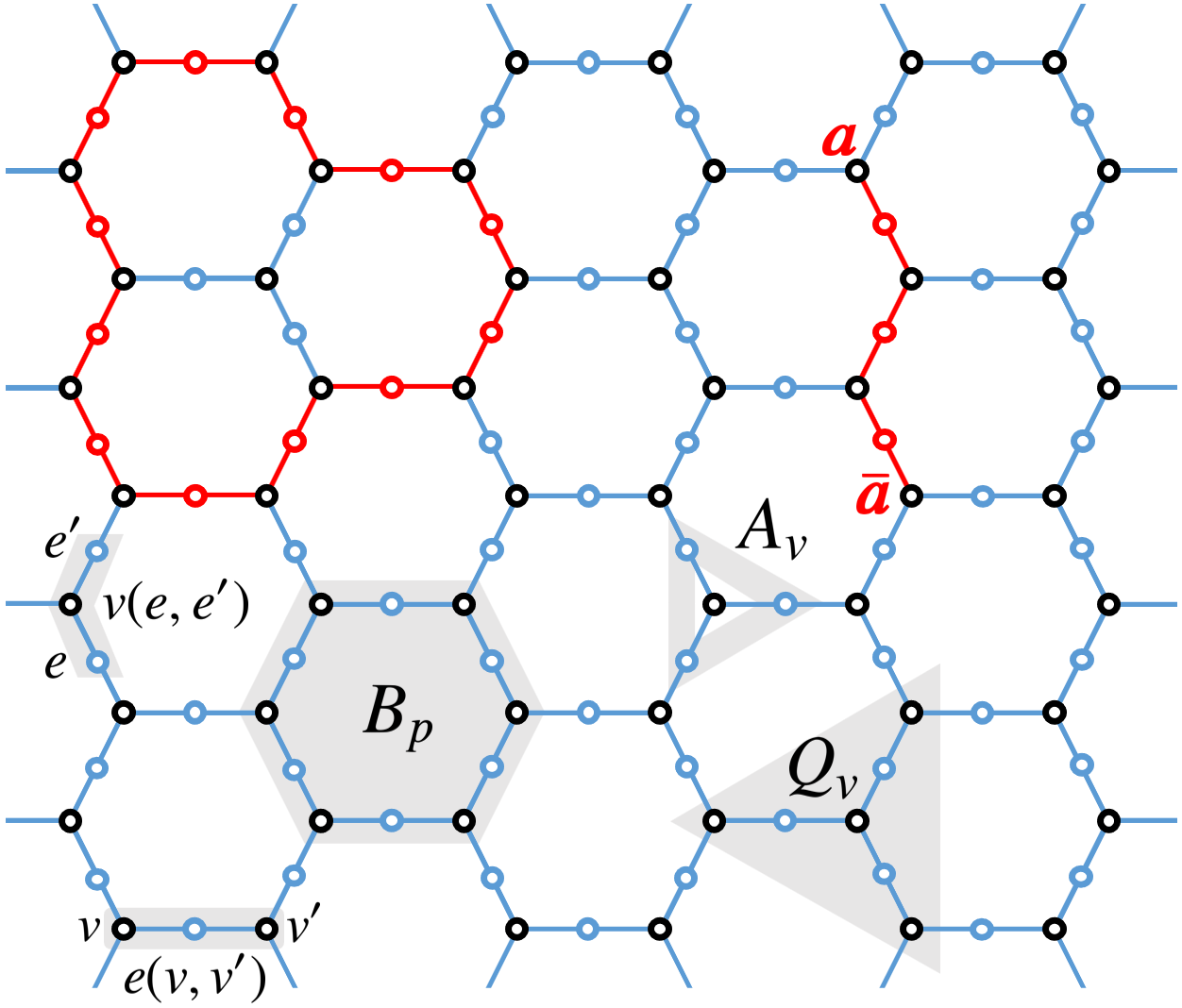}
    \caption{A snapshot of an excited state. The system is defined on a honeycomb lattice, the circles are two sets of physical qubits on the edges and vertices, respectively. Red (blue) circles represent edge qubits in the state $1$($0$), and black circles are vertex qubits. A loop in the toric code is formed by a string of edge qubits with state 1 along which the 1D SPT states are decorated. Ends of open loops (the open loop can deform freely except the endpoints) correspond to anyons $\pmb{a}$, which can be $\pmb{e}$ or $\pmb{f}$ anyons of the toric code. The qubits in the grey regions support the plaquette projector $B_p$, star projector $A_v$ and vertex projector $Q_v$ of the parent Hamiltonian~\eqref{2D_Hamiltonian}, respectively.}
    \label{Figure_1}
\end{figure}

Next, we consider the same honeycomb lattice on which the toric code ground state with string tension $\ket{\Psi(\eta)}$ is prepared on the qubits at the edges of the lattice. To decorate the loops, we add to each vertex $v$ a qubit as shown in Fig.~\ref{Figure_1}. 
The decoration is carried out with a simple procedure: whenever a loop is formed on the edges, we contract the MPS tensors~\eqref{eq:MPS(g)} on the vertices along the closed loop. The vertices away from the loops are set to the product state $\ket{+\cdots +}$. The resulting decorated 2D state $\ket{\Psi(g,\eta)}$ is thus a superposition of MPS-loop configurations and has a global $\mathbb{Z}_2^T$ symmetry generated by $K\prod_v X_v$, i.e. global spin flips on all vertices followed by complex conjugation.

At $g=1$ and $\eta = 1$ (no string tension), the ground state is a tensor product of the toric code ground state and a product state on all vertex qubits
\begin{equation}	|\Psi(g=1,\eta=1)\rangle=\ket{\Psi_{\text{TC}}} \otimes \left(\Motimes_v\ket{+}_v\right),
	\end{equation}
which has a trivial SET order, where the time-reversal symmetry fractionalizes trivially over the anyons of the toric code. We will simply refer to the phase it belongs to as the toric code (TC) phase.
At $g=-1$, the system can be obtained from the toric code limit by a constant-depth quantum circuit $|\Psi(-1,1)\rangle = U|\Psi(1,1)\rangle$, where $U$ is defined as	\begin{align}\label{Circuit_CCZ}
     U = \left(\prod_{\langle v,v^\prime\rangle} CCZ_{ vv^{\prime}e( v,v^\prime)}\right)\left(\prod_{ \langle e,e^\prime\rangle} CCZ_{ ee^{\prime} v( e,e^\prime) }\right),
	\end{align}
which is a 2D analogue of how we obtained $|\psi(-1)\rangle$ from $|\psi(1)\rangle$ in the 1D SPT model.
The first product goes over all distinct pairs of nearest neighbouring vertices with $\langle v,v^\prime\rangle = \langle v',v\rangle$, and the second product goes over all different pairs of nearest neighbouring edges $\langle e,e^\prime\rangle$. We use $e(v,v^\prime)$ (or $v( e,e^\prime)$) to denote the edge (or vertex) between the nearest neighbouring pair $\langle v,v^\prime\rangle$ (or $\langle e,e^\prime\rangle$), as shown in Fig.~\ref{Figure_1}. The $CCZ$ gate satisfies
\begin{equation}
 CCZ_{abc}=\begin{cases}
     -1, \mbox{ if all qubits at $a$, $b$, $c$ are 1, } \\
     1, \,\,\, \mbox{     otherwise}.
 \end{cases}
\end{equation}
The wavefunction $\ket{\Psi(-1,1)}$ is the fixed point for a non-trivial SET phase~\cite{garre:2021}, where the symmetry fractionalizes non-trivially over the $\pmb{e}$ and $\pmb{f}$ anyons of the toric code. We refer to the phase as SET-TC. 

\section{TNS representation}\label{PEPS_rep}

Away from the fixed points, the state $\ket{\Psi(g,\eta)}$ can be conveniently represented as a 2D TNS by decorating the MPS~\eqref{eq:MPS(g)} onto the single-line TNS. The resulting decorated single-line TNS, similar to the TNS in Eq.~\eqref{single_line_TNS}, consists of tensors with a bond dimension of $D = 3$, which are placed at the vertices and edges of the honeycomb lattice. 
The virtual degrees of freedom are spanned by the basis $\{|0),|1),|2)\}$. We apply a $\mathbb{Z}_2$ grading on this virtual space such that the parity of $|0)$ is even and the parity of $|1)$ and $|2)$ is odd, so the dimension of the odd parity subspace is $2$. 
    
\begin{figure}
    \centering
    \includegraphics[width=8cm]{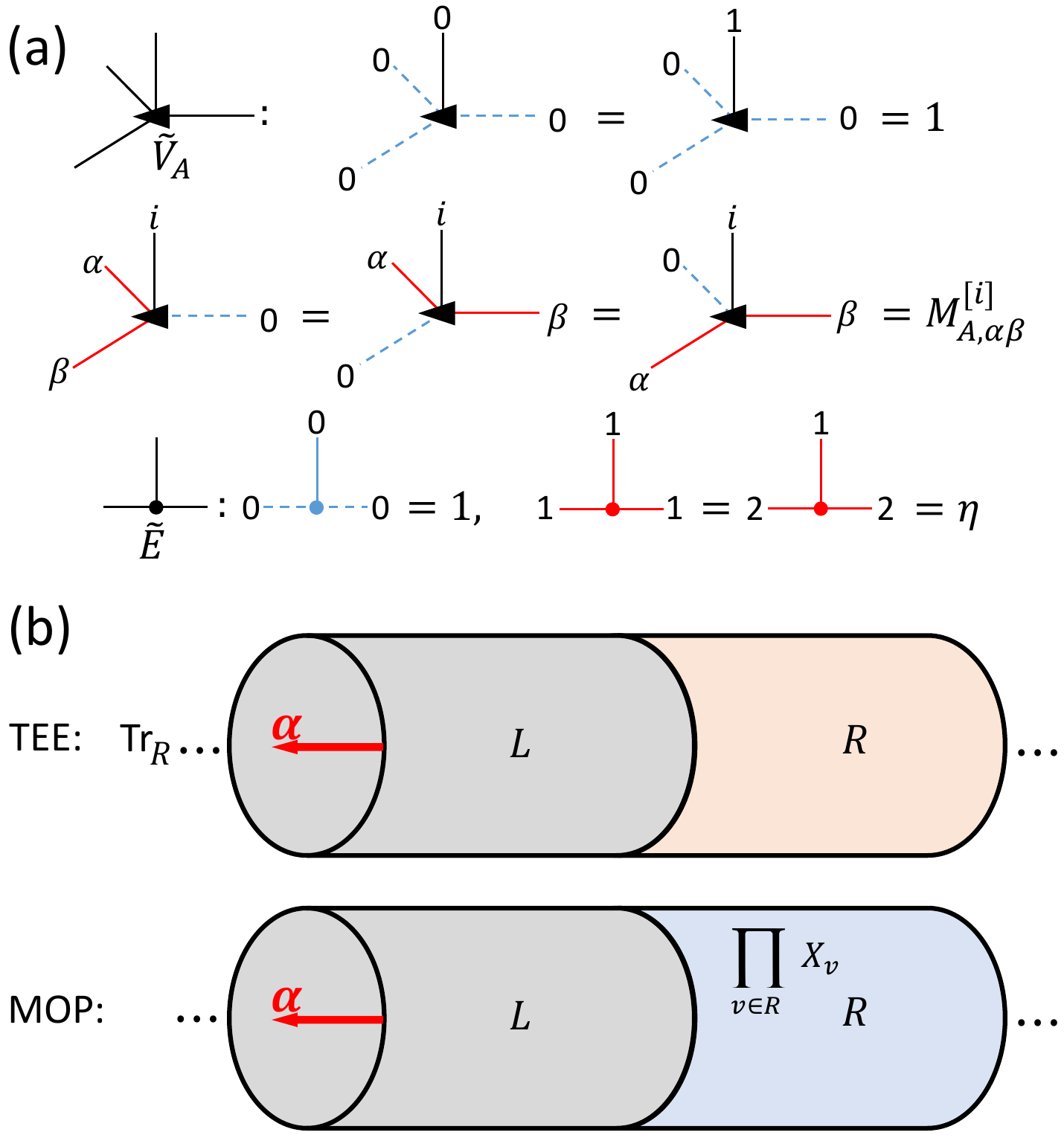}
    \caption{(a) The local tensors $\tilde{V}_A$ and $\tilde{E}$ of the decorated TNS and their non-zero entries, $\tilde{V}_B$ is obtained by replacing $M_A$ in $\tilde{V}_A$ with $M_B$. (b) The TEE and MOP on an infinitely long cylinder, where the anyon flux $\pmb{\alpha}$ penetrating inside the cylinder gives rise to the MES $\ket{\Psi_{\pmb{\alpha}}}$. The TEE comes from the reduced density matrix obtained by tracing out all physical qubits on half of the cylinder, and the MOP is obtained by applying the symmetry operator to the vertices on half of the cylinder and evaluating the expectation value.}
    \label{Figure_2}
\end{figure}

The decorated vertex tensor $\tilde{V}$ is schematically shown in Fig.~\ref{Figure_2}a. In contrast to the vertex tensor $V$ in Eq.~\eqref{TC_tensors}, the decorated vertex tensor $\tilde{V}$ has a physical leg of dimension $2$ corresponding to a vertex qubit. The $\mathbb{Z}_2$ Gauss law at the vertex tensor $\tilde{V}$ implies that either the vertex is not covered by any string or the vertex is covered by a closed loop segment. In the former case, the physical vertex qubit is $\sqrt{2}\ket{+}$ and the three virtual legs are $|0)$. In the latter case, the entries of $\tilde{V}$ given by the physical leg together with the two odd virtual legs are exactly defined by the MPS tensor $M^{[i]}$, as shown in Fig.~\ref{Figure_2}a. To construct the single-line TNS with a bond dimension $D = 3$, the MPS matrices $M^{[i]}$ used for the decoration have to be symmetric under the swapping of the two virtual indices (transpose).
This ensures that there is no ambiguity in the direction of contracting the MPS along a loop within the TNS~\footnote{Alternatively, we could decorate the MPS onto the double-line TNS of toric code~\cite{Gu:2008}, which can keep track of the direction of tensor contraction along a loop in the cost of a larger bond dimension.}.
While the original MPS matrices~\eqref{eq:MPS(g)} are not symmetric under transpose,
in Appendix~\ref{Appendix:Real_symmetric_tensors}, 
we utilize the gauge redundancy in the MPS representation to obtain a set of equivalent MPS tensors $M_{A}$ and $M_{B}$ in a two-site unit cell, which have the desired property.
Since the honeycomb lattice is a bipartite lattice, we use $M_{A}$ and $M_{B}$ to define two vertex tensors $\tilde{V}_A$ and $\tilde{V}_B$ for the two sublattices $A$ and $B$ of the honeycomb lattice, separately. In summary, the tensor $\tilde{V}_A(g)$ on the A sublattice is 
    
\begin{equation}\label{vertex_tensor}
    \tilde{V}^{[i]}_{A,\alpha\beta\gamma}(g)=\begin{cases}
    1, \quad\quad\quad\,\mbox{if } \alpha=\beta=\gamma=0;\\ 
    M^{[i]}_{A,\alpha\beta}(g),\mbox{ if }
    p(\alpha)=p(\beta)=1, \gamma=0;\\
     M^{[i]}_{A,\alpha\gamma}(g),\mbox{ if }
    p(\alpha)=p(\gamma)=1, \beta=0;\\
     M^{[i]}_{A,\beta\gamma}(g),\mbox{ if }
    p(\beta)=p(\gamma)=1, \alpha=0;\\
    0,\quad\quad\quad\,\mbox{otherwise},
    \end{cases}
    \end{equation}
where $p(\alpha)$ denotes the parity of $|\alpha)$. The construction works analogously for the tensor $\tilde{V}_B(g)$. 
    
The edge tensor $\tilde{E}$ of the decorated TNS is shown in Fig.~\ref{Figure_2}a and it maps the parity of the virtual degree of freedom to the physical degree of freedom and implements the string tension:
    \begin{equation}
    \tilde{E}^{[s]}_{ij}(\eta)=\eta^{s}\delta_{ij} \delta_{p(i),s}.
    \end{equation}
With these local tensors, the decorated  TNS can be constructed as 
\begin{equation}\label{eq:gs}
|\Psi(g,\eta)\rangle=\frac{1}{\sqrt{\mathcal{N}(g,\eta)}}\sum_{\{s_e,i_v\}}\text{tTr}\left(\Motimes_{v} \tilde{V}^{[i_{v}]}(g)\Motimes_e \tilde{E}^{[s_e]}(\eta)\right)|\{s_e,i_v\} \rangle,
\end{equation}
where $\tilde{V}$ can be $\tilde{V}_A$ or $\tilde{V}_B$ depending on which sublattice the vertex belongs to, and $\mathcal{N}(g,\eta)$ is the squared norm of the decorated TNS.

\section{Parent Hamiltonian}\label{Sec_parent_hamitonian}
So far, we have obtained a continuously parameterized family of TNS that interpolates between different fixed-point wavefunctions. We can also show that the states in Eq.~\eqref{eq:gs} are indeed ground states of a local Hamiltonian, which depends smoothly on the same set of parameters. More precisely, there exists a frustration-free, $\mathbb{Z}_2^T$-symmetric parent Hamiltonian that is a sum of local projectors
\begin{equation}\label{2D_Hamiltonian}
     H(g,\eta) = \sum_v A_v+\sum_p B_p(g,\eta) +\sum_v Q_v(g),
\end{equation}
where $g\in[-1,1]$ and $\eta>0$. Each vertex projector $A_v$, analogous to those in Eq.~\eqref{eq:fp_h}, projects onto the +1 eigenspace of the product of Pauli $Z$ around the vertex $v$. The plaquette projector $B_p(g,\eta)$ and the vertex projector $Q_v(g)$ act on the spins of a plaquette and around a vertex, respectively (see Fig.~\ref{Figure_1}). Let $v(e), v'(e)$ be the two vertices connected via the edge $e$, the projectors are explicitly given by
\begin{align}\label{eq:H}
    B_p(g,\eta) &= \frac{K_p}{2}\sech\left(\sum_{e\in p}\left[\tau(g) Z_e(1-Z_{v(e)}Z_{v'(e)})+\lambda(\eta) Z_e\right]\right),
    \nonumber \\
     Q_v(g) &= \frac{(1-A_v)M_v}{2}\sech\left(\tau(g)\sum_{e\in v} (1-Z_e)Z_{v(e)}Z_{v^{\prime}(e)}\right),
\end{align}
with 
\begin{align}
     K_p &= -\prod_{e\in p}X_e+\prod_{e\in p}e^{-\tau(g) Z_e(1-Z_{v(e)}Z_{v'(e)})}\eta^{-Z_e},
\nonumber \\
M_v &= -X_v+\prod_{e\in v}e^{-\tau(g) (1-Z_e)Z_{v(e)}Z_{v^{\prime}(e)}},
\end{align}
where $\lambda (\eta)= \log(\eta)$ and $\tau(g) = -\log(g)/4$. Although for $g\leq 0$, the complex-valued logarithmic function $\tau(g)$ encounters a singularity and branch points, the plaquette and the vertex projectors in Eq.~\eqref{eq:H} remain analytic in $g$ for $g\in (-1,1)$, i.e. all the singularities are removable. We present the details of the derivation in Appendix~\ref{sm:sec:parent_H}. 

At $g = 1$ and $\eta = 1$, we recover $B_p(1,1) = B_p$ as in Eq.~\eqref{eq:fp_h}. The vertex term $Q_v(1) = (1-A_v)(1-X_v)/2$ fixes the spin on the vertex $v$ to be in the state $\ket{+}$ in the ground state. The Hamiltonian is thus the same as the toric code Hamiltonian Eq.~\eqref{eq:fp_h} with the additional vertex terms. For $g = 1$ and $\eta>0$, when removing the $Q_v$ term, the Hamiltonian is a parent Hamiltonian for the toric code ground state with string tension shown in Eq.~\eqref{eq:TC_with_tension}. An alternative parent Hamiltonian is given in Ref.~\cite{Castelnovo:2008}. At $g=-1$ and $\eta = 1$, we recover the fixed-point Hamiltonian for the SET-TC phase: 
\begin{align}\label{sm:eq:set_h}
    B_p(-1,1) &= \frac{1}{2}\left(1-\prod_{e\in p} X_e e^{-i\pi Z_e(1-Z_{v(e)}Z_{v'(e)})/4}\right),
    \nonumber\\
    Q_v(-1) &=\frac{1-A_v}{2}
    \left(1-X_v\prod_{e\in v}e^{i\pi(1-Z_e)Z_{v(e)}Z_{v'(e)}/4}\right).
\end{align}
Note that the projector $(1-A_v)$ in $Q_v(-1)$ is necessary for $Q_v(-1)$ being Hermitian.

The Hamiltonian also has the duality $H(-g,\eta) = UH(g,\eta)U^{\dag} = e^{-i\pi H_{\text{pivot}}/8}H(g,\eta)  e^{i\pi H_{\text{pivot}}/8}$, where $U$ is a finite-depth local quantum circuit given in Eq.~\eqref{Circuit_CCZ} and $H_{\text{pivot}}$ is an example of a pivot Hamiltonian~\cite{Tantivasadakarn:2021}
\begin{equation}
    H_{\text{pivot}}=\sum_{e\in E}(1-Z_e)(1-Z_{v(e)}Z_{v'(e)}),
\end{equation}
where $E$ denotes the set of all the edges.
The Hamiltonians at $g>0$ and $g<0$ thus share the same spectrum. At the line $g = 0$, the Hamiltonian has an enhanced $U(1)$ pivot symmetry generated by $H_{\text{pivot}}$, i.e. $[\exp(i\theta H_{\text{pivot}}),H(0,\eta)]=0, \forall \theta\in\mathbb{R}$, see Appendix~\ref{pivot_symmetry} for the proof. The $U(1)$ symmetry manifests itself in the $O(2)$ criticality along the SET transition line, which we discuss in the next section (see Fig.~\ref{Phase_diagram_TEE_MOP}). 

By tuning the parameter $g$ from $-1$ to $+1$, the system can change from one SET phase to another SET phase. However, an intermediate phase generically exists between the two SET phases. The parameter $\eta$ can be tuned to avoid such an intermediate phase so that a direct transition between the two SET phases is possible.

\section{Phase diagram and order parameters}\label{sec:phase_diagram}

\begin{figure}
    \centering
    \includegraphics[width=8cm]{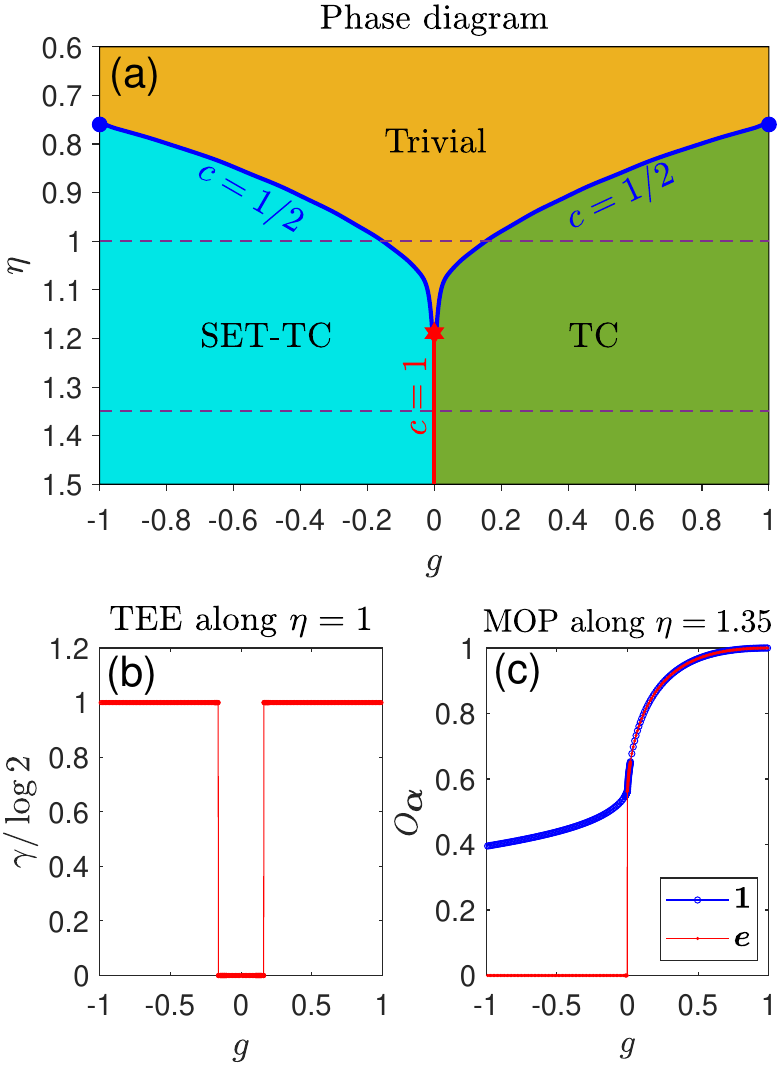}
    \caption{ (a) Phase diagram of the decorated TNS spanned by the string tension $\eta$ and a tuning parameter $g$. The central charge is denoted by $c$. The phase transitions along vertical lines at $g=\pm1$, highlighted by the blue dots, occur at $\eta=3^{-1/4}$. The tricritical point is at $(g,\eta) = (0,2^{1/4})$.  (b) The topological entanglement entropy of a minimally entangled state along $\eta=1$. (c) The membrane order parameters from the minimally entangled states $\pmb{1}$ and $\pmb{e}$. }
    \label{Phase_diagram_TEE_MOP}
\end{figure}

To obtain the phase diagram of the system, we extract the correlation length of the ground state by the corner transfer matrix renormalization group algorithm~\cite{Nishino:1996,Corboz:2014} (see Appendix~\ref{CTM_corr_length} for details), and the resulting phase diagram is shown in Fig~\ref{Phase_diagram_TEE_MOP}a. The system hosts three distinct phases, the SET-TC phase with $\mathbb{Z}_2$ topological order and a non-trivial $\mathbb{Z}_2^{T}$ symmetry fractionalization, the TC phase with $\mathbb{Z}_2$ topological order and trivial symmetry fractionalization, and a totally trivial phase without topological order. 
Note that the norm of each MPS loop inside the wavefunction $\ket{\Psi(g,\eta)}$ contributes weight to the amplitude of the configuration (an explicit expression for the amplitude is given in Appendix~\ref{map_to_classical_models}).

The universality class of the phase boundaries can be determined by mapping the squared norm of the decorated TNS to the partition function of classical statistical models. As shown in Appendix~\ref{map_to_classical_models},
along $g=\pm1$, the decorated TNS can be mapped to the 2D classical Ising model, the two critical points are located at $(g,\eta) = (\pm1,3^{-1/4})$. For $g\neq 0$ and $g\neq \pm 1$, the model is mapped to an anisotropic Ashkin-Teller model (see Appendix~\ref{2d SPT}).  The phase boundary between the TC (SET-TC) phase and the trivial phase is thus described by the $(2+0)$D Ising CFT with a central charge $c=1/2$. Along $g=0$, the decorated TNS can be mapped to the classical $O(2)$ loop model, which has a high-temperature gapped phase and a low-temperature critical phase described by the compactified free boson CFT with central charge $c=1$~\cite{Nienhuis:1982,Batchelor:1989,yellow_book_CFT}. The transition between low- and high-temperature phases at $\eta=2^{1/4}$ is of the Kosterlitz-Thouless (KT) type. Therefore, the phase boundary between the SET-TC and the TC phase, including the tricritical point, has a central charge $c=1$. 

We further characterize these phases using non-local order parameters. The presence of an intrinsic topological order can be detected by the topological entanglement entropy (TEE) \cite{Levin:2006,Kitaev_TEE:2006}. The entanglement entropy of a topological state satisfies $S\sim aN-\gamma$, where $\gamma$ is a universal correction called TEE, $a$ is a non-universal coefficient from the area law, and $N$ is the length of the entanglement bipartition. On a torus, the TEE has to be extracted from the minimally entangled states (MES)~\cite{Zhang:2012}, which are topologically degenerate ground states in a special basis such that the entanglement entropy is minimal. There are four MES $\ket{\Psi_{\pmb{\alpha}}}$ labelled by the anyons $\pmb{\alpha}=\pmb{1},\pmb{e},\pmb{m},\pmb{f}$.

Instead of the von Neumann entropy, we consider the  Renyi entropy, which is easier to calculate using tensor-network methods. As shown in Fig.~\ref{Figure_2}b, for a system on an infinitely long cylinder with circumference $N$, the 
 $n$-Renyi entropy is 
\begin{equation}\label{n-Renyi entropy}
S^{(n)}_{\pmb{\alpha}}=\frac{1}{1-n}\log\Tr(\rho^n_{\pmb{\alpha}}),\quad \rho_{\pmb{\alpha}}=\Tr_R \ket{\Psi_{\pmb{\alpha}}}\bra{\Psi_{\pmb{\alpha}}},
\end{equation}
where $\Tr_R$ is the partial trace over all physical degrees of freedom of the MES $\ket{\Psi_{\pmb{\alpha}}}$ on the right half of the infinite cylinder.
The topological Renyi entropy is independent of $n$~\cite{Flammia:2009}, we choose $n = 2$ for our calculation.
In Appendix~\ref{TEE_TN}, we exploit the tensor-network approach to calculate the TEE $\gamma$ directly in the limit $N\rightarrow\infty$ without extrapolation. The TEE obtained from a boundary MPS with bond dimension $\chi=20$ is shown in Fig~\ref{Phase_diagram_TEE_MOP}b. 
In the SET-TC phase and the TC phase, the four MES $|\Psi_{\pmb{\alpha}}\rangle$ give the same TEE $\gamma=\log 2$ as expected from the $\mathbb{Z}_2$ topological order. In the trivial phase, the ground state of the system becomes unique on a torus and the MES states are no longer well-defined. In this unique ground state, we indeed obtain $\gamma = 0$, indicating the absence of topological order.

As the SET-TC phase and the TC phase share the same TEE, we can further distinguish the two using the membrane order parameter (MOP), which captures the symmetry fractionalization pattern of SET phases~\cite{Huang:2014}. In our case, the system has an additional $\mathbb{Z}_2$ symmetry generated by a global spin flip on the vertices $\prod_v X_v$ (it follows from the additional global spin flip symmetry in the 1D model Eq.~\eqref{SPT_Hamiltonian}). This allows us to define a MOP on an infinitely long cylinder as
\begin{equation}\label{MOP}
     O_{\pmb{\alpha}}=\lim_{N\rightarrow \infty}
     \left(\bra{\Psi_{\pmb{\alpha}}}\prod_{v\in R }X_v\ket{\Psi_{\pmb{\alpha}}}\right)^{1/N},
 \end{equation}
 where $\ket{\Psi_{\pmb{\alpha}}}$ is an MES, $N$ is the circumference of the cylinder, and $R$ is the set of vertices of the right part of the cylinder. It can be shown that the MOP has a selection rule and dictates that $O_{\pmb{\alpha}}=0$ if the symmetry fractionalizes non-trivially on the anyon $\pmb{\alpha}$~\footnote{More precisely, using the additional symmetry $\prod_v X_v$ and the technique from Ref.~\cite{Pollmann:2012}, one can show that the MOP will vanish if only one of the symmetries $K\prod_{v}X_x$ or $K$ fractionalizes over the anyons. Our example belongs to the first case. The symmetry $K$ is not fractionalized, similar to the 1D SPT chain used for the decoration.}. 
 As shown in Appendix \ref{MOP_TN}, the calculation of the MOP is similar to that of the TEE, and we can use tensor-network methods to directly calculate $O_{\pmb{\alpha}}$ in the limit $N\rightarrow\infty$ without extrapolation. Fig.~\ref{Phase_diagram_TEE_MOP}c
shows the MOP obtained from a boundary MPS with $\chi=20$.
Since the symmetry fractionalization on $\pmb{1}$ and $\pmb{e}$ is identical to that on $\pmb{m}$ and $\pmb{f}$, we have $O_{\pmb{1}}=O_{\pmb{m}}$ and $O_{\pmb{e}}=O_{\pmb{f}}$. 
We only show $O_{\pmb{1}}$ and $O_{\pmb{e}}$ in Fig.~\ref{Phase_diagram_TEE_MOP}c. In the TC phase,  $O_{\pmb{1}}$ and $O_{\pmb{e}}$ are non-zero, implying no symmetry fractionalization on the anyons. In the SET-TC phase, $O_{\pmb{e}}$ vanishes, indicating that the symmetry fractionalizes on the $\pmb{e}$ and $\pmb{f}$ anyons. 

An alternative way to distinguish the SET-TC phase from the TC phase is by examining the entanglement spectrum. In the SET-TC phase, the time-reversal symmetry represented by $\mathcal{T}$ on $\rho_{\pmb{1}}$ and $\rho_{\pmb{m}}$ satisfies $\mathcal{T}^2=1$, whereas the time-reversal symmetry on $\rho_{\pmb{e}}$ and $\rho_{\pmb{f}}$ is represented projectively, i.e., $\mathcal{T}^2=-1$, due to symmetry fractionalization, as shown in Appendix~\ref{TEE_TN}. Therefore, from Kramers' theorem, each level of the entanglement spectra in the $\pmb{e}$ and $\pmb{f}$ sectors is even-fold degenerate in the SET-TC phase, which is an important feature inherited from 1D non-trivial SPT states~\cite{Pollmann:2009}.

\section{Discussion and outlook}\label{discussion}

In this work, we construct a family of 2D TNS that corresponds to the exact ground states of $\mathbb{Z}_2^T$-symmetric Hamiltonians. In particular, the system describes a direct continuous quantum phase transition between two distinct SET phases with $\mathbb{Z}_2^T$ time-reversal symmetry. Although we expect that these constructed ground states require fine tuning to be reached, they serve as a useful starting point for a more general understanding of the SET phase transitions.

Along the phase boundary separating the two SET phases, we obtain a particularly interesting class of toy states which are ground states of local Hamiltonians.
For example, one of these states is $\ket{\Psi\left(0,\sqrt{2}\right)}\propto\sum_C 2^{N(C)/2}\ket{C}$, where $C$ labels the configurations of closed loops decorated with GHZ states, and $N(C)$ denotes the total number of loops in $C$. The power-law decay of correlation functions is revealed by non-local operators~\cite{Freedman:2005}. Moreover, these states have an area-law entanglement entropy up to a subleading logarithmic correction~\cite{fradkin:2006}. They serve as interesting examples for studying topological critical phases~\cite{Freedman:2005,scaffidi:2017,verresen:2021}, whose universality is characterized by non-local correlators.

The phase diagram of the system can be further extended. As we discuss in Appendix~\ref{2d SPT}, by introducing Ising couplings to vertex spins, it is  possible to continuously tune the system along a tensor-network solvable path to ferromagnetic or antiferromagnetic phases, where the $\mathbb{Z}_2^T$ symmetry is spontaneously broken. By the quantum-classical mapping mentioned in Sec.~\ref{sec:phase_diagram}, the phase boundaries of these transitions can be shown to align with the critical regimes of an anisotropic Ashkin-Teller model. 

The construction can be straightforwardly generalized to enrich the double-semion model~\cite{Freedman:2004,levin:2005}. When restricted to the closed loop subspace, the toric code model and the double-semion model are related by a diagonal unitary transformation $U_{\text{TC-DS}}=\sum_{C}(-1)^{N(C)}\ket{C}\bra{C}$, where $C$ is a configuration of decorated loops. Because $U_{\text{TC-DS}}$ commutes with the decoration procedure (we state this more precisely in Appendix~\ref{sm:sec:parent_H}), the phase diagram in Fig.~\ref{Phase_diagram_TEE_MOP}a is preserved under the unitary transformation. In the non-trivial SET double-semion phase, the symmetry fractionalizes over the semions and the anti-semions. For the gauge group $\mathbb{Z}_2$ and the global symmetry $\mathbb{Z}_2^T$, the SET classification based on Abelian Chern-Simons theories is given by the third cohomology group $H^3(\mathbb{Z}_2\times \mathbb{Z}_2^T,U(1))=\mathbb{Z}_2\times \mathbb{Z}_2$~\cite{Essin:2013,Lu:2016}. Here the first $\mathbb{Z}_2$ index originates from the Dijkgraaf-Witten classification and it labels two topological orders described by the toric code and double-semion theories. The second $\mathbb{Z}_2$ labels different symmetry fractionalization patterns over the anyons under time-reversal symmetry. Our construction thus generates direct phase transitions between all of those with the same topological order. 

By a similar procedure, decorating the domain walls in 2D $\mathbb{Z}_2$ SPT phases gives rise to SPT phases protected by the symmetry $\mathbb{Z}_2\times\mathbb{Z}_2^T$~\cite{Chen:2014}. By a duality transformation, the SET-TC and the TC phases can be mapped to the 2D $\mathbb{Z}_2\times\mathbb{Z}_2^T$ SPT phases (see Ref.~\cite{Gu:2012} and Appendix~\ref{2d SPT}), the tensor-network solvable phase diagram Fig.~\ref{Phase_diagram_TEE_MOP}a is thus dual to a $\mathbb{Z}_2\times\mathbb{Z}_2^T$-protected phase diagram, where the two SET phases are replaced by two 2D $\mathbb{Z}_2\times\mathbb{Z}_2^T$ SPT phases and the trivial phase is replaced by a ferromagnetic phase in which the $\mathbb{Z}_2$ symmetry is spontaneously broken. 

A key ingredient for the construction is the existence of an MPS path that interpolates between the 1D SPT phases with a constant bond dimension. It will be interesting to apply the proposed construction to the generalization of such MPS paths, such as the MPS skeletons~\cite{Jones:2021}, to obtain a broader class of SET phases and their phase transitions. The simplicity of the TNS description of the ground states raises the question of whether these states admit an efficient quantum circuit representation and are easy to study on a quantum computer, similar to the 1D MPS path~\cite{Smith:2022}. While the SET fixed points may be efficiently prepared~\cite{Satzinger2021,Liu:2022}, the existence of an efficient state preparation near or at the critical points remains an intriguing open question.

\section{Acknowledgement}
Y.-J.L was supported by the Max Planck Gesellschaft (MPG) through the International Max Planck Research School for Quantum Science and Technology (IMPRS-QST). W.-T.X, Y.-J.L. and F.P. acknowledge support from the Deutsche Forschungsgemeinschaft (DFG, German Research Foundation) under Germany’s Excellence Strategy--EXC--2111--390814868 and DFG grants No. KN1254/1-2, KN1254/2-1, the European Research Council (ERC) under the European Union’s Horizon 2020 research and innovation programme (Grant Agreement No. 851161), as well as the Munich Quantum Valley, which is supported by the Bavarian state government with funds from the Hightech Agenda Bayern Plus.

\textbf{Data and materials availability}: Raw data and simulation codes are available in Zenodo~\cite{lukas_haller_2023_7886542}.
\bibliography{SET.bib}

\begin{thebibliography}{76}%
\makeatletter
\providecommand \@ifxundefined [1]{%
 \@ifx{#1\undefined}
}%
\providecommand \@ifnum [1]{%
 \ifnum #1\expandafter \@firstoftwo
 \else \expandafter \@secondoftwo
 \fi
}%
\providecommand \@ifx [1]{%
 \ifx #1\expandafter \@firstoftwo
 \else \expandafter \@secondoftwo
 \fi
}%
\providecommand \natexlab [1]{#1}%
\providecommand \enquote  [1]{``#1''}%
\providecommand \bibnamefont  [1]{#1}%
\providecommand \bibfnamefont [1]{#1}%
\providecommand \citenamefont [1]{#1}%
\providecommand \href@noop [0]{\@secondoftwo}%
\providecommand \href [0]{\begingroup \@sanitize@url \@href}%
\providecommand \@href[1]{\@@startlink{#1}\@@href}%
\providecommand \@@href[1]{\endgroup#1\@@endlink}%
\providecommand \@sanitize@url [0]{\catcode `\\12\catcode `\$12\catcode
  `\&12\catcode `\#12\catcode `\^12\catcode `\_12\catcode `\%12\relax}%
\providecommand \@@startlink[1]{}%
\providecommand \@@endlink[0]{}%
\providecommand \url  [0]{\begingroup\@sanitize@url \@url }%
\providecommand \@url [1]{\endgroup\@href {#1}{\urlprefix }}%
\providecommand \urlprefix  [0]{URL }%
\providecommand \Eprint [0]{\href }%
\providecommand \doibase [0]{https://doi.org/}%
\providecommand \selectlanguage [0]{\@gobble}%
\providecommand \bibinfo  [0]{\@secondoftwo}%
\providecommand \bibfield  [0]{\@secondoftwo}%
\providecommand \translation [1]{[#1]}%
\providecommand \BibitemOpen [0]{}%
\providecommand \bibitemStop [0]{}%
\providecommand \bibitemNoStop [0]{.\EOS\space}%
\providecommand \EOS [0]{\spacefactor3000\relax}%
\providecommand \BibitemShut  [1]{\csname bibitem#1\endcsname}%
\let\auto@bib@innerbib\@empty
\bibitem [{\citenamefont {Tsui}\ \emph {et~al.}(1982)\citenamefont {Tsui},
  \citenamefont {Stormer},\ and\ \citenamefont {Gossard}}]{tsui:1982}%
  \BibitemOpen
  \bibfield  {author} {\bibinfo {author} {\bibfnamefont {D.~C.}\ \bibnamefont
  {Tsui}}, \bibinfo {author} {\bibfnamefont {H.~L.}\ \bibnamefont {Stormer}},\
  and\ \bibinfo {author} {\bibfnamefont {A.~C.}\ \bibnamefont {Gossard}},\
  }\bibfield  {title} {\bibinfo {title} {{Two-dimensional magnetotransport in
  the extreme quantum limit}},\ }\href
  {https://doi.org/10.1103/PhysRevLett.48.1559} {\bibfield  {journal} {\bibinfo
   {journal} {Phys. Rev. Lett.}\ }\textbf {\bibinfo {volume} {48}},\ \bibinfo
  {pages} {1559} (\bibinfo {year} {1982})}\BibitemShut {NoStop}%
\bibitem [{\citenamefont {Dijkgraaf}\ and\ \citenamefont
  {Witten}(1990)}]{dijkgraaf:1990}%
  \BibitemOpen
  \bibfield  {author} {\bibinfo {author} {\bibfnamefont {R.}~\bibnamefont
  {Dijkgraaf}}\ and\ \bibinfo {author} {\bibfnamefont {E.}~\bibnamefont
  {Witten}},\ }\bibfield  {title} {\bibinfo {title} {Topological gauge theories
  and group cohomology},\ }\href
  {https://link.springer.com/content/pdf/10.1007/BF02096988.pdf} {\bibfield
  {journal} {\bibinfo  {journal} {Communications in Mathematical Physics}\
  }\textbf {\bibinfo {volume} {129}},\ \bibinfo {pages} {393} (\bibinfo {year}
  {1990})}\BibitemShut {NoStop}%
\bibitem [{\citenamefont {Wen}(1990)}]{WEN:1990}%
  \BibitemOpen
  \bibfield  {author} {\bibinfo {author} {\bibfnamefont {X.~G.}\ \bibnamefont
  {Wen}},\ }\bibfield  {title} {\bibinfo {title} {{Topological orders in rigid
  states}},\ }\href {https://doi.org/10.1142/s0217979290000139} {\bibfield
  {journal} {\bibinfo  {journal} {Int. J. Mod. Phys. B}\ }\textbf {\bibinfo
  {volume} {04}},\ \bibinfo {pages} {239} (\bibinfo {year} {1990})}\BibitemShut
  {NoStop}%
\bibitem [{\citenamefont {Wen}(2017)}]{Wen:2017}%
  \BibitemOpen
  \bibfield  {author} {\bibinfo {author} {\bibfnamefont {X.-G.}\ \bibnamefont
  {Wen}},\ }\bibfield  {title} {\bibinfo {title} {Colloquium: Zoo of
  quantum-topological phases of matter},\ }\href
  {https://doi.org/10.1103/RevModPhys.89.041004} {\bibfield  {journal}
  {\bibinfo  {journal} {Rev. Mod. Phys.}\ }\textbf {\bibinfo {volume} {89}},\
  \bibinfo {pages} {041004} (\bibinfo {year} {2017})}\BibitemShut {NoStop}%
\bibitem [{\citenamefont {Gu}\ and\ \citenamefont {Wen}(2009)}]{gu:2009}%
  \BibitemOpen
  \bibfield  {author} {\bibinfo {author} {\bibfnamefont {Z.-C.}\ \bibnamefont
  {Gu}}\ and\ \bibinfo {author} {\bibfnamefont {X.-G.}\ \bibnamefont {Wen}},\
  }\bibfield  {title} {\bibinfo {title} {Tensor-entanglement-filtering
  renormalization approach and symmetry-protected topological order},\ }\href
  {https://doi.org/10.1103/PhysRevB.80.155131} {\bibfield  {journal} {\bibinfo
  {journal} {Phys. Rev. B}\ }\textbf {\bibinfo {volume} {80}},\ \bibinfo
  {pages} {155131} (\bibinfo {year} {2009})}\BibitemShut {NoStop}%
\bibitem [{\citenamefont {Pollmann}\ \emph {et~al.}(2010)\citenamefont
  {Pollmann}, \citenamefont {Turner}, \citenamefont {Berg},\ and\ \citenamefont
  {Oshikawa}}]{Pollmann:2009}%
  \BibitemOpen
  \bibfield  {author} {\bibinfo {author} {\bibfnamefont {F.}~\bibnamefont
  {Pollmann}}, \bibinfo {author} {\bibfnamefont {A.~M.}\ \bibnamefont
  {Turner}}, \bibinfo {author} {\bibfnamefont {E.}~\bibnamefont {Berg}},\ and\
  \bibinfo {author} {\bibfnamefont {M.}~\bibnamefont {Oshikawa}},\ }\bibfield
  {title} {\bibinfo {title} {Entanglement spectrum of a topological phase in
  one dimension},\ }\href {https://doi.org/10.1103/PhysRevB.81.064439}
  {\bibfield  {journal} {\bibinfo  {journal} {Phys. Rev. B}\ }\textbf {\bibinfo
  {volume} {81}},\ \bibinfo {pages} {064439} (\bibinfo {year}
  {2010})}\BibitemShut {NoStop}%
\bibitem [{\citenamefont {Chen}\ \emph
  {et~al.}(2011{\natexlab{a}})\citenamefont {Chen}, \citenamefont {Gu},\ and\
  \citenamefont {Wen}}]{chen:2011b}%
  \BibitemOpen
  \bibfield  {author} {\bibinfo {author} {\bibfnamefont {X.}~\bibnamefont
  {Chen}}, \bibinfo {author} {\bibfnamefont {Z.-C.}\ \bibnamefont {Gu}},\ and\
  \bibinfo {author} {\bibfnamefont {X.-G.}\ \bibnamefont {Wen}},\ }\bibfield
  {title} {\bibinfo {title} {Classification of gapped symmetric phases in
  one-dimensional spin systems},\ }\href
  {https://doi.org/10.1103/PhysRevB.83.035107} {\bibfield  {journal} {\bibinfo
  {journal} {Phys. Rev. B}\ }\textbf {\bibinfo {volume} {83}},\ \bibinfo
  {pages} {035107} (\bibinfo {year} {2011}{\natexlab{a}})}\BibitemShut
  {NoStop}%
\bibitem [{\citenamefont {Schuch}\ \emph {et~al.}(2011)\citenamefont {Schuch},
  \citenamefont {P\'erez-Garc\'{\i}a},\ and\ \citenamefont
  {Cirac}}]{Pschuch:2011}%
  \BibitemOpen
  \bibfield  {author} {\bibinfo {author} {\bibfnamefont {N.}~\bibnamefont
  {Schuch}}, \bibinfo {author} {\bibfnamefont {D.}~\bibnamefont
  {P\'erez-Garc\'{\i}a}},\ and\ \bibinfo {author} {\bibfnamefont
  {I.}~\bibnamefont {Cirac}},\ }\bibfield  {title} {\bibinfo {title}
  {Classifying quantum phases using matrix product states and projected
  entangled pair states},\ }\href {https://doi.org/10.1103/PhysRevB.84.165139}
  {\bibfield  {journal} {\bibinfo  {journal} {Phys. Rev. B}\ }\textbf {\bibinfo
  {volume} {84}},\ \bibinfo {pages} {165139} (\bibinfo {year}
  {2011})}\BibitemShut {NoStop}%
\bibitem [{\citenamefont {Chen}\ \emph
  {et~al.}(2011{\natexlab{b}})\citenamefont {Chen}, \citenamefont {Gu},\ and\
  \citenamefont {Wen}}]{chen:2011a}%
  \BibitemOpen
  \bibfield  {author} {\bibinfo {author} {\bibfnamefont {X.}~\bibnamefont
  {Chen}}, \bibinfo {author} {\bibfnamefont {Z.-C.}\ \bibnamefont {Gu}},\ and\
  \bibinfo {author} {\bibfnamefont {X.-G.}\ \bibnamefont {Wen}},\ }\bibfield
  {title} {\bibinfo {title} {Complete classification of one-dimensional gapped
  quantum phases in interacting spin systems},\ }\href
  {https://doi.org/10.1103/PhysRevB.84.235128} {\bibfield  {journal} {\bibinfo
  {journal} {Phys. Rev. B}\ }\textbf {\bibinfo {volume} {84}},\ \bibinfo
  {pages} {235128} (\bibinfo {year} {2011}{\natexlab{b}})}\BibitemShut
  {NoStop}%
\bibitem [{\citenamefont {Pollmann}\ \emph {et~al.}(2012)\citenamefont
  {Pollmann}, \citenamefont {Berg}, \citenamefont {Turner},\ and\ \citenamefont
  {Oshikawa}}]{Pollmann:2012a}%
  \BibitemOpen
  \bibfield  {author} {\bibinfo {author} {\bibfnamefont {F.}~\bibnamefont
  {Pollmann}}, \bibinfo {author} {\bibfnamefont {E.}~\bibnamefont {Berg}},
  \bibinfo {author} {\bibfnamefont {A.~M.}\ \bibnamefont {Turner}},\ and\
  \bibinfo {author} {\bibfnamefont {M.}~\bibnamefont {Oshikawa}},\ }\bibfield
  {title} {\bibinfo {title} {Symmetry protection of topological phases in
  one-dimensional quantum spin systems},\ }\href
  {https://doi.org/10.1103/PhysRevB.85.075125} {\bibfield  {journal} {\bibinfo
  {journal} {Phys. Rev. B}\ }\textbf {\bibinfo {volume} {85}},\ \bibinfo
  {pages} {075125} (\bibinfo {year} {2012})}\BibitemShut {NoStop}%
\bibitem [{\citenamefont {Chen}\ \emph {et~al.}(2013)\citenamefont {Chen},
  \citenamefont {Gu}, \citenamefont {Liu},\ and\ \citenamefont
  {Wen}}]{chen:2013}%
  \BibitemOpen
  \bibfield  {author} {\bibinfo {author} {\bibfnamefont {X.}~\bibnamefont
  {Chen}}, \bibinfo {author} {\bibfnamefont {Z.-C.}\ \bibnamefont {Gu}},
  \bibinfo {author} {\bibfnamefont {Z.-X.}\ \bibnamefont {Liu}},\ and\ \bibinfo
  {author} {\bibfnamefont {X.-G.}\ \bibnamefont {Wen}},\ }\bibfield  {title}
  {\bibinfo {title} {Symmetry protected topological orders and the group
  cohomology of their symmetry group},\ }\href
  {https://doi.org/10.1103/PhysRevB.87.155114} {\bibfield  {journal} {\bibinfo
  {journal} {Phys. Rev. B}\ }\textbf {\bibinfo {volume} {87}},\ \bibinfo
  {pages} {155114} (\bibinfo {year} {2013})}\BibitemShut {NoStop}%
\bibitem [{\citenamefont {Essin}\ and\ \citenamefont
  {Hermele}(2013)}]{Essin:2013}%
  \BibitemOpen
  \bibfield  {author} {\bibinfo {author} {\bibfnamefont {A.~M.}\ \bibnamefont
  {Essin}}\ and\ \bibinfo {author} {\bibfnamefont {M.}~\bibnamefont
  {Hermele}},\ }\bibfield  {title} {\bibinfo {title} {Classifying
  fractionalization: Symmetry classification of gapped {${\mathbb{Z}}_{2}$}
  spin liquids in two dimensions},\ }\href
  {https://doi.org/10.1103/PhysRevB.87.104406} {\bibfield  {journal} {\bibinfo
  {journal} {Phys. Rev. B}\ }\textbf {\bibinfo {volume} {87}},\ \bibinfo
  {pages} {104406} (\bibinfo {year} {2013})}\BibitemShut {NoStop}%
\bibitem [{\citenamefont {Mesaros}\ and\ \citenamefont
  {Ran}(2013)}]{Mesaros:2013}%
  \BibitemOpen
  \bibfield  {author} {\bibinfo {author} {\bibfnamefont {A.}~\bibnamefont
  {Mesaros}}\ and\ \bibinfo {author} {\bibfnamefont {Y.}~\bibnamefont {Ran}},\
  }\bibfield  {title} {\bibinfo {title} {Classification of symmetry enriched
  topological phases with exactly solvable models},\ }\href
  {https://doi.org/10.1103/PhysRevB.87.155115} {\bibfield  {journal} {\bibinfo
  {journal} {Phys. Rev. B}\ }\textbf {\bibinfo {volume} {87}},\ \bibinfo
  {pages} {155115} (\bibinfo {year} {2013})}\BibitemShut {NoStop}%
\bibitem [{\citenamefont {Laughlin}(1983)}]{laughlin:1983}%
  \BibitemOpen
  \bibfield  {author} {\bibinfo {author} {\bibfnamefont {R.~B.}\ \bibnamefont
  {Laughlin}},\ }\bibfield  {title} {\bibinfo {title} {Anomalous quantum hall
  effect: An incompressible quantum fluid with fractionally charged
  excitations},\ }\href {https://doi.org/10.1103/PhysRevLett.50.1395}
  {\bibfield  {journal} {\bibinfo  {journal} {Phys. Rev. Lett.}\ }\textbf
  {\bibinfo {volume} {50}},\ \bibinfo {pages} {1395} (\bibinfo {year}
  {1983})}\BibitemShut {NoStop}%
\bibitem [{\citenamefont {Tarantino}\ \emph {et~al.}(2016)\citenamefont
  {Tarantino}, \citenamefont {Lindner},\ and\ \citenamefont
  {Fidkowski}}]{Tarantino:2016}%
  \BibitemOpen
  \bibfield  {author} {\bibinfo {author} {\bibfnamefont {N.}~\bibnamefont
  {Tarantino}}, \bibinfo {author} {\bibfnamefont {N.~H.}\ \bibnamefont
  {Lindner}},\ and\ \bibinfo {author} {\bibfnamefont {L.}~\bibnamefont
  {Fidkowski}},\ }\bibfield  {title} {\bibinfo {title} {Symmetry
  fractionalization and twist defects},\ }\href
  {https://doi.org/10.1088/1367-2630/18/3/035006} {\bibfield  {journal}
  {\bibinfo  {journal} {New Journal of Physics}\ }\textbf {\bibinfo {volume}
  {18}},\ \bibinfo {pages} {035006} (\bibinfo {year} {2016})}\BibitemShut
  {NoStop}%
\bibitem [{\citenamefont {Lu}\ and\ \citenamefont
  {Vishwanath}(2016)}]{Lu:2016}%
  \BibitemOpen
  \bibfield  {author} {\bibinfo {author} {\bibfnamefont {Y.-M.}\ \bibnamefont
  {Lu}}\ and\ \bibinfo {author} {\bibfnamefont {A.}~\bibnamefont
  {Vishwanath}},\ }\bibfield  {title} {\bibinfo {title} {Classification and
  properties of symmetry-enriched topological phases: Chern-simons approach
  with applications to ${Z}_{2}$ spin liquids},\ }\href
  {https://doi.org/10.1103/PhysRevB.93.155121} {\bibfield  {journal} {\bibinfo
  {journal} {Phys. Rev. B}\ }\textbf {\bibinfo {volume} {93}},\ \bibinfo
  {pages} {155121} (\bibinfo {year} {2016})}\BibitemShut {NoStop}%
\bibitem [{\citenamefont {Heinrich}\ \emph {et~al.}(2016)\citenamefont
  {Heinrich}, \citenamefont {Burnell}, \citenamefont {Fidkowski},\ and\
  \citenamefont {Levin}}]{Heinrich:2016}%
  \BibitemOpen
  \bibfield  {author} {\bibinfo {author} {\bibfnamefont {C.}~\bibnamefont
  {Heinrich}}, \bibinfo {author} {\bibfnamefont {F.}~\bibnamefont {Burnell}},
  \bibinfo {author} {\bibfnamefont {L.}~\bibnamefont {Fidkowski}},\ and\
  \bibinfo {author} {\bibfnamefont {M.}~\bibnamefont {Levin}},\ }\bibfield
  {title} {\bibinfo {title} {Symmetry-enriched string nets: Exactly solvable
  models for set phases},\ }\href {https://doi.org/10.1103/PhysRevB.94.235136}
  {\bibfield  {journal} {\bibinfo  {journal} {Phys. Rev. B}\ }\textbf {\bibinfo
  {volume} {94}},\ \bibinfo {pages} {235136} (\bibinfo {year}
  {2016})}\BibitemShut {NoStop}%
\bibitem [{\citenamefont {Cheng}\ \emph {et~al.}(2017)\citenamefont {Cheng},
  \citenamefont {Gu}, \citenamefont {Jiang},\ and\ \citenamefont
  {Qi}}]{Meng:2017}%
  \BibitemOpen
  \bibfield  {author} {\bibinfo {author} {\bibfnamefont {M.}~\bibnamefont
  {Cheng}}, \bibinfo {author} {\bibfnamefont {Z.-C.}\ \bibnamefont {Gu}},
  \bibinfo {author} {\bibfnamefont {S.}~\bibnamefont {Jiang}},\ and\ \bibinfo
  {author} {\bibfnamefont {Y.}~\bibnamefont {Qi}},\ }\bibfield  {title}
  {\bibinfo {title} {Exactly solvable models for symmetry-enriched topological
  phases},\ }\href {https://doi.org/10.1103/PhysRevB.96.115107} {\bibfield
  {journal} {\bibinfo  {journal} {Phys. Rev. B}\ }\textbf {\bibinfo {volume}
  {96}},\ \bibinfo {pages} {115107} (\bibinfo {year} {2017})}\BibitemShut
  {NoStop}%
\bibitem [{\citenamefont {Barkeshli}\ \emph {et~al.}(2019)\citenamefont
  {Barkeshli}, \citenamefont {Bonderson}, \citenamefont {Cheng},\ and\
  \citenamefont {Wang}}]{Barkeshli:2019}%
  \BibitemOpen
  \bibfield  {author} {\bibinfo {author} {\bibfnamefont {M.}~\bibnamefont
  {Barkeshli}}, \bibinfo {author} {\bibfnamefont {P.}~\bibnamefont
  {Bonderson}}, \bibinfo {author} {\bibfnamefont {M.}~\bibnamefont {Cheng}},\
  and\ \bibinfo {author} {\bibfnamefont {Z.}~\bibnamefont {Wang}},\ }\bibfield
  {title} {\bibinfo {title} {Symmetry fractionalization, defects, and gauging
  of topological phases},\ }\href {https://doi.org/10.1103/PhysRevB.100.115147}
  {\bibfield  {journal} {\bibinfo  {journal} {Phys. Rev. B}\ }\textbf {\bibinfo
  {volume} {100}},\ \bibinfo {pages} {115147} (\bibinfo {year}
  {2019})}\BibitemShut {NoStop}%
\bibitem [{\citenamefont {Aasen}\ \emph {et~al.}(2021)\citenamefont {Aasen},
  \citenamefont {Bonderson},\ and\ \citenamefont {Knapp}}]{aasen:2021}%
  \BibitemOpen
  \bibfield  {author} {\bibinfo {author} {\bibfnamefont {D.}~\bibnamefont
  {Aasen}}, \bibinfo {author} {\bibfnamefont {P.}~\bibnamefont {Bonderson}},\
  and\ \bibinfo {author} {\bibfnamefont {C.}~\bibnamefont {Knapp}},\ }\bibfield
   {title} {\bibinfo {title} {Characterization and classification of fermionic
  symmetry enriched topological phases},\ }\href
  {https://arxiv.org/abs/2109.10911} {\bibfield  {journal} {\bibinfo  {journal}
  {arXiv preprint arXiv:2109.10911}\ } (\bibinfo {year} {2021})}\BibitemShut
  {NoStop}%
\bibitem [{\citenamefont {Barkeshli}\ \emph {et~al.}(2022)\citenamefont
  {Barkeshli}, \citenamefont {Chen}, \citenamefont {Hsin},\ and\ \citenamefont
  {Manjunath}}]{barkeshli:2022}%
  \BibitemOpen
  \bibfield  {author} {\bibinfo {author} {\bibfnamefont {M.}~\bibnamefont
  {Barkeshli}}, \bibinfo {author} {\bibfnamefont {Y.-A.}\ \bibnamefont {Chen}},
  \bibinfo {author} {\bibfnamefont {P.-S.}\ \bibnamefont {Hsin}},\ and\
  \bibinfo {author} {\bibfnamefont {N.}~\bibnamefont {Manjunath}},\ }\bibfield
  {title} {\bibinfo {title} {Classification of $(2+1)$d invertible fermionic
  topological phases with symmetry},\ }\href
  {https://doi.org/10.1103/PhysRevB.105.235143} {\bibfield  {journal} {\bibinfo
   {journal} {Phys. Rev. B}\ }\textbf {\bibinfo {volume} {105}},\ \bibinfo
  {pages} {235143} (\bibinfo {year} {2022})}\BibitemShut {NoStop}%
\bibitem [{\citenamefont {Bais}\ and\ \citenamefont
  {Slingerland}(2009)}]{bais:2009}%
  \BibitemOpen
  \bibfield  {author} {\bibinfo {author} {\bibfnamefont {F.~A.}\ \bibnamefont
  {Bais}}\ and\ \bibinfo {author} {\bibfnamefont {J.~K.}\ \bibnamefont
  {Slingerland}},\ }\bibfield  {title} {\bibinfo {title} {Condensate-induced
  transitions between topologically ordered phases},\ }\href
  {https://doi.org/10.1103/PhysRevB.79.045316} {\bibfield  {journal} {\bibinfo
  {journal} {Phys. Rev. B}\ }\textbf {\bibinfo {volume} {79}},\ \bibinfo
  {pages} {045316} (\bibinfo {year} {2009})}\BibitemShut {NoStop}%
\bibitem [{\citenamefont {Garre-Rubio}\ \emph {et~al.}(2017)\citenamefont
  {Garre-Rubio}, \citenamefont {Iblisdir},\ and\ \citenamefont
  {P\'erez-Garc\'{\i}a}}]{Garre-Rubio:2017}%
  \BibitemOpen
  \bibfield  {author} {\bibinfo {author} {\bibfnamefont {J.}~\bibnamefont
  {Garre-Rubio}}, \bibinfo {author} {\bibfnamefont {S.}~\bibnamefont
  {Iblisdir}},\ and\ \bibinfo {author} {\bibfnamefont {D.}~\bibnamefont
  {P\'erez-Garc\'{\i}a}},\ }\bibfield  {title} {\bibinfo {title} {Symmetry
  reduction induced by anyon condensation: A tensor network approach},\ }\href
  {https://doi.org/10.1103/PhysRevB.96.155123} {\bibfield  {journal} {\bibinfo
  {journal} {Phys. Rev. B}\ }\textbf {\bibinfo {volume} {96}},\ \bibinfo
  {pages} {155123} (\bibinfo {year} {2017})}\BibitemShut {NoStop}%
\bibitem [{\citenamefont {Williamson}\ \emph {et~al.}(2017)\citenamefont
  {Williamson}, \citenamefont {Bultinck},\ and\ \citenamefont
  {Verstraete}}]{williamson:2017}%
  \BibitemOpen
  \bibfield  {author} {\bibinfo {author} {\bibfnamefont {D.~J.}\ \bibnamefont
  {Williamson}}, \bibinfo {author} {\bibfnamefont {N.}~\bibnamefont
  {Bultinck}},\ and\ \bibinfo {author} {\bibfnamefont {F.}~\bibnamefont
  {Verstraete}},\ }\bibfield  {title} {\bibinfo {title} {Symmetry-enriched
  topological order in tensor networks: Defects, gauging and anyon
  condensation},\ }\href {https://doi.org/10.48550/arXiv.1711.07982} {\bibfield
   {journal} {\bibinfo  {journal} {arXiv preprint arXiv:1711.07982}\ }
  (\bibinfo {year} {2017})}\BibitemShut {NoStop}%
\bibitem [{\citenamefont {Levin}\ and\ \citenamefont
  {Gu}(2012{\natexlab{a}})}]{levin:2012}%
  \BibitemOpen
  \bibfield  {author} {\bibinfo {author} {\bibfnamefont {M.}~\bibnamefont
  {Levin}}\ and\ \bibinfo {author} {\bibfnamefont {Z.-C.}\ \bibnamefont {Gu}},\
  }\bibfield  {title} {\bibinfo {title} {Braiding statistics approach to
  symmetry-protected topological phases},\ }\href
  {https://doi.org/10.1103/PhysRevB.86.115109} {\bibfield  {journal} {\bibinfo
  {journal} {Phys. Rev. B}\ }\textbf {\bibinfo {volume} {86}},\ \bibinfo
  {pages} {115109} (\bibinfo {year} {2012}{\natexlab{a}})}\BibitemShut
  {NoStop}%
\bibitem [{\citenamefont {Kitaev}(2003)}]{Kitaev:2003}%
  \BibitemOpen
  \bibfield  {author} {\bibinfo {author} {\bibfnamefont {A.~Y.}\ \bibnamefont
  {Kitaev}},\ }\bibfield  {title} {\bibinfo {title} {{Fault-tolerant quantum
  computation by anyons}},\ }\href
  {https://doi.org/10.1016/S0003-4916(02)00018-0} {\bibfield  {journal}
  {\bibinfo  {journal} {Ann. Phys. (N. Y).}\ }\textbf {\bibinfo {volume}
  {303}},\ \bibinfo {pages} {2} (\bibinfo {year} {2003})}\BibitemShut {NoStop}%
\bibitem [{\citenamefont {Chen}\ \emph {et~al.}(2014)\citenamefont {Chen},
  \citenamefont {Lu},\ and\ \citenamefont {Vishwanath}}]{Chen:2014}%
  \BibitemOpen
  \bibfield  {author} {\bibinfo {author} {\bibfnamefont {X.}~\bibnamefont
  {Chen}}, \bibinfo {author} {\bibfnamefont {Y.-M.}\ \bibnamefont {Lu}},\ and\
  \bibinfo {author} {\bibfnamefont {A.}~\bibnamefont {Vishwanath}},\ }\bibfield
   {title} {\bibinfo {title} {Symmetry-protected topological phases from
  decorated domain walls},\ }\href {https://doi.org/10.1038/ncomms4507}
  {\bibfield  {journal} {\bibinfo  {journal} {Nature communications}\ }\textbf
  {\bibinfo {volume} {5}},\ \bibinfo {pages} {3507} (\bibinfo {year}
  {2014})}\BibitemShut {NoStop}%
\bibitem [{\citenamefont {Li}\ \emph {et~al.}(2014)\citenamefont {Li},
  \citenamefont {Yang}, \citenamefont {Cheng}, \citenamefont {Liu},\ and\
  \citenamefont {Tu}}]{Li:2014}%
  \BibitemOpen
  \bibfield  {author} {\bibinfo {author} {\bibfnamefont {W.}~\bibnamefont
  {Li}}, \bibinfo {author} {\bibfnamefont {S.}~\bibnamefont {Yang}}, \bibinfo
  {author} {\bibfnamefont {M.}~\bibnamefont {Cheng}}, \bibinfo {author}
  {\bibfnamefont {Z.-X.}\ \bibnamefont {Liu}},\ and\ \bibinfo {author}
  {\bibfnamefont {H.-H.}\ \bibnamefont {Tu}},\ }\bibfield  {title} {\bibinfo
  {title} {Topology and criticality in the resonating
  affleck-kennedy-lieb-tasaki loop spin liquid states},\ }\href
  {https://doi.org/10.1103/PhysRevB.89.174411} {\bibfield  {journal} {\bibinfo
  {journal} {Phys. Rev. B}\ }\textbf {\bibinfo {volume} {89}},\ \bibinfo
  {pages} {174411} (\bibinfo {year} {2014})}\BibitemShut {NoStop}%
\bibitem [{\citenamefont {Huang}\ \emph {et~al.}(2014)\citenamefont {Huang},
  \citenamefont {Chen},\ and\ \citenamefont {Pollmann}}]{Huang:2014}%
  \BibitemOpen
  \bibfield  {author} {\bibinfo {author} {\bibfnamefont {C.-Y.}\ \bibnamefont
  {Huang}}, \bibinfo {author} {\bibfnamefont {X.}~\bibnamefont {Chen}},\ and\
  \bibinfo {author} {\bibfnamefont {F.}~\bibnamefont {Pollmann}},\ }\bibfield
  {title} {\bibinfo {title} {Detection of symmetry-enriched topological
  phases},\ }\href {https://doi.org/10.1103/PhysRevB.90.045142} {\bibfield
  {journal} {\bibinfo  {journal} {Phys. Rev. B}\ }\textbf {\bibinfo {volume}
  {90}},\ \bibinfo {pages} {045142} (\bibinfo {year} {2014})}\BibitemShut
  {NoStop}%
\bibitem [{\citenamefont {Ben-Zion}\ \emph {et~al.}(2016)\citenamefont
  {Ben-Zion}, \citenamefont {Das},\ and\ \citenamefont
  {McGreevy}}]{Ben-Zion:2016}%
  \BibitemOpen
  \bibfield  {author} {\bibinfo {author} {\bibfnamefont {D.}~\bibnamefont
  {Ben-Zion}}, \bibinfo {author} {\bibfnamefont {D.}~\bibnamefont {Das}},\ and\
  \bibinfo {author} {\bibfnamefont {J.}~\bibnamefont {McGreevy}},\ }\bibfield
  {title} {\bibinfo {title} {Exactly solvable models of spin liquids with
  spinons, and of three-dimensional topological paramagnets},\ }\href
  {https://doi.org/10.1103/PhysRevB.93.155147} {\bibfield  {journal} {\bibinfo
  {journal} {Phys. Rev. B}\ }\textbf {\bibinfo {volume} {93}},\ \bibinfo
  {pages} {155147} (\bibinfo {year} {2016})}\BibitemShut {NoStop}%
\bibitem [{\citenamefont {Maeshima}\ \emph {et~al.}(2001)\citenamefont
  {Maeshima}, \citenamefont {Hieida}, \citenamefont {Akutsu}, \citenamefont
  {Nishino},\ and\ \citenamefont {Okunishi}}]{maeshima:2001}%
  \BibitemOpen
  \bibfield  {author} {\bibinfo {author} {\bibfnamefont {N.}~\bibnamefont
  {Maeshima}}, \bibinfo {author} {\bibfnamefont {Y.}~\bibnamefont {Hieida}},
  \bibinfo {author} {\bibfnamefont {Y.}~\bibnamefont {Akutsu}}, \bibinfo
  {author} {\bibfnamefont {T.}~\bibnamefont {Nishino}},\ and\ \bibinfo {author}
  {\bibfnamefont {K.}~\bibnamefont {Okunishi}},\ }\bibfield  {title} {\bibinfo
  {title} {Vertical density matrix algorithm: A higher-dimensional numerical
  renormalization scheme based on the tensor product state ansatz},\ }\href
  {https://doi.org/10.1103/PhysRevE.64.016705} {\bibfield  {journal} {\bibinfo
  {journal} {Phys. Rev. E}\ }\textbf {\bibinfo {volume} {64}},\ \bibinfo
  {pages} {016705} (\bibinfo {year} {2001})}\BibitemShut {NoStop}%
\bibitem [{\citenamefont {Verstraete}\ and\ \citenamefont
  {Cirac}(2004)}]{verstraete2004}%
  \BibitemOpen
  \bibfield  {author} {\bibinfo {author} {\bibfnamefont {F.}~\bibnamefont
  {Verstraete}}\ and\ \bibinfo {author} {\bibfnamefont {J.~I.}\ \bibnamefont
  {Cirac}},\ }\bibfield  {title} {\bibinfo {title} {Renormalization algorithms
  for quantum-many body systems in two and higher dimensions},\ }\href
  {https://doi.org/10.48550/arXiv.cond-mat/0407066} {\bibfield  {journal}
  {\bibinfo  {journal} {arXiv preprint cond-mat/0407066}\ } (\bibinfo {year}
  {2004})}\BibitemShut {NoStop}%
\bibitem [{\citenamefont {Levin}\ and\ \citenamefont {Wen}(2006)}]{Levin:2006}%
  \BibitemOpen
  \bibfield  {author} {\bibinfo {author} {\bibfnamefont {M.}~\bibnamefont
  {Levin}}\ and\ \bibinfo {author} {\bibfnamefont {X.-G.}\ \bibnamefont
  {Wen}},\ }\bibfield  {title} {\bibinfo {title} {Detecting topological order
  in a ground state wave function},\ }\href
  {https://doi.org/10.1103/PhysRevLett.96.110405} {\bibfield  {journal}
  {\bibinfo  {journal} {Phys. Rev. Lett.}\ }\textbf {\bibinfo {volume} {96}},\
  \bibinfo {pages} {110405} (\bibinfo {year} {2006})}\BibitemShut {NoStop}%
\bibitem [{\citenamefont {Kitaev}\ and\ \citenamefont
  {Preskill}(2006)}]{Kitaev_TEE:2006}%
  \BibitemOpen
  \bibfield  {author} {\bibinfo {author} {\bibfnamefont {A.}~\bibnamefont
  {Kitaev}}\ and\ \bibinfo {author} {\bibfnamefont {J.}~\bibnamefont
  {Preskill}},\ }\bibfield  {title} {\bibinfo {title} {Topological entanglement
  entropy},\ }\href {https://doi.org/10.1103/PhysRevLett.96.110404} {\bibfield
  {journal} {\bibinfo  {journal} {Phys. Rev. Lett.}\ }\textbf {\bibinfo
  {volume} {96}},\ \bibinfo {pages} {110404} (\bibinfo {year}
  {2006})}\BibitemShut {NoStop}%
\bibitem [{\citenamefont {Tantivasadakarn}\ \emph {et~al.}(2023)\citenamefont
  {Tantivasadakarn}, \citenamefont {Thorngren}, \citenamefont {Vishwanath},\
  and\ \citenamefont {Verresen}}]{Tantivasadakarn:2021}%
  \BibitemOpen
  \bibfield  {author} {\bibinfo {author} {\bibfnamefont {N.}~\bibnamefont
  {Tantivasadakarn}}, \bibinfo {author} {\bibfnamefont {R.}~\bibnamefont
  {Thorngren}}, \bibinfo {author} {\bibfnamefont {A.}~\bibnamefont
  {Vishwanath}},\ and\ \bibinfo {author} {\bibfnamefont {R.}~\bibnamefont
  {Verresen}},\ }\bibfield  {title} {\bibinfo {title} {Pivot hamiltonians as
  generators of symmetry and entanglement},\ }\href
  {https://scipost.org/SciPostPhys.14.2.012} {\bibfield  {journal} {\bibinfo
  {journal} {SciPost Physics}\ }\textbf {\bibinfo {volume} {14}},\ \bibinfo
  {pages} {012} (\bibinfo {year} {2023})}\BibitemShut {NoStop}%
\bibitem [{\citenamefont {Rokhsar}\ and\ \citenamefont
  {Kivelson}(1988)}]{rokhsar:1988}%
  \BibitemOpen
  \bibfield  {author} {\bibinfo {author} {\bibfnamefont {D.~S.}\ \bibnamefont
  {Rokhsar}}\ and\ \bibinfo {author} {\bibfnamefont {S.~A.}\ \bibnamefont
  {Kivelson}},\ }\bibfield  {title} {\bibinfo {title} {Superconductivity and
  the quantum hard-core dimer gas},\ }\href
  {https://doi.org/10.1103/PhysRevLett.61.2376} {\bibfield  {journal} {\bibinfo
   {journal} {Phys. Rev. Lett.}\ }\textbf {\bibinfo {volume} {61}},\ \bibinfo
  {pages} {2376} (\bibinfo {year} {1988})}\BibitemShut {NoStop}%
\bibitem [{\citenamefont {Ardonne}\ \emph {et~al.}(2004)\citenamefont
  {Ardonne}, \citenamefont {Fendley},\ and\ \citenamefont
  {Fradkin}}]{Ardonne:2004}%
  \BibitemOpen
  \bibfield  {author} {\bibinfo {author} {\bibfnamefont {E.}~\bibnamefont
  {Ardonne}}, \bibinfo {author} {\bibfnamefont {P.}~\bibnamefont {Fendley}},\
  and\ \bibinfo {author} {\bibfnamefont {E.}~\bibnamefont {Fradkin}},\
  }\bibfield  {title} {\bibinfo {title} {Topological order and conformal
  quantum critical points},\ }\href
  {https://doi.org/https://doi.org/10.1016/j.aop.2004.01.004} {\bibfield
  {journal} {\bibinfo  {journal} {Annals of Physics}\ }\textbf {\bibinfo
  {volume} {310}},\ \bibinfo {pages} {493} (\bibinfo {year}
  {2004})}\BibitemShut {NoStop}%
\bibitem [{\citenamefont {Carr}(2010)}]{castelnovo:2010}%
  \BibitemOpen
  \bibfield  {author} {\bibinfo {author} {\bibfnamefont {L.}~\bibnamefont
  {Carr}},\ }\href {https://doi.org/10.1201/b10273-10} {\emph {\bibinfo {title}
  {Understanding quantum phase transitions}}}\ (\bibinfo  {publisher} {CRC
  press},\ \bibinfo {year} {2010})\ pp.\ \bibinfo {pages}
  {119--142}\BibitemShut {NoStop}%
\bibitem [{\citenamefont {Verstraete}\ \emph {et~al.}(2006)\citenamefont
  {Verstraete}, \citenamefont {Wolf}, \citenamefont {Perez-Garcia},\ and\
  \citenamefont {Cirac}}]{Verstraete:2006}%
  \BibitemOpen
  \bibfield  {author} {\bibinfo {author} {\bibfnamefont {F.}~\bibnamefont
  {Verstraete}}, \bibinfo {author} {\bibfnamefont {M.~M.}\ \bibnamefont
  {Wolf}}, \bibinfo {author} {\bibfnamefont {D.}~\bibnamefont {Perez-Garcia}},\
  and\ \bibinfo {author} {\bibfnamefont {J.~I.}\ \bibnamefont {Cirac}},\
  }\bibfield  {title} {\bibinfo {title} {Criticality, the area law, and the
  computational power of projected entangled pair states},\ }\href
  {https://doi.org/10.1103/PhysRevLett.96.220601} {\bibfield  {journal}
  {\bibinfo  {journal} {Phys. Rev. Lett.}\ }\textbf {\bibinfo {volume} {96}},\
  \bibinfo {pages} {220601} (\bibinfo {year} {2006})}\BibitemShut {NoStop}%
\bibitem [{\citenamefont {Xu}\ and\ \citenamefont {Zhang}(2018)}]{xu:2018}%
  \BibitemOpen
  \bibfield  {author} {\bibinfo {author} {\bibfnamefont {W.-T.}\ \bibnamefont
  {Xu}}\ and\ \bibinfo {author} {\bibfnamefont {G.-M.}\ \bibnamefont {Zhang}},\
  }\bibfield  {title} {\bibinfo {title} {Tensor network state approach to
  quantum topological phase transitions and their criticalities of
  {${\mathbb{Z}}_{2}$} topologically ordered states},\ }\href
  {https://doi.org/10.1103/PhysRevB.98.165115} {\bibfield  {journal} {\bibinfo
  {journal} {Phys. Rev. B}\ }\textbf {\bibinfo {volume} {98}},\ \bibinfo
  {pages} {165115} (\bibinfo {year} {2018})}\BibitemShut {NoStop}%
\bibitem [{\citenamefont {Zhu}\ and\ \citenamefont {Zhang}(2019)}]{zhu:2019}%
  \BibitemOpen
  \bibfield  {author} {\bibinfo {author} {\bibfnamefont {G.-Y.}\ \bibnamefont
  {Zhu}}\ and\ \bibinfo {author} {\bibfnamefont {G.-M.}\ \bibnamefont
  {Zhang}},\ }\bibfield  {title} {\bibinfo {title} {Gapless coulomb state
  emerging from a self-dual topological tensor-network state},\ }\href
  {https://doi.org/10.1103/PhysRevLett.122.176401} {\bibfield  {journal}
  {\bibinfo  {journal} {Phys. Rev. Lett.}\ }\textbf {\bibinfo {volume} {122}},\
  \bibinfo {pages} {176401} (\bibinfo {year} {2019})}\BibitemShut {NoStop}%
\bibitem [{\citenamefont {Xu}\ \emph {et~al.}(2020)\citenamefont {Xu},
  \citenamefont {Zhang},\ and\ \citenamefont {Zhang}}]{xu:2020}%
  \BibitemOpen
  \bibfield  {author} {\bibinfo {author} {\bibfnamefont {W.-T.}\ \bibnamefont
  {Xu}}, \bibinfo {author} {\bibfnamefont {Q.}~\bibnamefont {Zhang}},\ and\
  \bibinfo {author} {\bibfnamefont {G.-M.}\ \bibnamefont {Zhang}},\ }\bibfield
  {title} {\bibinfo {title} {Tensor network approach to phase transitions of a
  non-abelian topological phase},\ }\href
  {https://doi.org/10.1103/PhysRevLett.124.130603} {\bibfield  {journal}
  {\bibinfo  {journal} {Phys. Rev. Lett.}\ }\textbf {\bibinfo {volume} {124}},\
  \bibinfo {pages} {130603} (\bibinfo {year} {2020})}\BibitemShut {NoStop}%
\bibitem [{\citenamefont {Zhang}\ \emph {et~al.}(2020)\citenamefont {Zhang},
  \citenamefont {Xu}, \citenamefont {Wang},\ and\ \citenamefont
  {Zhang}}]{zhang:2020}%
  \BibitemOpen
  \bibfield  {author} {\bibinfo {author} {\bibfnamefont {Q.}~\bibnamefont
  {Zhang}}, \bibinfo {author} {\bibfnamefont {W.-T.}\ \bibnamefont {Xu}},
  \bibinfo {author} {\bibfnamefont {Z.-Q.}\ \bibnamefont {Wang}},\ and\
  \bibinfo {author} {\bibfnamefont {G.-M.}\ \bibnamefont {Zhang}},\ }\bibfield
  {title} {\bibinfo {title} {Non-hermitian effects of the intrinsic signs in
  topologically ordered wavefunctions},\ }\href
  {https://doi.org/10.1038/s42005-020-00479-y} {\bibfield  {journal} {\bibinfo
  {journal} {Communications Physics}\ }\textbf {\bibinfo {volume} {3}},\
  \bibinfo {pages} {209} (\bibinfo {year} {2020})}\BibitemShut {NoStop}%
\bibitem [{\citenamefont {Xu}\ and\ \citenamefont {Schuch}(2021)}]{Xu:2021}%
  \BibitemOpen
  \bibfield  {author} {\bibinfo {author} {\bibfnamefont {W.-T.}\ \bibnamefont
  {Xu}}\ and\ \bibinfo {author} {\bibfnamefont {N.}~\bibnamefont {Schuch}},\
  }\bibfield  {title} {\bibinfo {title} {Characterization of topological phase
  transitions from a non-abelian topological state and its galois conjugate
  through condensation and confinement order parameters},\ }\href
  {https://doi.org/10.1103/PhysRevB.104.155119} {\bibfield  {journal} {\bibinfo
   {journal} {Phys. Rev. B}\ }\textbf {\bibinfo {volume} {104}},\ \bibinfo
  {pages} {155119} (\bibinfo {year} {2021})}\BibitemShut {NoStop}%
\bibitem [{\citenamefont {Xu}\ \emph {et~al.}(2022)\citenamefont {Xu},
  \citenamefont {Garre-Rubio},\ and\ \citenamefont {Schuch}}]{Xu:2022}%
  \BibitemOpen
  \bibfield  {author} {\bibinfo {author} {\bibfnamefont {W.-T.}\ \bibnamefont
  {Xu}}, \bibinfo {author} {\bibfnamefont {J.}~\bibnamefont {Garre-Rubio}},\
  and\ \bibinfo {author} {\bibfnamefont {N.}~\bibnamefont {Schuch}},\
  }\bibfield  {title} {\bibinfo {title} {Complete characterization of
  non-abelian topological phase transitions and detection of anyon splitting
  with projected entangled pair states},\ }\href
  {https://doi.org/10.1103/PhysRevB.106.205139} {\bibfield  {journal} {\bibinfo
   {journal} {Phys. Rev. B}\ }\textbf {\bibinfo {volume} {106}},\ \bibinfo
  {pages} {205139} (\bibinfo {year} {2022})}\BibitemShut {NoStop}%
\bibitem [{\citenamefont {Freedman}\ \emph {et~al.}(2004)\citenamefont
  {Freedman}, \citenamefont {Nayak}, \citenamefont {Shtengel}, \citenamefont
  {Walker},\ and\ \citenamefont {Wang}}]{Freedman:2004}%
  \BibitemOpen
  \bibfield  {author} {\bibinfo {author} {\bibfnamefont {M.}~\bibnamefont
  {Freedman}}, \bibinfo {author} {\bibfnamefont {C.}~\bibnamefont {Nayak}},
  \bibinfo {author} {\bibfnamefont {K.}~\bibnamefont {Shtengel}}, \bibinfo
  {author} {\bibfnamefont {K.}~\bibnamefont {Walker}},\ and\ \bibinfo {author}
  {\bibfnamefont {Z.}~\bibnamefont {Wang}},\ }\bibfield  {title} {\bibinfo
  {title} {A class of p,t-invariant topological phases of interacting
  electrons},\ }\href
  {https://doi.org/https://doi.org/10.1016/j.aop.2004.01.006} {\bibfield
  {journal} {\bibinfo  {journal} {Ann. Phys.}\ }\textbf {\bibinfo {volume}
  {310}},\ \bibinfo {pages} {428} (\bibinfo {year} {2004})}\BibitemShut
  {NoStop}%
\bibitem [{\citenamefont {Levin}\ and\ \citenamefont {Wen}(2005)}]{levin:2005}%
  \BibitemOpen
  \bibfield  {author} {\bibinfo {author} {\bibfnamefont {M.~A.}\ \bibnamefont
  {Levin}}\ and\ \bibinfo {author} {\bibfnamefont {X.-G.}\ \bibnamefont
  {Wen}},\ }\bibfield  {title} {\bibinfo {title} {{String-net condensation: A
  physical mechanism for topological phases}},\ }\href
  {https://doi.org/10.1103/PhysRevB.71.045110} {\bibfield  {journal} {\bibinfo
  {journal} {Phys. Rev. B}\ }\textbf {\bibinfo {volume} {71}},\ \bibinfo
  {pages} {045110} (\bibinfo {year} {2005})}\BibitemShut {NoStop}%
\bibitem [{\citenamefont {Wolf}\ \emph {et~al.}(2006)\citenamefont {Wolf},
  \citenamefont {Ortiz}, \citenamefont {Verstraete},\ and\ \citenamefont
  {Cirac}}]{Wolf:2006-MPS_phase_transition}%
  \BibitemOpen
  \bibfield  {author} {\bibinfo {author} {\bibfnamefont {M.~M.}\ \bibnamefont
  {Wolf}}, \bibinfo {author} {\bibfnamefont {G.}~\bibnamefont {Ortiz}},
  \bibinfo {author} {\bibfnamefont {F.}~\bibnamefont {Verstraete}},\ and\
  \bibinfo {author} {\bibfnamefont {J.~I.}\ \bibnamefont {Cirac}},\ }\bibfield
  {title} {\bibinfo {title} {Quantum phase transitions in matrix product
  systems},\ }\href {https://doi.org/10.1103/PhysRevLett.97.110403} {\bibfield
  {journal} {\bibinfo  {journal} {Phys. Rev. Lett.}\ }\textbf {\bibinfo
  {volume} {97}},\ \bibinfo {pages} {110403} (\bibinfo {year}
  {2006})}\BibitemShut {NoStop}%
\bibitem [{\citenamefont {Jones}\ \emph {et~al.}(2021)\citenamefont {Jones},
  \citenamefont {Bibo}, \citenamefont {Jobst}, \citenamefont {Pollmann},
  \citenamefont {Smith},\ and\ \citenamefont {Verresen}}]{Jones:2021}%
  \BibitemOpen
  \bibfield  {author} {\bibinfo {author} {\bibfnamefont {N.~G.}\ \bibnamefont
  {Jones}}, \bibinfo {author} {\bibfnamefont {J.}~\bibnamefont {Bibo}},
  \bibinfo {author} {\bibfnamefont {B.}~\bibnamefont {Jobst}}, \bibinfo
  {author} {\bibfnamefont {F.}~\bibnamefont {Pollmann}}, \bibinfo {author}
  {\bibfnamefont {A.}~\bibnamefont {Smith}},\ and\ \bibinfo {author}
  {\bibfnamefont {R.}~\bibnamefont {Verresen}},\ }\bibfield  {title} {\bibinfo
  {title} {Skeleton of matrix-product-state-solvable models connecting
  topological phases of matter},\ }\href
  {https://doi.org/10.1103/PhysRevResearch.3.033265} {\bibfield  {journal}
  {\bibinfo  {journal} {Phys. Rev. Research}\ }\textbf {\bibinfo {volume}
  {3}},\ \bibinfo {pages} {033265} (\bibinfo {year} {2021})}\BibitemShut
  {NoStop}%
\bibitem [{\citenamefont {Castelnovo}\ and\ \citenamefont
  {Chamon}(2008)}]{Castelnovo:2008}%
  \BibitemOpen
  \bibfield  {author} {\bibinfo {author} {\bibfnamefont {C.}~\bibnamefont
  {Castelnovo}}\ and\ \bibinfo {author} {\bibfnamefont {C.}~\bibnamefont
  {Chamon}},\ }\bibfield  {title} {\bibinfo {title} {Quantum topological phase
  transition at the microscopic level},\ }\href
  {https://doi.org/10.1103/PhysRevB.77.054433} {\bibfield  {journal} {\bibinfo
  {journal} {Phys. Rev. B}\ }\textbf {\bibinfo {volume} {77}},\ \bibinfo
  {pages} {054433} (\bibinfo {year} {2008})}\BibitemShut {NoStop}%
\bibitem [{\citenamefont {Castelnovo}\ \emph {et~al.}(2009)\citenamefont
  {Castelnovo}, \citenamefont {Trebst},\ and\ \citenamefont
  {Troyer}}]{castelnovo:2009}%
  \BibitemOpen
  \bibfield  {author} {\bibinfo {author} {\bibfnamefont {C.}~\bibnamefont
  {Castelnovo}}, \bibinfo {author} {\bibfnamefont {S.}~\bibnamefont {Trebst}},\
  and\ \bibinfo {author} {\bibfnamefont {M.}~\bibnamefont {Troyer}},\
  }\bibfield  {title} {\bibinfo {title} {Topological order and quantum
  criticality},\ }\href {https://arxiv.org/abs/0912.3272v2} {\bibfield
  {journal} {\bibinfo  {journal} {arXiv preprint arXiv:0912.3272}\ } (\bibinfo
  {year} {2009})}\BibitemShut {NoStop}%
\bibitem [{\citenamefont {Gu}\ \emph {et~al.}(2009)\citenamefont {Gu},
  \citenamefont {Levin}, \citenamefont {Swingle},\ and\ \citenamefont
  {Wen}}]{Gu_PEPS_rep_2009}%
  \BibitemOpen
  \bibfield  {author} {\bibinfo {author} {\bibfnamefont {Z.-C.}\ \bibnamefont
  {Gu}}, \bibinfo {author} {\bibfnamefont {M.}~\bibnamefont {Levin}}, \bibinfo
  {author} {\bibfnamefont {B.}~\bibnamefont {Swingle}},\ and\ \bibinfo {author}
  {\bibfnamefont {X.-G.}\ \bibnamefont {Wen}},\ }\bibfield  {title} {\bibinfo
  {title} {Tensor-product representations for string-net condensed states},\
  }\href {https://doi.org/10.1103/PhysRevB.79.085118} {\bibfield  {journal}
  {\bibinfo  {journal} {Phys. Rev. B}\ }\textbf {\bibinfo {volume} {79}},\
  \bibinfo {pages} {085118} (\bibinfo {year} {2009})}\BibitemShut {NoStop}%
\bibitem [{\citenamefont {Garre-Rubio}\ \emph {et~al.}(2021)\citenamefont
  {Garre-Rubio}, \citenamefont {Iqbal},\ and\ \citenamefont
  {Stephen}}]{garre:2021}%
  \BibitemOpen
  \bibfield  {author} {\bibinfo {author} {\bibfnamefont {J.}~\bibnamefont
  {Garre-Rubio}}, \bibinfo {author} {\bibfnamefont {M.}~\bibnamefont {Iqbal}},\
  and\ \bibinfo {author} {\bibfnamefont {D.~T.}\ \bibnamefont {Stephen}},\
  }\bibfield  {title} {\bibinfo {title} {String order parameters for symmetry
  fractionalization in an enriched toric code},\ }\href
  {https://doi.org/10.1103/PhysRevB.103.125104} {\bibfield  {journal} {\bibinfo
   {journal} {Phys. Rev. B}\ }\textbf {\bibinfo {volume} {103}},\ \bibinfo
  {pages} {125104} (\bibinfo {year} {2021})}\BibitemShut {NoStop}%
\bibitem [{\citenamefont {Gu}\ \emph {et~al.}(2008)\citenamefont {Gu},
  \citenamefont {Levin},\ and\ \citenamefont {Wen}}]{Gu:2008}%
  \BibitemOpen
  \bibfield  {author} {\bibinfo {author} {\bibfnamefont {Z.-C.}\ \bibnamefont
  {Gu}}, \bibinfo {author} {\bibfnamefont {M.}~\bibnamefont {Levin}},\ and\
  \bibinfo {author} {\bibfnamefont {X.-G.}\ \bibnamefont {Wen}},\ }\bibfield
  {title} {\bibinfo {title} {Tensor-entanglement renormalization group approach
  as a unified method for symmetry breaking and topological phase
  transitions},\ }\href {https://doi.org/10.1103/PhysRevB.78.205116} {\bibfield
   {journal} {\bibinfo  {journal} {Phys. Rev. B}\ }\textbf {\bibinfo {volume}
  {78}},\ \bibinfo {pages} {205116} (\bibinfo {year} {2008})}\BibitemShut
  {NoStop}%
\bibitem [{\citenamefont {Nishino}\ and\ \citenamefont
  {Okunishi}(1996)}]{Nishino:1996}%
  \BibitemOpen
  \bibfield  {author} {\bibinfo {author} {\bibfnamefont {T.}~\bibnamefont
  {Nishino}}\ and\ \bibinfo {author} {\bibfnamefont {K.}~\bibnamefont
  {Okunishi}},\ }\bibfield  {title} {\bibinfo {title} {Corner transfer matrix
  renormalization group method},\ }\href {https://doi.org/10.1143/JPSJ.65.891}
  {\bibfield  {journal} {\bibinfo  {journal} {Journal of the Physical Society
  of Japan}\ }\textbf {\bibinfo {volume} {65}},\ \bibinfo {pages} {891}
  (\bibinfo {year} {1996})}\BibitemShut {NoStop}%
\bibitem [{\citenamefont {Corboz}\ \emph {et~al.}(2014)\citenamefont {Corboz},
  \citenamefont {Rice},\ and\ \citenamefont {Troyer}}]{Corboz:2014}%
  \BibitemOpen
  \bibfield  {author} {\bibinfo {author} {\bibfnamefont {P.}~\bibnamefont
  {Corboz}}, \bibinfo {author} {\bibfnamefont {T.~M.}\ \bibnamefont {Rice}},\
  and\ \bibinfo {author} {\bibfnamefont {M.}~\bibnamefont {Troyer}},\
  }\bibfield  {title} {\bibinfo {title} {Competing states in the $t$-$j$ model:
  Uniform $d$-wave state versus stripe state},\ }\href
  {https://doi.org/10.1103/PhysRevLett.113.046402} {\bibfield  {journal}
  {\bibinfo  {journal} {Phys. Rev. Lett.}\ }\textbf {\bibinfo {volume} {113}},\
  \bibinfo {pages} {046402} (\bibinfo {year} {2014})}\BibitemShut {NoStop}%
\bibitem [{\citenamefont {Nienhuis}(1982)}]{Nienhuis:1982}%
  \BibitemOpen
  \bibfield  {author} {\bibinfo {author} {\bibfnamefont {B.}~\bibnamefont
  {Nienhuis}},\ }\bibfield  {title} {\bibinfo {title} {Exact critical point and
  critical exponents of $\mathrm{O}(n)$ models in two dimensions},\ }\href
  {https://doi.org/10.1103/PhysRevLett.49.1062} {\bibfield  {journal} {\bibinfo
   {journal} {Phys. Rev. Lett.}\ }\textbf {\bibinfo {volume} {49}},\ \bibinfo
  {pages} {1062} (\bibinfo {year} {1982})}\BibitemShut {NoStop}%
\bibitem [{\citenamefont {Batchelor}\ and\ \citenamefont
  {Bl\"ote}(1989)}]{Batchelor:1989}%
  \BibitemOpen
  \bibfield  {author} {\bibinfo {author} {\bibfnamefont {M.~T.}\ \bibnamefont
  {Batchelor}}\ and\ \bibinfo {author} {\bibfnamefont {H.~W.~J.}\ \bibnamefont
  {Bl\"ote}},\ }\bibfield  {title} {\bibinfo {title} {Conformal invariance and
  critical behavior of the o(n) model on the honeycomb lattice},\ }\href
  {https://doi.org/10.1103/PhysRevB.39.2391} {\bibfield  {journal} {\bibinfo
  {journal} {Phys. Rev. B}\ }\textbf {\bibinfo {volume} {39}},\ \bibinfo
  {pages} {2391} (\bibinfo {year} {1989})}\BibitemShut {NoStop}%
\bibitem [{\citenamefont {Philippe Di~Francesco}(1997)}]{yellow_book_CFT}%
  \BibitemOpen
  \bibfield  {author} {\bibinfo {author} {\bibfnamefont {D.~S.}\ \bibnamefont
  {Philippe Di~Francesco}, \bibfnamefont {Pierre~Mathieu}},\ }\href@noop {}
  {\emph {\bibinfo {title} {Conformal Field Theory}}}\ (\bibinfo  {publisher}
  {Springer-Verlag New York, Inc},\ \bibinfo {year} {1997})\BibitemShut
  {NoStop}%
\bibitem [{\citenamefont {Zhang}\ \emph {et~al.}(2012)\citenamefont {Zhang},
  \citenamefont {Grover}, \citenamefont {Turner}, \citenamefont {Oshikawa},\
  and\ \citenamefont {Vishwanath}}]{Zhang:2012}%
  \BibitemOpen
  \bibfield  {author} {\bibinfo {author} {\bibfnamefont {Y.}~\bibnamefont
  {Zhang}}, \bibinfo {author} {\bibfnamefont {T.}~\bibnamefont {Grover}},
  \bibinfo {author} {\bibfnamefont {A.}~\bibnamefont {Turner}}, \bibinfo
  {author} {\bibfnamefont {M.}~\bibnamefont {Oshikawa}},\ and\ \bibinfo
  {author} {\bibfnamefont {A.}~\bibnamefont {Vishwanath}},\ }\bibfield  {title}
  {\bibinfo {title} {Quasiparticle statistics and braiding from ground-state
  entanglement},\ }\href {https://doi.org/10.1103/PhysRevB.85.235151}
  {\bibfield  {journal} {\bibinfo  {journal} {Phys. Rev. B}\ }\textbf {\bibinfo
  {volume} {85}},\ \bibinfo {pages} {235151} (\bibinfo {year}
  {2012})}\BibitemShut {NoStop}%
\bibitem [{\citenamefont {Flammia}\ \emph {et~al.}(2009)\citenamefont
  {Flammia}, \citenamefont {Hamma}, \citenamefont {Hughes},\ and\ \citenamefont
  {Wen}}]{Flammia:2009}%
  \BibitemOpen
  \bibfield  {author} {\bibinfo {author} {\bibfnamefont {S.~T.}\ \bibnamefont
  {Flammia}}, \bibinfo {author} {\bibfnamefont {A.}~\bibnamefont {Hamma}},
  \bibinfo {author} {\bibfnamefont {T.~L.}\ \bibnamefont {Hughes}},\ and\
  \bibinfo {author} {\bibfnamefont {X.-G.}\ \bibnamefont {Wen}},\ }\bibfield
  {title} {\bibinfo {title} {Topological entanglement r\'enyi entropy and
  reduced density matrix structure},\ }\href
  {https://doi.org/10.1103/PhysRevLett.103.261601} {\bibfield  {journal}
  {\bibinfo  {journal} {Phys. Rev. Lett.}\ }\textbf {\bibinfo {volume} {103}},\
  \bibinfo {pages} {261601} (\bibinfo {year} {2009})}\BibitemShut {NoStop}%
\bibitem [{\citenamefont {Pollmann}\ and\ \citenamefont
  {Turner}(2012)}]{Pollmann:2012}%
  \BibitemOpen
  \bibfield  {author} {\bibinfo {author} {\bibfnamefont {F.}~\bibnamefont
  {Pollmann}}\ and\ \bibinfo {author} {\bibfnamefont {A.~M.}\ \bibnamefont
  {Turner}},\ }\bibfield  {title} {\bibinfo {title} {Detection of
  symmetry-protected topological phases in one dimension},\ }\href
  {https://doi.org/10.1103/PhysRevB.86.125441} {\bibfield  {journal} {\bibinfo
  {journal} {Phys. Rev. B}\ }\textbf {\bibinfo {volume} {86}},\ \bibinfo
  {pages} {125441} (\bibinfo {year} {2012})}\BibitemShut {NoStop}%
\bibitem [{\citenamefont {Freedman}\ \emph {et~al.}(2005)\citenamefont
  {Freedman}, \citenamefont {Nayak},\ and\ \citenamefont
  {Shtengel}}]{Freedman:2005}%
  \BibitemOpen
  \bibfield  {author} {\bibinfo {author} {\bibfnamefont {M.}~\bibnamefont
  {Freedman}}, \bibinfo {author} {\bibfnamefont {C.}~\bibnamefont {Nayak}},\
  and\ \bibinfo {author} {\bibfnamefont {K.}~\bibnamefont {Shtengel}},\
  }\bibfield  {title} {\bibinfo {title} {Line of critical points in $2+1$
  dimensions: Quantum critical loop gases and non-abelian gauge theory},\
  }\href {https://doi.org/10.1103/PhysRevLett.94.147205} {\bibfield  {journal}
  {\bibinfo  {journal} {Phys. Rev. Lett.}\ }\textbf {\bibinfo {volume} {94}},\
  \bibinfo {pages} {147205} (\bibinfo {year} {2005})}\BibitemShut {NoStop}%
\bibitem [{\citenamefont {Fradkin}\ and\ \citenamefont
  {Moore}(2006)}]{fradkin:2006}%
  \BibitemOpen
  \bibfield  {author} {\bibinfo {author} {\bibfnamefont {E.}~\bibnamefont
  {Fradkin}}\ and\ \bibinfo {author} {\bibfnamefont {J.~E.}\ \bibnamefont
  {Moore}},\ }\bibfield  {title} {\bibinfo {title} {Entanglement entropy of 2d
  conformal quantum critical points: Hearing the shape of a quantum drum},\
  }\href {https://doi.org/10.1103/PhysRevLett.97.050404} {\bibfield  {journal}
  {\bibinfo  {journal} {Phys. Rev. Lett.}\ }\textbf {\bibinfo {volume} {97}},\
  \bibinfo {pages} {050404} (\bibinfo {year} {2006})}\BibitemShut {NoStop}%
\bibitem [{\citenamefont {Scaffidi}\ \emph {et~al.}(2017)\citenamefont
  {Scaffidi}, \citenamefont {Parker},\ and\ \citenamefont
  {Vasseur}}]{scaffidi:2017}%
  \BibitemOpen
  \bibfield  {author} {\bibinfo {author} {\bibfnamefont {T.}~\bibnamefont
  {Scaffidi}}, \bibinfo {author} {\bibfnamefont {D.~E.}\ \bibnamefont
  {Parker}},\ and\ \bibinfo {author} {\bibfnamefont {R.}~\bibnamefont
  {Vasseur}},\ }\bibfield  {title} {\bibinfo {title} {Gapless
  symmetry-protected topological order},\ }\href
  {https://doi.org/10.1103/PhysRevX.7.041048} {\bibfield  {journal} {\bibinfo
  {journal} {Phys. Rev. X}\ }\textbf {\bibinfo {volume} {7}},\ \bibinfo {pages}
  {041048} (\bibinfo {year} {2017})}\BibitemShut {NoStop}%
\bibitem [{\citenamefont {Verresen}\ \emph {et~al.}(2021)\citenamefont
  {Verresen}, \citenamefont {Thorngren}, \citenamefont {Jones},\ and\
  \citenamefont {Pollmann}}]{verresen:2021}%
  \BibitemOpen
  \bibfield  {author} {\bibinfo {author} {\bibfnamefont {R.}~\bibnamefont
  {Verresen}}, \bibinfo {author} {\bibfnamefont {R.}~\bibnamefont {Thorngren}},
  \bibinfo {author} {\bibfnamefont {N.~G.}\ \bibnamefont {Jones}},\ and\
  \bibinfo {author} {\bibfnamefont {F.}~\bibnamefont {Pollmann}},\ }\bibfield
  {title} {\bibinfo {title} {Gapless topological phases and symmetry-enriched
  quantum criticality},\ }\href {https://doi.org/10.1103/PhysRevX.11.041059}
  {\bibfield  {journal} {\bibinfo  {journal} {Phys. Rev. X}\ }\textbf {\bibinfo
  {volume} {11}},\ \bibinfo {pages} {041059} (\bibinfo {year}
  {2021})}\BibitemShut {NoStop}%
\bibitem [{\citenamefont {Levin}\ and\ \citenamefont
  {Gu}(2012{\natexlab{b}})}]{Gu:2012}%
  \BibitemOpen
  \bibfield  {author} {\bibinfo {author} {\bibfnamefont {M.}~\bibnamefont
  {Levin}}\ and\ \bibinfo {author} {\bibfnamefont {Z.-C.}\ \bibnamefont {Gu}},\
  }\bibfield  {title} {\bibinfo {title} {Braiding statistics approach to
  symmetry-protected topological phases},\ }\href
  {https://doi.org/10.1103/PhysRevB.86.115109} {\bibfield  {journal} {\bibinfo
  {journal} {Phys. Rev. B}\ }\textbf {\bibinfo {volume} {86}},\ \bibinfo
  {pages} {115109} (\bibinfo {year} {2012}{\natexlab{b}})}\BibitemShut
  {NoStop}%
\bibitem [{\citenamefont {Smith}\ \emph {et~al.}(2022)\citenamefont {Smith},
  \citenamefont {Jobst}, \citenamefont {Green},\ and\ \citenamefont
  {Pollmann}}]{Smith:2022}%
  \BibitemOpen
  \bibfield  {author} {\bibinfo {author} {\bibfnamefont {A.}~\bibnamefont
  {Smith}}, \bibinfo {author} {\bibfnamefont {B.}~\bibnamefont {Jobst}},
  \bibinfo {author} {\bibfnamefont {A.~G.}\ \bibnamefont {Green}},\ and\
  \bibinfo {author} {\bibfnamefont {F.}~\bibnamefont {Pollmann}},\ }\bibfield
  {title} {\bibinfo {title} {Crossing a topological phase transition with a
  quantum computer},\ }\href
  {https://doi.org/10.1103/PhysRevResearch.4.L022020} {\bibfield  {journal}
  {\bibinfo  {journal} {Phys. Rev. Res.}\ }\textbf {\bibinfo {volume} {4}},\
  \bibinfo {pages} {L022020} (\bibinfo {year} {2022})}\BibitemShut {NoStop}%
\bibitem [{\citenamefont {Satzinger}\ \emph {et~al.}(2021)\citenamefont
  {Satzinger}, \citenamefont {Liu}, \citenamefont {Smith}, \citenamefont
  {Knapp}, \citenamefont {Newman}, \citenamefont {Jones}, \citenamefont {Chen},
  \citenamefont {Quintana}, \citenamefont {Mi}, \citenamefont {Dunsworth} \emph
  {et~al.}}]{Satzinger2021}%
  \BibitemOpen
  \bibfield  {author} {\bibinfo {author} {\bibfnamefont {K.~J.}\ \bibnamefont
  {Satzinger}}, \bibinfo {author} {\bibfnamefont {Y.-J.}\ \bibnamefont {Liu}},
  \bibinfo {author} {\bibfnamefont {A.}~\bibnamefont {Smith}}, \bibinfo
  {author} {\bibfnamefont {C.}~\bibnamefont {Knapp}}, \bibinfo {author}
  {\bibfnamefont {M.}~\bibnamefont {Newman}}, \bibinfo {author} {\bibfnamefont
  {C.}~\bibnamefont {Jones}}, \bibinfo {author} {\bibfnamefont
  {Z.}~\bibnamefont {Chen}}, \bibinfo {author} {\bibfnamefont {C.}~\bibnamefont
  {Quintana}}, \bibinfo {author} {\bibfnamefont {X.}~\bibnamefont {Mi}},
  \bibinfo {author} {\bibfnamefont {A.}~\bibnamefont {Dunsworth}}, \emph
  {et~al.},\ }\bibfield  {title} {\bibinfo {title} {Realizing topologically
  ordered states on a quantum processor},\ }\href
  {https://doi.org/10.1126/science.abi8378} {\bibfield  {journal} {\bibinfo
  {journal} {Science}\ }\textbf {\bibinfo {volume} {374}},\ \bibinfo {pages}
  {1237} (\bibinfo {year} {2021})}\BibitemShut {NoStop}%
\bibitem [{\citenamefont {Liu}\ \emph {et~al.}(2022)\citenamefont {Liu},
  \citenamefont {Shtengel}, \citenamefont {Smith},\ and\ \citenamefont
  {Pollmann}}]{Liu:2022}%
  \BibitemOpen
  \bibfield  {author} {\bibinfo {author} {\bibfnamefont {Y.-J.}\ \bibnamefont
  {Liu}}, \bibinfo {author} {\bibfnamefont {K.}~\bibnamefont {Shtengel}},
  \bibinfo {author} {\bibfnamefont {A.}~\bibnamefont {Smith}},\ and\ \bibinfo
  {author} {\bibfnamefont {F.}~\bibnamefont {Pollmann}},\ }\bibfield  {title}
  {\bibinfo {title} {Methods for simulating string-net states and anyons on a
  digital quantum computer},\ }\href
  {https://doi.org/10.1103/PRXQuantum.3.040315} {\bibfield  {journal} {\bibinfo
   {journal} {PRX Quantum}\ }\textbf {\bibinfo {volume} {3}},\ \bibinfo {pages}
  {040315} (\bibinfo {year} {2022})}\BibitemShut {NoStop}%
\bibitem [{\citenamefont {Haller}\ \emph {et~al.}(2023)\citenamefont {Haller},
  \citenamefont {Xu}, \citenamefont {Liu},\ and\ \citenamefont
  {Pollmann}}]{lukas_haller_2023_7886542}%
  \BibitemOpen
  \bibfield  {author} {\bibinfo {author} {\bibfnamefont {L.}~\bibnamefont
  {Haller}}, \bibinfo {author} {\bibfnamefont {W.-T.}\ \bibnamefont {Xu}},
  \bibinfo {author} {\bibfnamefont {Y.-J.}\ \bibnamefont {Liu}},\ and\ \bibinfo
  {author} {\bibfnamefont {F.}~\bibnamefont {Pollmann}},\ }\href
  {https://doi.org/10.5281/zenodo.7886542} {\bibinfo {title} {{Quantum Phase
  Transition between Symmetry Enriched Topological Phases in Tensor-Network
  States}}} (\bibinfo {year} {2023})\BibitemShut {NoStop}%
\bibitem [{\citenamefont {{\c{S}}ahino{\u{g}}lu}\ \emph
  {et~al.}(2021)\citenamefont {{\c{S}}ahino{\u{g}}lu}, \citenamefont
  {Williamson}, \citenamefont {Bultinck}, \citenamefont {Mari{\"e}n},
  \citenamefont {Haegeman}, \citenamefont {Schuch},\ and\ \citenamefont
  {Verstraete}}]{sah2021}%
  \BibitemOpen
  \bibfield  {author} {\bibinfo {author} {\bibfnamefont {M.~B.}\ \bibnamefont
  {{\c{S}}ahino{\u{g}}lu}}, \bibinfo {author} {\bibfnamefont {D.}~\bibnamefont
  {Williamson}}, \bibinfo {author} {\bibfnamefont {N.}~\bibnamefont
  {Bultinck}}, \bibinfo {author} {\bibfnamefont {M.}~\bibnamefont
  {Mari{\"e}n}}, \bibinfo {author} {\bibfnamefont {J.}~\bibnamefont
  {Haegeman}}, \bibinfo {author} {\bibfnamefont {N.}~\bibnamefont {Schuch}},\
  and\ \bibinfo {author} {\bibfnamefont {F.}~\bibnamefont {Verstraete}},\
  }\bibfield  {title} {\bibinfo {title} {Characterizing topological order with
  matrix product operators},\ }\href
  {https://doi.org/10.1007/s00023-020-00992-4} {\bibfield  {journal} {\bibinfo
  {journal} {Annales Henri Poincar{\'e}}\ }\textbf {\bibinfo {volume} {22}},\
  \bibinfo {pages} {563} (\bibinfo {year} {2021})}\BibitemShut {NoStop}%
\bibitem [{\citenamefont {Williamson}\ \emph {et~al.}(2016)\citenamefont
  {Williamson}, \citenamefont {Bultinck}, \citenamefont {Mari\"en},
  \citenamefont {\ifmmode \mbox{\c{S}}\else
  \c{S}\fi{}ahino\ifmmode~\breve{g}\else \u{g}\fi{}lu}, \citenamefont
  {Haegeman},\ and\ \citenamefont {Verstraete}}]{will:2016}%
  \BibitemOpen
  \bibfield  {author} {\bibinfo {author} {\bibfnamefont {D.~J.}\ \bibnamefont
  {Williamson}}, \bibinfo {author} {\bibfnamefont {N.}~\bibnamefont
  {Bultinck}}, \bibinfo {author} {\bibfnamefont {M.}~\bibnamefont {Mari\"en}},
  \bibinfo {author} {\bibfnamefont {M.~B.}\ \bibnamefont {\ifmmode
  \mbox{\c{S}}\else \c{S}\fi{}ahino\ifmmode~\breve{g}\else \u{g}\fi{}lu}},
  \bibinfo {author} {\bibfnamefont {J.}~\bibnamefont {Haegeman}},\ and\
  \bibinfo {author} {\bibfnamefont {F.}~\bibnamefont {Verstraete}},\ }\bibfield
   {title} {\bibinfo {title} {Matrix product operators for symmetry-protected
  topological phases: Gauging and edge theories},\ }\href
  {https://doi.org/10.1103/PhysRevB.94.205150} {\bibfield  {journal} {\bibinfo
  {journal} {Phys. Rev. B}\ }\textbf {\bibinfo {volume} {94}},\ \bibinfo
  {pages} {205150} (\bibinfo {year} {2016})}\BibitemShut {NoStop}%
\bibitem [{\citenamefont {Li}\ \emph {et~al.}(2020)\citenamefont {Li},
  \citenamefont {Yang}, \citenamefont {Xie}, \citenamefont {Tu}, \citenamefont
  {Liao},\ and\ \citenamefont {Xiang}}]{Zi-Qian:2020}%
  \BibitemOpen
  \bibfield  {author} {\bibinfo {author} {\bibfnamefont {Z.-Q.}\ \bibnamefont
  {Li}}, \bibinfo {author} {\bibfnamefont {L.-P.}\ \bibnamefont {Yang}},
  \bibinfo {author} {\bibfnamefont {Z.~Y.}\ \bibnamefont {Xie}}, \bibinfo
  {author} {\bibfnamefont {H.-H.}\ \bibnamefont {Tu}}, \bibinfo {author}
  {\bibfnamefont {H.-J.}\ \bibnamefont {Liao}},\ and\ \bibinfo {author}
  {\bibfnamefont {T.}~\bibnamefont {Xiang}},\ }\bibfield  {title} {\bibinfo
  {title} {Critical properties of the two-dimensional $q$-state clock model},\
  }\href {https://doi.org/10.1103/PhysRevE.101.060105} {\bibfield  {journal}
  {\bibinfo  {journal} {Phys. Rev. E}\ }\textbf {\bibinfo {volume} {101}},\
  \bibinfo {pages} {060105} (\bibinfo {year} {2020})}\BibitemShut {NoStop}%
\bibitem [{\citenamefont {Kr{\v{c}}m{\'a}r}\ \emph {et~al.}(2016)\citenamefont
  {Kr{\v{c}}m{\'a}r}, \citenamefont {Gendiar},\ and\ \citenamefont
  {Nishino}}]{krvcmar_2016_phase}%
  \BibitemOpen
  \bibfield  {author} {\bibinfo {author} {\bibfnamefont {R.}~\bibnamefont
  {Kr{\v{c}}m{\'a}r}}, \bibinfo {author} {\bibfnamefont {A.}~\bibnamefont
  {Gendiar}},\ and\ \bibinfo {author} {\bibfnamefont {T.}~\bibnamefont
  {Nishino}},\ }\bibfield  {title} {\bibinfo {title} {Phase transition of the
  six-state clock model observed from the entanglement entropy},\ }\href@noop
  {} {\bibfield  {journal} {\bibinfo  {journal} {arXiv preprint
  arXiv:1612.07611}\ } (\bibinfo {year} {2016})}\BibitemShut {NoStop}%
\bibitem [{\citenamefont {Cirac}\ \emph {et~al.}(2011)\citenamefont {Cirac},
  \citenamefont {Poilblanc}, \citenamefont {Schuch},\ and\ \citenamefont
  {Verstraete}}]{Cirac:2011}%
  \BibitemOpen
  \bibfield  {author} {\bibinfo {author} {\bibfnamefont {J.~I.}\ \bibnamefont
  {Cirac}}, \bibinfo {author} {\bibfnamefont {D.}~\bibnamefont {Poilblanc}},
  \bibinfo {author} {\bibfnamefont {N.}~\bibnamefont {Schuch}},\ and\ \bibinfo
  {author} {\bibfnamefont {F.}~\bibnamefont {Verstraete}},\ }\bibfield  {title}
  {\bibinfo {title} {Entanglement spectrum and boundary theories with projected
  entangled-pair states},\ }\href {https://doi.org/10.1103/PhysRevB.83.245134}
  {\bibfield  {journal} {\bibinfo  {journal} {Phys. Rev. B}\ }\textbf {\bibinfo
  {volume} {83}},\ \bibinfo {pages} {245134} (\bibinfo {year}
  {2011})}\BibitemShut {NoStop}%
\end{thebibliography}%

\appendix

\section{Parent Hamiltonians away from the fixed points}\label{sm:sec:parent_H}
In this section, we construct a frustration-free parent Hamiltonian, whose ground states are exactly the decorated TNS. The parent Hamiltonian depends smoothly on the TNS parameters.
\subsection{Warm-up: 1D case}
Let us illustrate the idea of the parent Hamiltonian construction with a 1D example, which will later be generalized to the 2D case. 
One can verify that the MPS $\ket{\psi(g)}$ in Eq.~\eqref{MPS_with_g} that smoothly depends on some parameter $g\in[-1,1]$ can be re-parameterized as an imaginary-time evolved state. More precisely, when $g\in(0,1]$ it satisfies that $\ket{\psi(g)}=\ket{\phi(\tau(g))}$, where $\tau(g) = -\log(g)/4$ and the 1D imaginary-time evolved state is given by
\begin{equation}\label{sm:eq:img_time_gs1d}
    \ket{\phi(\tau)} \propto e^{\tau\sum_{i} Z_iZ_{i+1}}\ket{++\cdots +}, 
\end{equation}
By a direct substitution, we find the relation
\begin{equation}\label{relation_1d}
    K_i\ket{\phi(\tau)} =0,\quad \forall i;\quad K_i = -X_i + e^{-2\tau Z_{i-1}Z_i-2\tau Z_iZ_{i+1}}. 
\end{equation}
Notice that $K_i$ satisfies
\begin{equation}\label{eq:pos1}
    K_i^2 = 2\cosh\left(2\tau Z_{i-1}Z_i+2\tau Z_iZ_{i+1}\right)K_i, 
\end{equation}
and
\begin{equation}\label{eq:pos2}
    \left[\cosh\left(2\tau Z_{i-1}Z_i+2\tau Z_iZ_{i+1}\right), K_i\right]=0.
\end{equation}
This suggests that we can define a projector
\begin{equation}
    P_i = \frac{1}{2}\sech\left(2\tau Z_{i-1}Z_i+2\tau Z_iZ_{i+1}\right)K_i, \label{sm:eq:proj}
\end{equation}
such that $P_i^2 = P_i$ and $P_i\ket{\phi(\tau)} = 0$. One choice of a local parent Hamiltonian for $\tau\geq 0$ is, therefore, $h = \sum_i P_i$ with a ground state energy of zero.

To obtain a Hamiltonian smooth in $g\in[-1,1]$, we evaluate $P_i$ in Eq.~\eqref{sm:eq:proj} in terms of $g$, this yields
\begin{align}
    &2(1+g^2)P_i - 2(1+g^2)
    \nonumber\\
    &= -g_xX_i-\frac{g_{zz}}{2}(Z_{i-1}Z_i+Z_iZ_{i+1})+g_{zxz}Z_{i-1}X_iZ_{i+1},
\end{align}
where $g_x = (1+g)^2, g_{zz} = 2(1-g^2)$ and $g_{zxz} = (1-g)^2$.
A parent Hamiltonian analytic in $g$ is therefore given by
\begin{align}
    H(g) & = 2(1+g^2)\sum_i (P_i-1)
    \nonumber\\
    &= -g_x\sum_iX_i-g_{zz}\sum_iZ_{i}Z_{i+1}+g_{zxz}\sum_iZ_{i-1}X_iZ_{i+1},
\end{align}
with a ground state energy density of $-2(1+g^2)$. This Hamiltonian is exactly Eq.~\eqref{SPT_Hamiltonian} found in Ref.~\cite{Wolf:2006-MPS_phase_transition}. While in the derivation we assume $g\in(0,1]$, since all the functions depend analytically on $g$ for $g\in \{a+i\epsilon|a\in\mathbb{R},\epsilon\in(-1,1)\}$, by analytic continuation $H(g)$ remains to be a valid parent Hamiltonian for $g\in [-1,1]$.

\subsection{2D parent Hamiltonian}
Since each configuration in the 2D wavefunction consists of loops of 1D chains~\eqref{sm:eq:img_time_gs1d}, the 2D ground state also admits a representation in terms of imaginary time evolution starting from the fixed point of the TC phase 
\begin{equation}\label{2D_ITE}
    \ket{\Psi(g,\eta)} \propto \left(\prod_{e\in E}e^{\tau(g)\left[(1-Z_{e})Z_{v(e)}Z_{v'(e)}+ Z_e\right]/2}\eta^{Z_e/2}\right)\ket{\Psi(1,1)},
\end{equation}
where $\tau(g)=-\log (g)/4\geq 0$ and $g\in (0,1]$. 
Note that the alternative interpretation implies that the decoration (imaginary time evolution) commutes with any operators diagonal in the computational basis, including the unitary transformation $U_{\text{TC-DS}}$ discussed in Sec.~\ref{discussion} that maps between the toric code ground state and the double-semion ground state. A phase diagram of the same structure as Fig.~\ref{Phase_diagram_TEE_MOP}a can, therefore, also be obtained by enriching the double-semion model.

Analogously to the 1D case shown in Eq.~\eqref{relation_1d}, it can be verified that
\begin{align}\label{sm:eq:b_zero}
    K_p &\ket{\Psi(g,\eta)} = 0,\quad \forall p;\notag\\
    K_p &= -\prod_{e\in p}X_e+\prod_{e\in p}e^{-\tau(g) Z_e(1-Z_{v(e)}Z_{v'(e)})}\eta^{-Z_e}.
\end{align}
Similar to Eq.~\eqref{eq:pos1} and~\eqref{eq:pos2}, we can obtain a local plaquette projector for each plaquette $p$
\begin{equation}\label{sm:eq:b_proj}
    B_p(g,\eta) = \frac{1}{2}\sech\left(\sum_{e\in p}\left[\tau(g) Z_e(1-Z_{v(e)}Z_{v'(e)})+\lambda(\eta) Z_e\right]\right)K_p,
\end{equation}
where $\lambda (\eta)= \log(\eta)$.
The ground state $\ket{\Psi(g,\eta)}$ satisfies $B_p(g,\eta)\ket{\Psi(g,\eta)}=0$ for all $p$. Recall that $\ket{\Psi(g,\eta)}$ can be understood as a linear combination of closed-loop configurations weighted by some loop tension, where each loop is the 1D MPS state that depends smoothly on a parameter for $g\in [-1,1]$. 

The operators 
$$\sech\left(\tau\sum_{e\in p} Z_e(1-Z_{v(e)}Z_{v'(e)})+\lambda Z_e\right),$$ 
and $$\sech\left(\sum_{e\in p}\left[\tau Z_e(1-Z_{v(e)}Z_{v'(e)})+\lambda Z_e\right]\right)\prod_{e\in p}e^{-\tau Z_e(1-Z_{v(e)}Z_{v'(e)})-\lambda Z_e},$$
are both diagonal in the computational basis with diagonal elements of the form $1/\cosh(4n_{1}\tau+2n_2\lambda)$ and $e^{-4n_{1}\tau-2n_2\lambda}/\cosh(4n_{1}\tau+2n_2\lambda)$
for some integers $n_1,n_2\in[-3,3]$.
Inserting the re-parameterization $\tau(g) = -\log(g)/4$, the matrix elements can be written as 
\begin{align}
    \frac{1}{\cosh(4n_2\tau+2n_2\lambda)} &= \frac{2g^{n_1}}{\eta^{2n_2}+g^{2n_1}\eta^{-2n_2}},
    \nonumber \\
     \frac{e^{-4n_{1}\tau-2n_2\lambda}}{\cosh(4n_2\tau+2n_2\lambda)} &= \frac{2g^{2n_1}\eta^{-2n_2}}{\eta^{2n_2}+g^{2n_1}\eta^{-2n_2}},
\end{align}
which are analytic functions of $g$ for all $\eta>0$ and $g = a+i\epsilon$, where $a,\epsilon\in\mathbb{R}$ and $|\epsilon|<\delta(\lambda)$. Here $\delta(\lambda)$ is the positive real number that corresponds to the smallest distance between the real line and the zeros of $\cosh(4n_{1}\tau+2n_2\lambda)$ in the complex plane. Therefore, the projector Eq.~\eqref{sm:eq:b_proj} can be analytically continued to $\eta>0$ and $g\in[-1,1]$. For $g<0$, the logarithmic function $\tau(g)$ will encounter a branch cut. As we have shown, all the singularities are removable regardless of how the function is defined across the branch cut.

A similar analysis can be performed for the vertex operators. We have the relation
\begin{align}\label{sm:eq:v_zero}
     (1-&A_v)M_v\ket{\Psi(g,\eta)} = 0,\quad \forall v;\\
     M_v &= -X_v+\prod_{e\in v}e^{-\tau(g) (1-Z_e)Z_{v(e)}Z_{v^{\prime}(e)}}.
\end{align}
Note that we include an additional projector $(1-A_v)$ to project out the terms that violate the closed-loop constraint.
The vertex projector at vertex $v$ is given by
\begin{equation}\label{sm:eq:v_proj}
     Q_v(g) = \frac{(1-A_v)}{2}\sech\left(\tau(g)\sum_{e\in v} (1-Z_e)Z_{v(e)}Z_{v^{\prime}(e)}\right)M_v.
\end{equation}
Similar to the plaquette projectors, inserting $\tau(g) = -\log(g)/4$ in Eq.~\eqref{sm:eq:v_proj} results in a form analytic for $g$ close to the real axis,  allowing us to analytically continue the function to $g\in[-1,1]$. We can define the analytically continued projector for $g\in[-1,1]$.

Therefore, a parent Hamiltonian for $\eta\geq 0$ and $g\in[-1,1]$ is, as claimed in the main text,
\begin{equation}
    H(g, \eta) = \sum_vA_v+\sum_pB_p(g,\eta) +\sum_vQ_v(g),
\end{equation}
with a ground state energy of zero. At the fixed points $(g,\eta)=(\pm1,1)$, we recover the fixed-point Hamiltonians as given in the main text. As a consistency check of the analytic continuation, using the relation Eq.~\eqref{pivot2CCZ}, it can be shown that the imaginary time-evolved state Eq.~\eqref{2D_ITE} satisfies the relation $\ket{\Psi(-g,\eta)} = U\ket{\Psi(g,\eta)}$, where $g\in[-1,1]$ and $U$ is the quantum circuit defined in Eq.~\eqref{Circuit_CCZ}. It follows that the analytically continued state~\eqref{2D_ITE} is proportional to the TNS defined in the main text when $g<0$.

Indeed, the parent Hamiltonian is not unique. For example, Eq.~\eqref{sm:eq:b_zero} and~\eqref{sm:eq:v_zero} suggest that we may use $(g^3K_p)^2$ and $(1-A_v)(gM_v)^2$ to construct another parent Hamiltonian that depends smoothly on $g$ and has a ground state energy of zero. Nonetheless, the parent Hamiltonians will share the same low-energy physics. 

\section{$U(1)$ pivot symmetry at $g=0$}\label{pivot_symmetry}
In Ref.~\cite{Tantivasadakarn:2021}, it is found that the 1D Ising-cluster model shown in Eq.~\eqref{SPT_Hamiltonian} has a $U(1)$ pivot symmetry at $g=0$. In this subsection, we show that the 2D Hamiltonian in Eq.~\eqref{2D_Hamiltonian} also has a $U(1)$ pivot symmetry at $g=0$. We first derive the $U(1)$ pivot symmetry for the 1D case as a warm-up. 
\subsection{$U(1)$ pivot symmetry for 1D Hamiltonian}
We start from a local term $P_{i}(g=0)$ for the 1D parent Hamiltonian shown in Eq.~\eqref{sm:eq:proj}. At $g=0$, $\tau\rightarrow\infty$, it can be derived that
\begin{equation}
\mathcal{P}_{i}\equiv\lim_{\tau\rightarrow\infty} \text{sech}(2\tau Z_{i-1}Z_i+2\tau Z_{i}Z_{i+1})=\frac{1-Z_{i-1}Z_{i+1}}{2}, 
\end{equation}
and $P_i(g=0)=\mathcal{P}_iK_i/2$, $[\mathcal{P}_i,K_i]=0$. Moreover, notice that $\mathcal{P}_{i}(Z_{i-1}Z_{i}+Z_{i}Z_{i+1})=0$ and $P_i$ only acts on three sites, we have
\begin{equation}
\left[P_j(g=0),\sum_i Z_iZ_{i+1}\right]=0,\quad \forall j.
\end{equation}
Above equation implies that the generator of the $U(1)$ pivot symmetry can be defined as $H_{\text{Ising}}=\sum_{i}Z_iZ_{i+1}$, such that
\begin{equation}
[H_{\text{Ising}},H(g=0)]=0,\quad H(g=0)=\sum_{i}P_i(g=0).
\end{equation} 
The $U(1)$ pivot symmetry is $U_{\text{pivot}}(\theta)=e^{i\theta H_{\text{Ising}}}, \theta\in \mathbb{R}$. The Hamiltonian at $g=0$ is invariant under $U_{\text{pivot}}(\theta)$:
\begin{equation}
U_{\text{pivot}}(\theta)H(g=0)U^{\dagger}_{\text{pivot}}(\theta)=H(g=0),\quad ,\forall \theta. 
\end{equation}

When $\theta=\pi/4$, one can check that 
\begin{eqnarray}
U_{\text{pivot}}\left(\frac{\pi}{4}\right)& =&e^{\frac{\pi i}{4}\sum_{n}Z_nZ_{n+1}}=e^{\frac{\pi i}{4}\sum_{n}(1-2s_n)(1-2s_{n+1})}\notag\\
&=&e^{\frac{\pi i N}{4}}e^{-\pi i\sum_{n}s_n}e^{\pi i \sum_{n}s_{n}s_{n+1}}
=e^{\frac{\pi i N}{4}}\prod_nZ_n\prod_nCZ_{n,n+1},\notag
\end{eqnarray}
where $N$ is the length of the 1D chain and $s_i=(1-Z_i)/2$  is the transformation from Ising spins $Z_i=\pm1$ to qubits $s_i=0,1$. 
Therefore the pivot symmetry at $\theta=\pi/4$ transforms between the trivial and non-trivial SPT state:
\begin{equation}
U_{\text{pivot}}\left(\frac{\pi}{4}\right)H(g)U^{\dagger}_{\text{pivot}}\left(\frac{\pi}{4}\right)=H(-g).
\end{equation}

\subsection{$U(1)$ pivot symmetry for 2D parent Hamiltonian}
For the 2D case, the $U(1)$ pivot symmetry can be derived similarly. We begin from $B_p(g,\eta)$ shown in Eq.~\eqref{sm:eq:b_proj}. At $g=0$, $\tau\rightarrow\infty$, we have
\begin{equation}
\mathcal{P}_{p}\equiv\lim_{\tau\rightarrow\infty}\text{sech}\left(\tau G_p+\lambda(\eta)\sum_{e\in p} Z_e\right)=\text{sech}\left(\sum_{e\in p}\lambda(\eta) Z_e\right)\delta_{G_p,0},\notag\\
\end{equation}
where
\begin{equation}
G_p=\sum_{e\in p}Z_e(1-Z_{v(e)}Z_{v^{\prime}(e)}).
\end{equation}
Therefore, $B_p(g=0,\eta)=\mathcal{P}_p K_p/2$, $[\mathcal{P}_p,K_p]=0$. Moreover, using $\mathcal{P}_pG_p=0$, it follows that
\begin{equation}\label{commutator_B_p}
\left[B_p(g=0,\eta),\sum_{e\in E}Z_e(1-Z_{v(e)}Z_{v^{\prime}(e)})\right]=0,\quad \forall p.
\end{equation}

We can deal with the vertex terms $Q_{v}(g)$ shown in Eq.~\eqref{sm:eq:v_proj} similarly. At $g=0$, $\tau\rightarrow\infty$, another projector can be derived:
\begin{eqnarray}
\mathcal{P}_{v}\equiv\lim_{\tau\rightarrow\infty}\text{sech}\left(\tau G_v\right)=\delta_{G_v,0}, \,\, G_v=\sum_{e\in v}(1-Z_e)Z_{v(e)}Z_{v^\prime(e)}.
\end{eqnarray}
Analogous to the derivation of Eq.~\eqref{commutator_B_p}, we find 
\begin{equation}
    \left[Q_v(g = 0), \sum_{e\in E}(1-Z_e)Z_{v(e)}Z_{v^\prime(e)}\right]=0.
\end{equation}

To construct the $U(1)$ symmetry generator, we make use of the additional observation that
\begin{equation}
    \left[B_p(g,\eta),\sum_{e\in E}(1-Z_{v(e)}Z_{v^{\prime}(e)})\right]=\left[Q_v(g),\sum_{e\in E}(1-Z_e)\right]=0.
\end{equation}
Therefore, the generator $H_{\text{pivot}}^{(\text{2D})}$ of the $U(1)$ pivot symmetry can be constructed as:
\begin{equation}
H_{\text{pivot}}^{(\text{2D})}=\sum_{e}(1-Z_e)(1-Z_{v(e)}Z_{v^{\prime}(e)}),\quad [H_{\text{pivot}}^{(\text{2D})},H(g=0,\eta)]=0.
\end{equation}
The 2D parent Hamiltonian $H(g=0,\eta)$ is invariant under the transformation $U_{\text{pivot}}^{(\text{2D})}(\theta)=\exp(i\theta H_{\text{pivot}}^{(\text{2D})})$:
\begin{equation}
U_{\text{pivot}}^{(\text{2D})}(\theta)H(g=0,\eta) U_{\text{pivot}}^{(\text{2D})\dagger}(\theta)=H(g=0, \eta), \quad\theta\in \mathbb{R}.
\end{equation}

Analogous to the 1D case, at $\theta=\pi/8$, the $U(1)$ pivot symmetry realizes a unitary transformation: 
\begin{equation}
U_{\text{pivot}}^{(\text{2D})}\left(\frac{\pi}{8}\right)H(g,\eta) U_{\text{pivot}}^{(\text{2D})\dagger}\left(\frac{\pi}{8}\right)=H(-g,\eta).
\end{equation}
Via the transformation from Ising spins to qubits $s_e=(1-Z_e)/2$ and $s_v=(1-Z_v)/2$, $U_{\text{pivot}}^{(\text{2D})}(\pi/8)$ can be expressed in terms of $CCZ$ gates: 
\begin{eqnarray}\label{pivot2CCZ}
U_{\text{pivot}}^{(\text{2D})}\left(\frac{\pi}{8}\right)&=&\exp\left[\frac{\pi i}{8}\sum_{e}(1-Z_e)(1-Z_{v(e)}Z_{v^{\prime}(e)})\right]\notag\\
&=&\exp\left[\frac{\pi i}{2}\sum_{e}s_e\left(s_{v(e)}+s_{v^{\prime}(e)}-2s_{v(e)}s_{v^{\prime}(e)}\right)\right]\notag\\
&=&\exp\left[\frac{\pi i}{2}\sum_{e}s_e\left(s_{v(e)}+s_{v^{\prime}(e)}\right)\right]\prod_{e}CCZ_{v(e)v^\prime(e)e}\notag\\
&=&\prod_{v}\exp\left(\frac{i\pi}{2} s_v\sum_{e\in v}s_{e}\right)\prod_{e}CCZ_{v(e)v^\prime(e)e}\notag\\
&=&\prod_{\langle ee^{\prime}\rangle}\exp\left(i\pi s_{v(e,e^\prime)}s_{e}s_{e^{\prime}}\right)\prod_e CCZ_{v(e)v^\prime(e)e}\notag\\
&=&\prod_{\langle ee^{\prime}\rangle}CCZ_{v(e,e^{\prime})ee^{\prime}}\prod_{\langle vv^{\prime}\rangle}CCZ_{vv^\prime e(v,v^{\prime})}.
\end{eqnarray}
The second to last line is obtained by substituting the relation $\sum_{e\in v}s_{e}=2\sum_{\langle ee^{\prime}\rangle\in v}s_{e}s_{e^{\prime}}$, which is only valid in the closed-loop subspace, into the third last line. We prove that $U_{\text{pivot}}^{(\text{2D})}(\pi/8)$ is equivalent to the unitary transformation~\eqref{Circuit_CCZ} in the main text.

\section{Mapping the SET TNS norm to a partition function}\label{map_to_classical_models}
In this Appendix, we show that along $g=\pm1$, the decorated TNS can be mapped to the 2D classical Ising model, and along $g=0$ they can be mapped to the 2D classical $O(2)$ loop model. The essence of the quantum-classical mapping is identifying the squared norm of the decorated TNS with the partition function of an exactly solved 2D classical statistical model.

When decorating the MPS onto the loops of the toric code, the norm of the MPS, which depends on the length of the MPS, will affect the coefficients in the 2D decorated wavefunction. We first derive the norm of the MPS defined in Eqs.~\eqref{MPS_with_g} and \eqref{eq:MPS(g)}. The transfer operator can be defined from the MPS tensor
\begin{equation}\label{MPS_transfer_operator}
  T=\sum_{i} M^{[i]}\otimes \bar{M}^{[i]},
\end{equation}
whose eigenvalues are $(1+g,1-g,0,0)$. The squared norm of the periodic MPS~\eqref{MPS_with_g} with a length $L$ is
$\mathcal{N}(g)=(1+g)^L+(1-g)^L$. 

\begin{figure}
    \centering
    \includegraphics[width=8cm]{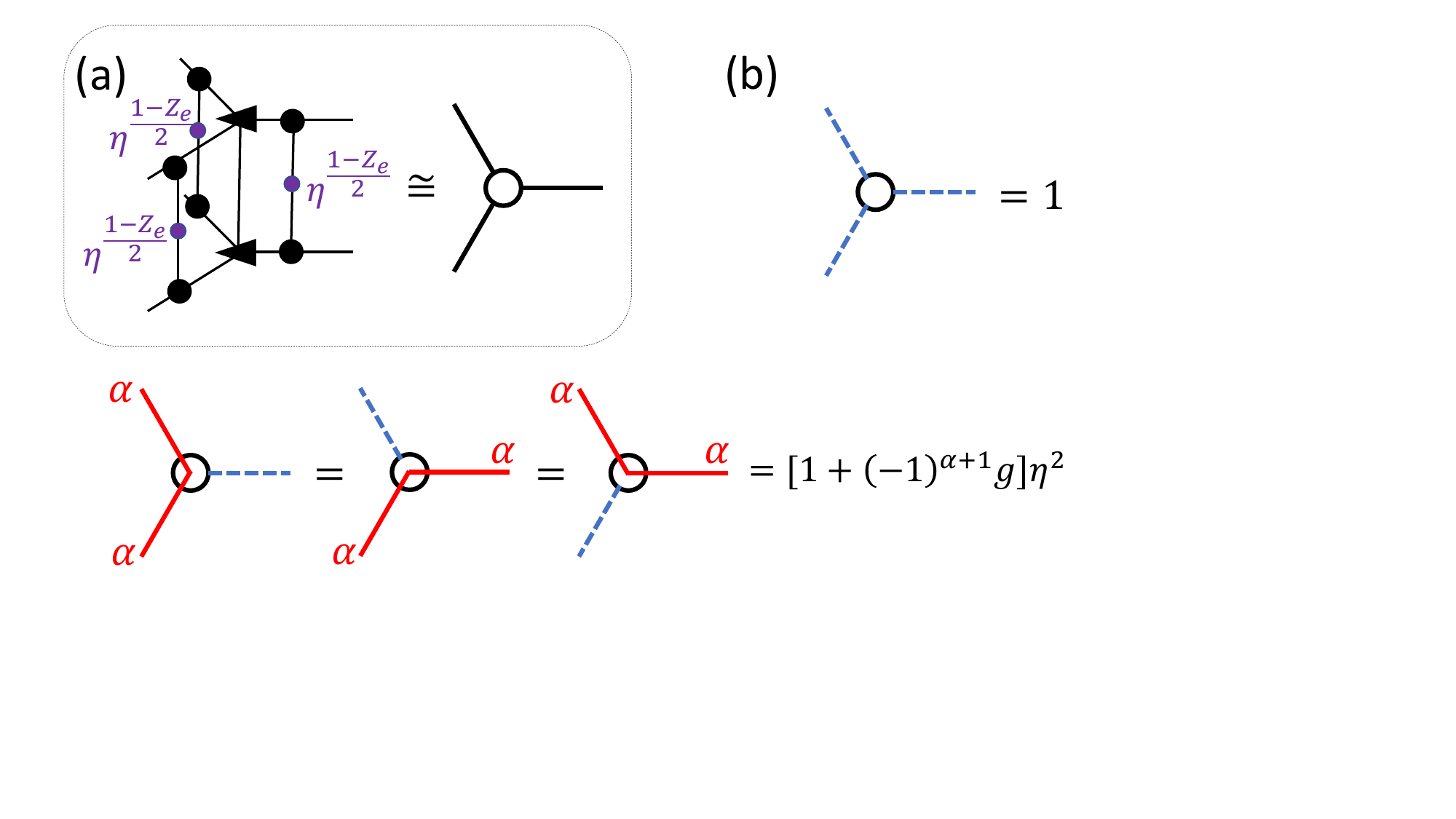}
    \caption{(a) Reduction of the double tensor bond dimension from $D^2=9$ to $3$. (b) The non-zero entries of the reduced double tensor, where blue dashed lines represent the one-dimensional odd bond and red solid lines represent the two-dimensional even bond.}
    \label{compressed_double_tensor}
\end{figure}

Then, let us consider the norm of the decorated TNS, which is a tensor network generated by the double tensor in Fig.~\ref{compressed_double_tensor}a. We duplicate the physical degrees of freedom at the edges so that the tensor looks more symmetric. 
Because the virtual degrees of freedom in the bra and ket layers have the same parity, we can reduce the bond dimension of the double tensor from $D^2=9$ to $5$. The bond dimension $5$ is a direct sum of a $1$-dimensional even bond and a $4$-dimensional odd bond. The $4$-dimensional odd bonds support the MPS transfer operator~\eqref{MPS_transfer_operator}. However, since the MPS transfer operator has two zero eigenvalues, we can further reduce the dimension of an even bond from $4$ to $2$ by diagonalizing the MPS transfer operator. Finally, the bond dimension of the double tensor is reduced to $3$ and its non-zero entries are given in Fig.~\ref{compressed_double_tensor}b.

From the reduced double tensor shown in Fig.~\ref{compressed_double_tensor}, there are two kinds of loops with labels $\alpha=1,2$ and loop tension $(1\pm g)\eta^2$, respectively. Therefore, the squared norm of the decorated TNS~\eqref{eq:gs} is given by
\begin{equation}\label{PEPS_norm}
 \mathcal{N}(g,\eta)=2^{N_v}\sum_{C}\prod_{c\in C}\left[\left(\frac{\eta^2+g\eta^2}{2}\right)^{l_c}+\left(\frac{\eta^2-g\eta^2}{2}\right)^{l_c}\right],
\end{equation}
where $N_v$ is total number of vertices, $C$ is a given closed loop configuration, and $c\in C$ is a closed loop in $C$, and $l_{\text{c}}$ is the length of a given loop $c$.

When $g=0,\pm1$, the squared norm of the decorated TNS becomes the partition function of the classical $O(n)$ loop models\cite{Nienhuis:1982,yellow_book_CFT}
\begin{equation}
  \mathcal{Z}(n,K)=\sum_{C}n^{N(C)}K^{L(C)},
\end{equation}
where $N(C)$ is the total number of loops in $C$, $L(C)$ is the total length of all loops in $C$, $n$ is called the loop fugacity and $K$ is the loop tension. The position of the critical point $K_c$ and the central charge $c$ at the critical point are\cite{Batchelor:1989}
\begin{eqnarray}
  K_c&=&(2+\sqrt{2-n})^{-1/2}, \quad c=1-\frac{6(h-1)^2}{h}, \notag\\ h&=&-\frac{1}{\pi}\arccos(-\frac{n}{2})+1.
\end{eqnarray}

When $g=\pm1$, the squared norm~\eqref{PEPS_norm} of the decorated TNS is equivalent to the partition function of the $O(1)$ loop model
\begin{equation}
\mathcal{N}(g=\pm1,\eta)\propto\sum_{C}\eta^{2L(C)}= \mathcal{Z}(1,\eta^2),
\end{equation}
which is also equivalent to the Ising model on a triangular lattice. The critical point is at $\eta_c=3^{-1/4}\approx 0.7598$ and $c=1/2$.
When $g=0$, the squared norm~\eqref{PEPS_norm} of the decorated TNS is equivalent to the partition function of the $O(2)$ loop model:
\begin{equation} 
\mathcal{N}(g=0,\eta)\propto\sum_{C}2^{N(C)}\left(\eta^2/2\right)^{L(C)}=\mathcal{Z}\left(2,\eta^2/2\right).
\end{equation}
It is well known that the $O(2)$ loop model is qualitatively equivalent to the classical XY model. The critical point $\eta_c=2^{1/4}\approx1.189$ is a Kosterlitz-Thouless phase transition point with central charge $c=1$. When $\eta<\eta_c$, it is the gapped dilute loop phase. When $\eta>\eta_c$, it is the gapless dense loop phase described by a compactified free boson CFT with $c=1$. The $O(2)$ symmetry of the loop model is consistent with the $U(1)$ pivot symmetry of the parent Hamiltonian shown in Appendix~\ref{pivot_symmetry}. 

\section{2D $\mathbb{Z}_2\times\mathbb{Z}_2^T$ SPT states and corresponding partition function}\label{2d SPT}
A duality transformation exists between the 2D trivial (non-trivial) $\mathbb{Z}_2$ SPT model and the toric code (double-semion) model~\cite{Gu:2012}, which is given by
\begin{equation}\label{duality_transformation}
Z_e=Z_{p(e)}Z_{p^\prime(e)},\quad\prod_{e\in p}X_e=X_p,
\end{equation}
where the $Z_p,X_p$ are Pauli operators located at plaquettes and $p(e), p^{\prime}(e)$ are two plaquettes adjacent to edge $e$. 
 Applying the duality transformation to the imaginary time evolved wavefunction~\eqref{2D_ITE} describing $\mathbb{Z}_2^T$ SET phase transitions gives rise to the following wavefunction describing $\mathbb{Z}_2\times\mathbb{Z}_2^T$ SPT phase transitions:
\begin{equation}\label{2D_SPT_state}
\ket{\Psi_{\text{SPT}}}\propto \prod_{e\in E}e^{\frac{\tau}{2} Z_{v(e)}Z_{v^{\prime}(e)}+(\frac{\tau}{2}-\lambda)Z_{p(e)}Z_{p^{\prime}(e)}-\frac{\tau}{2}Z_{v(e)}Z_{v^{\prime}(e)}Z_{p(e)}Z_{p^{\prime}(e)}}\ket{+}_v\ket{+}_p,
\end{equation}
where $\ket{+}_v$ ($\ket{+}_p$) is a product state $\ket{++\cdots+}$ of all vertex (plaquette) qubits. The duality transformation preserves the structure of the phase diagram, as shown in Ref.~\cite{xu:2018}. The TC phase is mapped to the trivial $\mathbb{Z}_2\times\mathbb{Z}_2^{T}$ SPT phase and the SET-TC phase is mapped to a non-trivial $\mathbb{Z}_2\times\mathbb{Z}_2^{T}$ SPT phase.  The trivial phase of the phase diagram shown in Fig.~\ref{Phase_diagram_TEE_MOP}a is mapped to the symmetry broken phase, in which the $\mathbb{Z}_2$ spin flip symmetry of plaquette spins is spontaneously broken.

The squared norm of the wavefunction~\eqref{2D_SPT_state} can be expressed as
\begin{equation}
    ||\ket{\Psi_{\text{SPT}}}||^2\propto \sum_{\{Z_p,Z_v\}}\prod_{e\in E}e^{\tau Z_{v(e)}Z_{v^{\prime}(e)}+(\tau-2\lambda)Z_{p(e)}Z_{p^{\prime}(e)}-\tau Z_{v(e)}Z_{v^{\prime}(e)}Z_{p(e)}Z_{p^{\prime}(e)}}.
\end{equation}
It can be interpreted as the partition function of the Ashkin-Teller model, which consists of two coupled Ising models, one has Ising spins $\{Z_v\}$ on the honeycomb lattice and the other has Ising spins $\{Z_p\}$ on the triangular lattice. This partition function is equivalent to the partition function~\eqref{PEPS_norm}. This suggests that we can also add the additional deformation $\prod_e e^{\beta Z_{v(e)}Z_{v^{\prime}(e)}}$ to the original SET model and obtain a ferromagnetic phase or antiferromagnetic phase in which the $\mathbb{Z}_2^T$ symmetry is broken spontaneously.

\section{Real symmetric tensors and symmetry fractionalization}
\label{Appendix:Real_symmetric_tensors}

Here, we show that the MPS tensors have a real and symmetric form under exchanging two virtual indices such that the single-line tensors of the decorated TNS are real and have a bond dimension $D=3$. This lowers the numerical cost. We apply a gauge transformation to MPS tensors in the 2-site unit cells
\begin{equation}
 \vcenter{\hbox{\includegraphics[width=8cm]{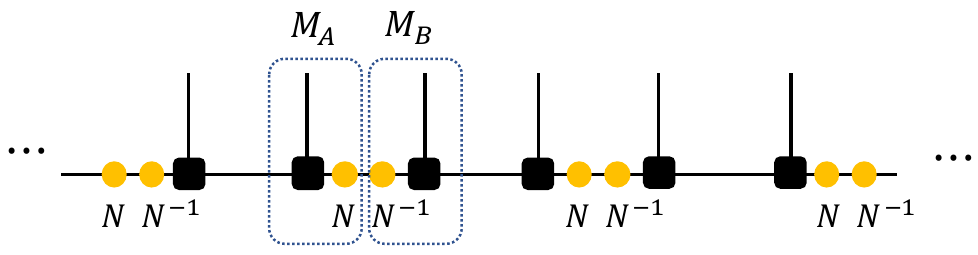}}}, 
\end{equation}
where the gauge transformation is given by
\begin{equation}
    N=\left(
      \begin{array}{cc}
        g & -g \\
        -g & 1 \\
      \end{array}
    \right), \quad N^{-1}=\frac{1}{g(1-g)}\left( \begin{array}{cc}
        1 & g \\
        g & g \\
      \end{array}\right).
\end{equation}
The real and symmetric MPS tensors in a 2-site unit cell are
\begin{eqnarray}
  M_{A}^{[0]}&=&M^{[0]} N=\left(
      \begin{array}{cc}
        0 & 0 \\
        0 & 1-g \\
      \end{array}
    \right), \notag\\ M_{A}^{[1]}&=&M^{[1]} N=\left(
      \begin{array}{cc}
        g(1-g) & 0 \\
        0 & 0 \\
      \end{array}
    \right),\notag\\
    M_{B}^{[0]}&=&N^{-1} M^{[0]}=\frac{1}{g(1-g)}\left(
                \begin{array}{cc}
                  g & g \\
                  g & g \\
                \end{array}
              \right),\notag\\
     M_{B}^{[1]}&=&N^{-1} M^{[1]}=\frac{1}{g(1-g)}\left(
                \begin{array}{cc}
                  1 & g \\
                  g & g^2 \\
                \end{array}
              \right).
\end{eqnarray}
  At $g=0$ and $1$, because the MPS is non-injective, the gauge transformation $N$ is not well-defined. 
 
 Next, we consider the $\mathbb{Z}_2^T$ symmetry of the MPS tensors $M_A$ and $M_B$ in a unit cell. By applying the $\mathbb{Z}_2^T$ symmetry to the MPS, we find
\begin{eqnarray}\label{MPS_transform}
    \sum_{i}(X)_{ij}\bar{M}_{A}^{[i]}&=&\text{sign}(g)U M_{A}^{[j]} U^{T},\notag\\  \sum_{i}(X)_{ij}\bar{M}_{B}^{[i]}&=&\text{sign}(g)(U^{T})^{-1} M_{B}^{[j]} U^{-1},\notag\\
    \sum_{ik}(X)_{ij}(X)_{kl}\bar{M}_{A}^{[i]}\bar{M}_{B}^{[k]}&=&U M_{A}^{[j]}M_{B}^{[l]} U^{-1},
\end{eqnarray}
where 
\begin{equation}\label{defination_U}
    U=\left(\begin{array}{cc}
        0 & \text{sign}(g)\sqrt{|g|} \\
        1/\sqrt{|g|} & 0
    \end{array}\right),\quad U^{-1}=\text{sign}(g)U.
\end{equation}
$U$ is the representation of the symmetry operator on the virtual level. Because $U\bar{U}=\text{sign}(g)$, it is a projective representation when $g<0$.

With the MPS tensors $M_A$ and $M_B$, we can construct the tensors of the decorated TNS shown in Fig.~\ref{Figure_2}a and Eq.~\eqref{vertex_tensor}. The tensors of the decorated TNS have two symmetries, one originates from the topological order and the other comes from the $\mathbb{Z}_2^T$ symmetry. The symmetry from the topological order is
\begin{equation}
\vcenter{\hbox{\includegraphics[width=8cm]{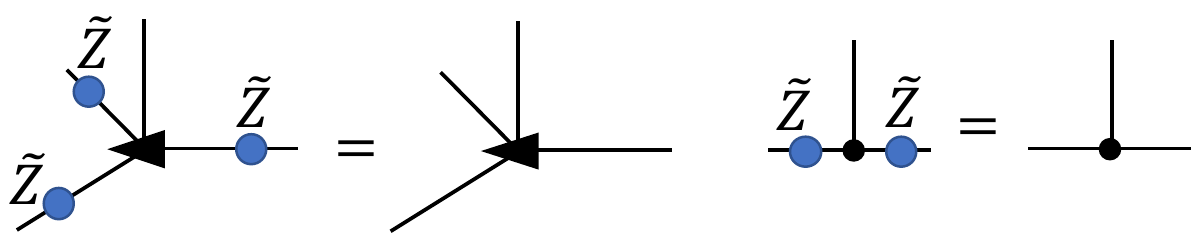}}},
\end{equation}
where $\Tilde{Z}=1\oplus (-\mathbbm{1})$. Because of Eq.~\eqref{MPS_transform}, it can be found that applying the $\mathbb{Z}_2^T$ symmetry to the decorated TNS tensor gives rise to  
\begin{equation}\label{tilde_V}
 \vcenter{\hbox{\includegraphics[width=8.5cm]{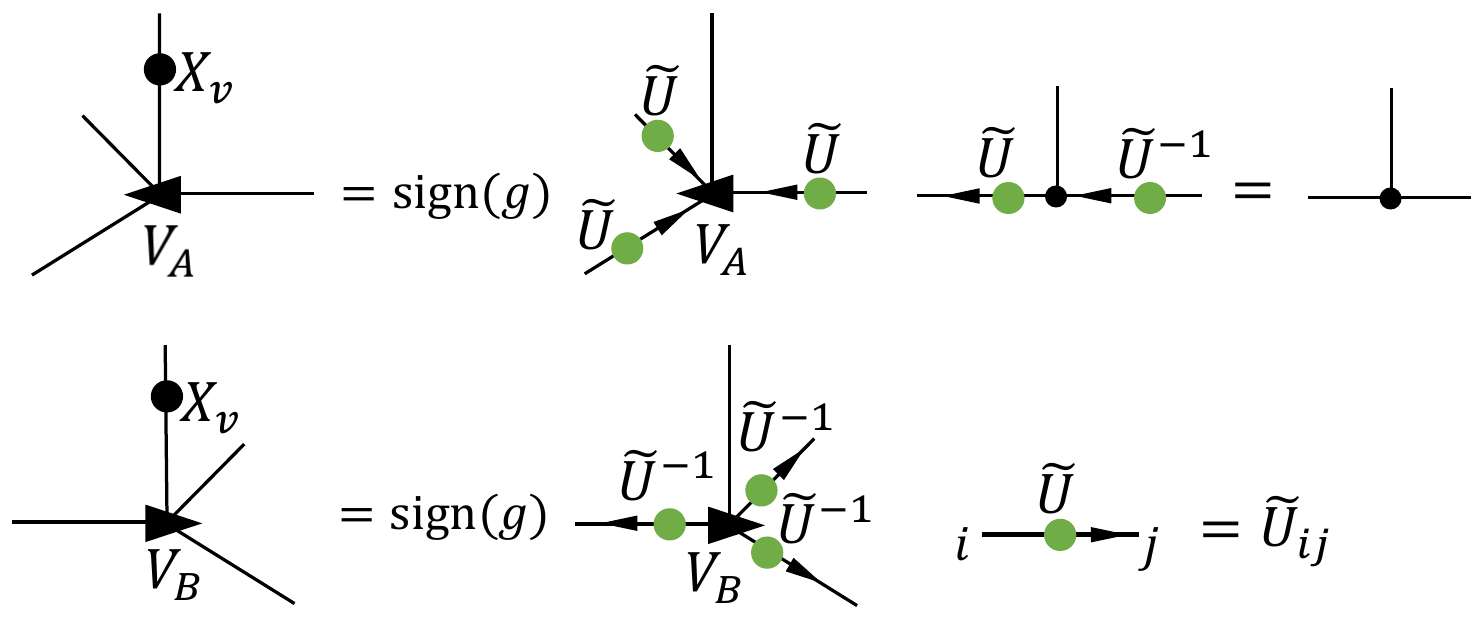}}},
\end{equation}
where $\tilde{U}=1\oplus U$. Because $\Tilde{U}$ is not a symmetric matrix, we use arrows to differentiate its row and column indices. Considering that in a unit cell, sign$(g)$ will be cancelled, we have
\begin{equation}
 \vcenter{\hbox{\includegraphics[width=8.5cm]{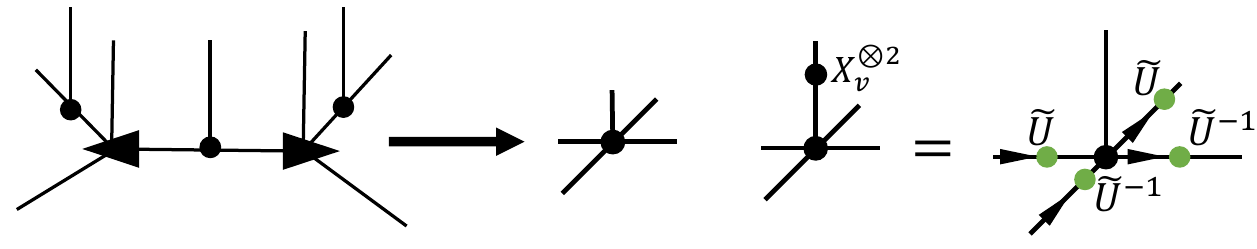}}},
\end{equation}
where $X_v^{\otimes 2}$ acts on two physical degrees of freedom of two vertices. Applying the $\mathbb{Z}_2^T$ symmetry twice, it can be found that $\tilde{U}\bar{\tilde{U}}=\Tilde{Z}$. Because a single-line tensor carrying an anyon $\pmb{e}$ satisfies   
\begin{equation}
 \vcenter{\hbox{\includegraphics[width=4.5cm]{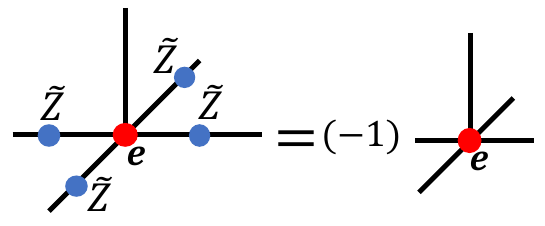}}},
\end{equation}
applying the $\mathbb{Z}_2^{T}$ symmetry twice on an $\pmb{e}$ anyon gives rise to a minus sign. The $\mathbb{Z}_2^{T}$ symmetry  fractionalizes on the $\pmb{e}$ anyons (and also on the $\pmb{f}$ anyons). 

\textcolor{black}{Note that the decorated TNS can be made to satisfy the MPO-injectivity~\cite{sah2021} by grouping the edge and vertex tensors appropriately. The set of virtual matrix-product-operator (MPO) symmetries corresponding to the action of the physical symmetry group, including the product MPO symmetry consisting of $\bar{U}$, encodes the universal labels of the quantum phase of the system~\cite{will:2016, williamson:2017}. }

\section{CTMRG and correlation length}\label{CTM_corr_length}
In this subsection, we show the basic idea of the CTMRG algorithm and the results of the correlation length.  At first, we use a simplified notation of the double tensors
\begin{equation}
 \vcenter{\hbox{\includegraphics[width=3cm]{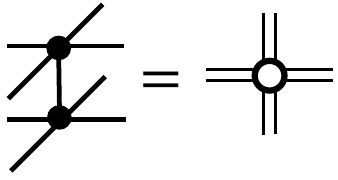}}}.
 \end{equation}
Since the above tensor is not symmetric under exchanging left and right (or upper and lower) indices, the transfer operator of the decorated TNS is non-Hermitian. Therefore, we approximate the environment of the blocked double tensor in terms of four edge tensors and four corner tensors with a bond dimension $\chi$~\cite{Nishino:1996}:
\begin{equation}\label{CTMRG} \vcenter{\hbox{\includegraphics[width=7cm]{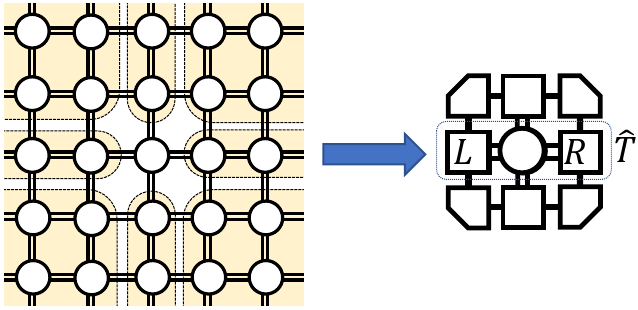}}}.
 \end{equation}
These edge tensors and corner tensors can be obtained using the CTMRG algorithm~\cite{Corboz:2014}. The correlation length $\xi_i=-1/\log(\hat{t}_i/\hat{t}_0)$ can be calculated from the largest eigenvalue $\hat{t}_0$ and the $(i+1)$-th largest eigenvalue $\hat{t}_i$ of the transfer operator $\hat{T}$ shown in Eq.~\eqref{CTMRG}. 

\begin{figure}
    \centering
    \includegraphics[width=8.6cm]{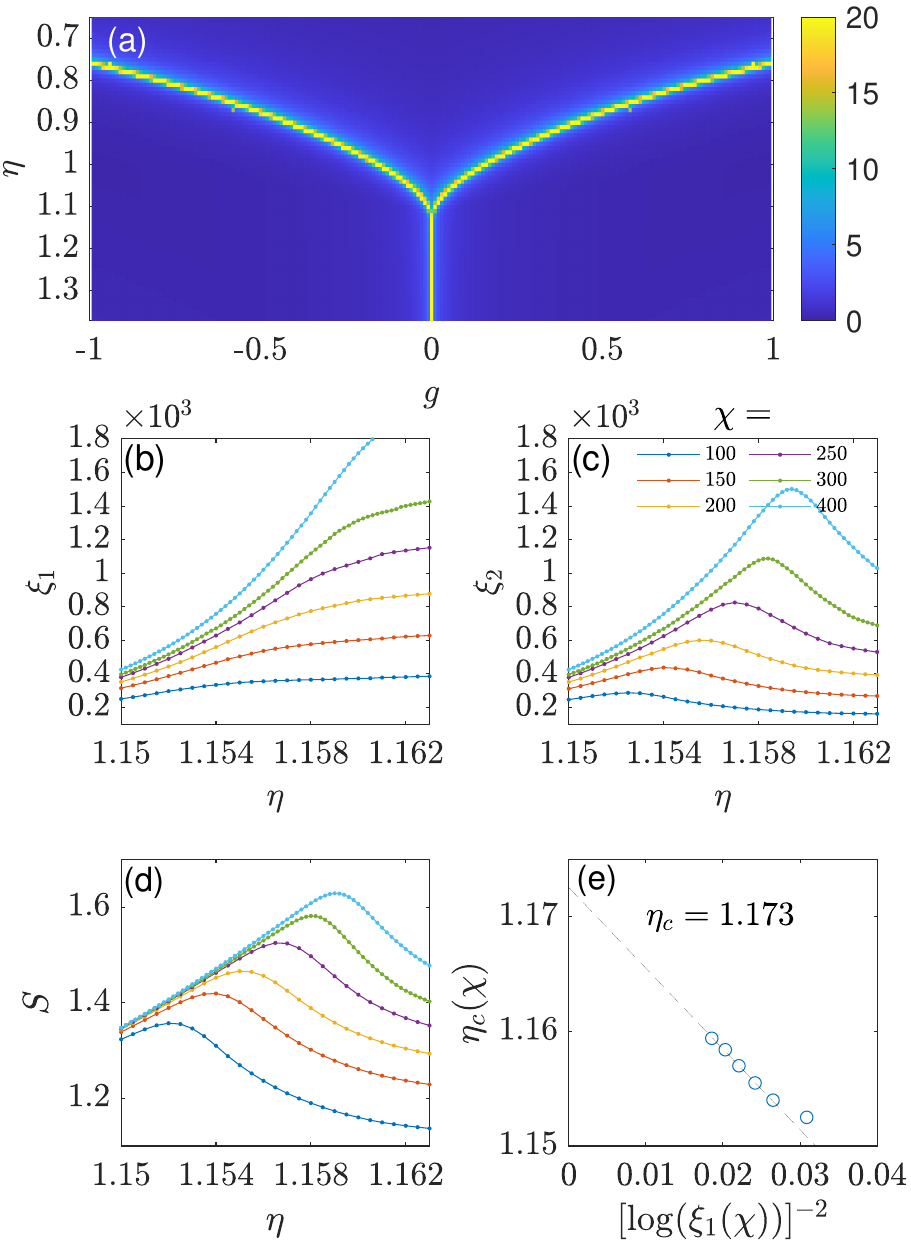}
    \caption{\textcolor{black}{(a) The correlation length $\xi_1$ obtained from $\hat{T}$ shown in Eq.~\eqref{CTMRG}. (b) The correlation length $\xi_1$ obtained from $\hat{T}$ along $g=0$. (c) The correlation length $\xi_2$ obtained from $\hat{T}$ along $g=0$. (d) The entanglement entropy $S$ obtained from corner matrices. (e) Extrapolating the position of the tricritical point. $\eta_c(\chi)$ is obtained from the peaks in (c).} }
    \label{correlation_length}
\end{figure}

We scan the whole phase diagram by calculating the correlation length $\xi_1$ using the CTM environment with bond dimension $\chi=20$. The results shown in Fig.~\ref{correlation_length}a  clearly indicate the phase boundaries. We notice that the position of the tricritical point obtained from the correlation length is not very close to the exact result $(g,\eta)=(0,2^{1/4})\approx(0,1.1892)$. This is reasonable because it is notoriously hard to numerically determine the KT phase transition point. The reason is that there is a logarithmic correction to the position of the KT phase transition point due to the finite bond dimension $\chi$~\cite{Zi-Qian:2020}:
\begin{equation}\label{extrapolate}
\eta_c(\chi)=\eta_c+a \log[\xi_1^{-2}(\chi)],
\end{equation}
where $\xi_1(\chi)$ is the correlation length from a finite bond dimension $\chi$, $\eta_c(\chi)$ is the location of the phase transition from a finite $\chi$, and $a$ is a constant. \textcolor{black}{We can calculate the correlation length $\xi_i(\chi)$ along $g=0$ for various large bond dimensions $\chi$ using the reduced tensor shown in Fig.~\ref{compressed_double_tensor}. As shown in Fig.~\ref{correlation_length}b, no signature of the phase transition can be found in $\xi_1(\chi)$ and we can not determine $\eta_c(\chi)$. However, we find that $\xi_2(\chi)$ exhibits peaks, which move towards the exact critical point with increasing $\chi$ (see Fig.~\ref{correlation_length}c), indicating that it could be used to determine $\eta_c(\chi)$. An alternative way to determine $\eta_c(\chi)$ is to use the entanglement entropy $S$ from boundary MPS or corner tensors of the CTMRG environment~\cite{krvcmar_2016_phase,Zi-Qian:2020}. As shown in Fig.~\ref{correlation_length}d, the locations of the peaks in $\xi_2$ and $S$ coincide, the differences are smaller than 0.0005.}  Using Eq.~\eqref{extrapolate}, the position of the tricritical point can be extrapolated, and the result is shown in Fig.~\ref{compressed_double_tensor}e, indicating that a larger bond dimension is needed to get a more accurate result.

\section{Calculation of membrane order parameters using tensor networks}\label{MOP_TN}

In this Appendix, we show how to simplify the calculation of the MOP shown in Eq.~\eqref{MOP} using tensor networks. We define a modified double tensor that sandwiches the symmetry operator $X_v^{\otimes 2}$:  
  \begin{equation}
 \vcenter{\hbox{\includegraphics[width=3cm]{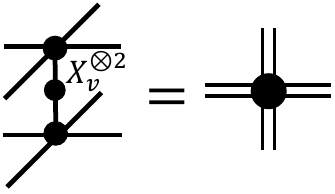}}}.
 \end{equation}
Since there is no canonical form,  a given 2D TNS is usually unnormalized, and the MOP has to be expressed as a ratio of two tensor networks. The tensor network in the numerator of the ratio is
 \begin{equation}\label{numerator}
 \vcenter{\hbox{\includegraphics[width=6cm]{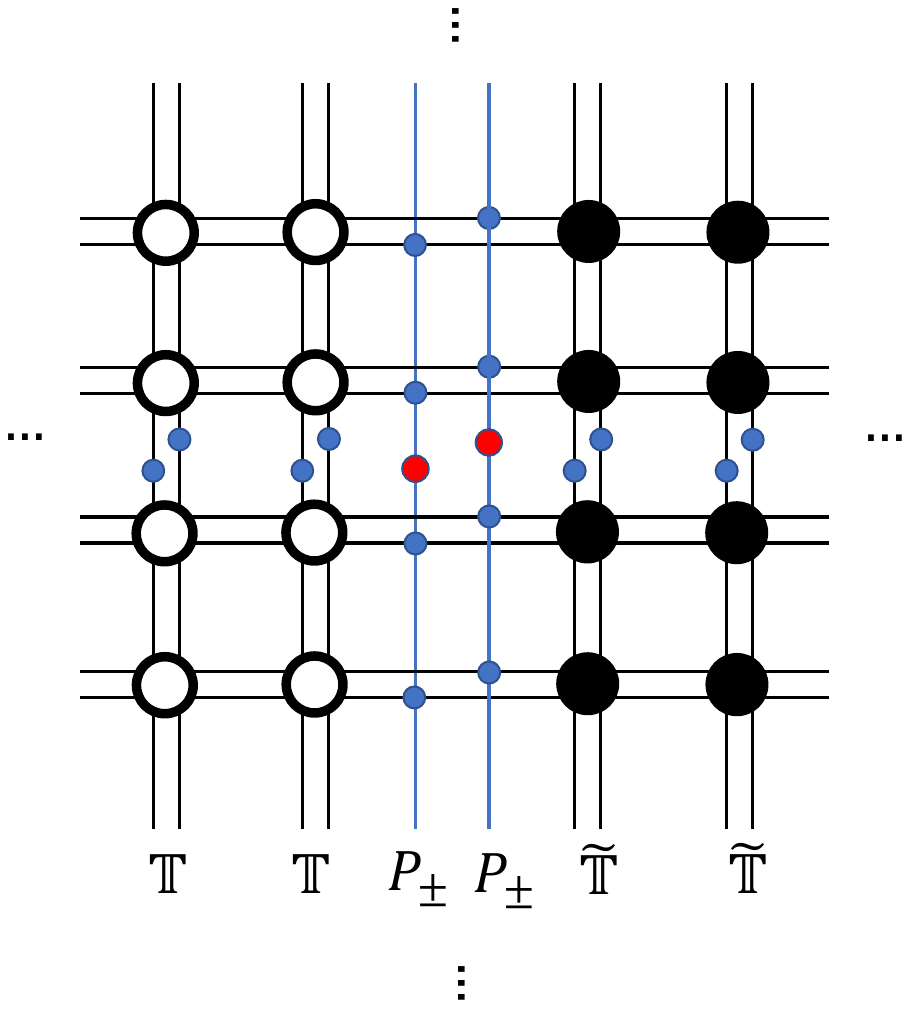}}},
 \end{equation}
and the tensor network in the denominator of the ratio  represents the norm of the decorated TNS:
  \begin{equation}\label{denominator}
 \vcenter{\hbox{\includegraphics[width=6cm]{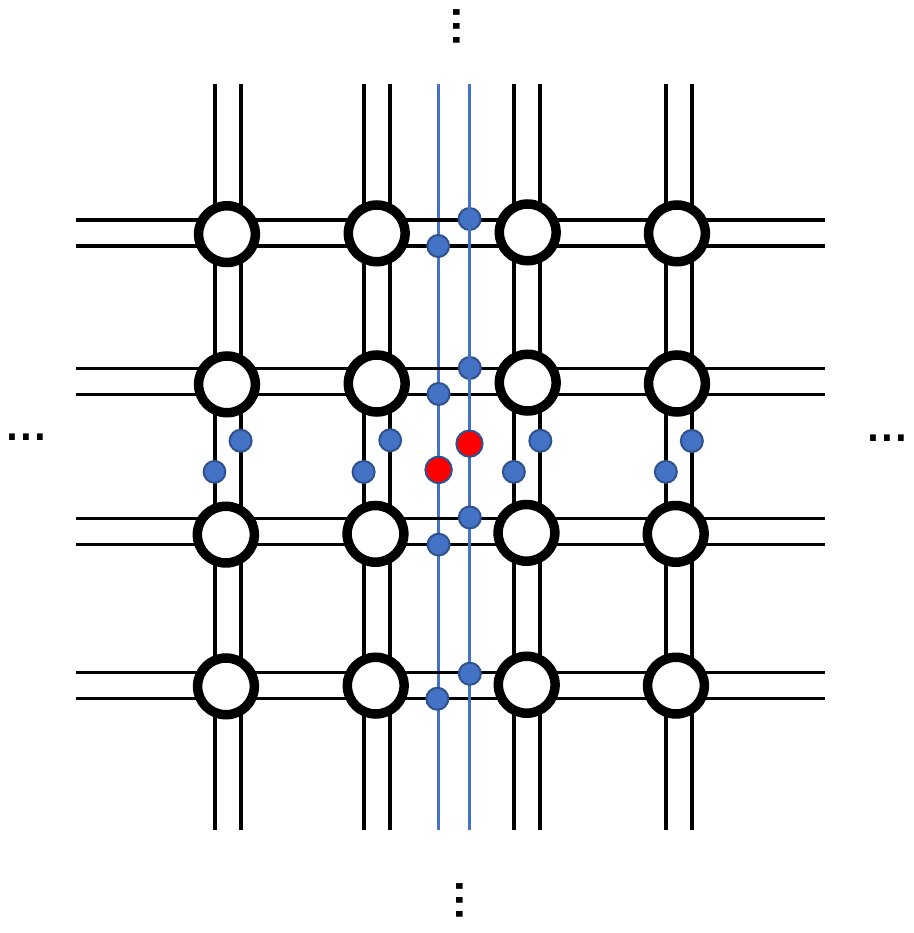}}}.
 \end{equation}
 The entries of the tensors generating the vertical matrix product operator (MPO) are 
 \begin{equation}
     \vcenter{\hbox{\includegraphics[width=1.5cm]{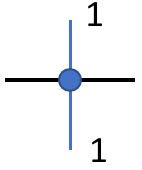}}}=\mathbbm{1}, \quad \quad \quad \vcenter{\hbox{\includegraphics[width=1.5cm]{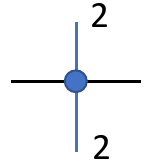}}}=\tilde{Z},\notag\\ 
\end{equation}
and the matrices inserted along the horizontal lines are
\begin{equation}
     \vcenter{\hbox{\includegraphics[width=0.3cm]{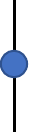}}}=\begin{cases}
     \mathbbm{1},\mbox{ if } \pmb{\alpha}=\pmb{1} \mbox{ or }\pmb{e},\\
     \tilde{Z},\mbox{ if } \pmb{\alpha}=\pmb{m} \mbox{ or }\pmb{f},
     \end{cases},\vcenter{\hbox{\includegraphics[width=0.3cm]{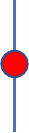}}}=\begin{cases}
     \mathbbm{1}/2,\mbox{ if } \pmb{\alpha}=\pmb{1} \mbox{ or }\pmb{m},\\
     Z/2,\mbox{ if } \pmb{\alpha}=\pmb{e} \mbox{ or }\pmb{f}.
     \end{cases}\notag
 \end{equation}
The vertical MPO and horizontal matrices are used to generate MES in the bra and ket layers. Explicitly, a
vertical MPO is a projector 
\begin{equation}\label{projectors}
P_{\pm}=\frac{1}{2}(\mathbbm{1}^{\otimes N}\pm\tilde{Z}^{\otimes N}),\quad P_{\pm}^2=P_{\pm},
\end{equation}
where $P_{+}$ ($P_{-}$) corresponds to the red dot being $\mathbbm{1}/2$ ($\tilde{Z}/2$), respectively, and $N$ is the circumference of the cylinder.

Then we can contract the tensor networks for the numerator and denominator. We define the left fixed point $\sigma_L$ and the right fixed points $\sigma_R$ of the transfer operators $\mathbb{T}$ (see Eq.~\eqref{numerator}), as well as the left fixed point $\tilde{\sigma}_L$ and the right fixed points $\tilde{\sigma}_R$ of the transfer operator $\tilde{\mathbb{T}}$ (also see Eq.~\eqref{numerator}). These fixed points can be approximated by the MPS
  \begin{equation}\label{fixed_points}
 \vcenter{\hbox{\includegraphics[width=7cm]{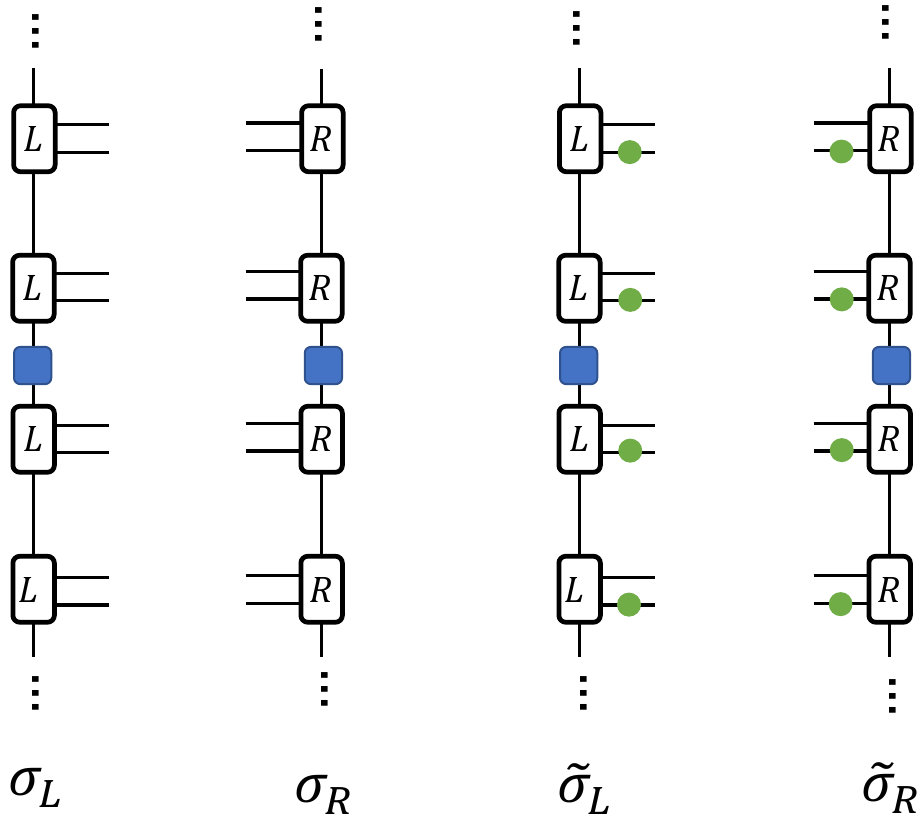}}},
 \end{equation}
 where the tensor $L$ and $R$ come from the edge tensors of the CTM environment shown in Eq.~\eqref{CTMRG}, the tensors represented by the green dots are $\tilde{U}=1\oplus U$ and $U$ is defined in Eq.~\ref{defination_U}. The fixed points $\tilde{\sigma}_L$ and $\tilde{\sigma}_R$ of $\mathbb{\tilde{T}}$ are derived from the fixed points $\sigma_L$ and $\sigma
 _R$ of $\mathbb{T}$ using Eq.~\eqref{tilde_V}. The matrices represented by blue boxes in Eq.~\eqref{fixed_points} come from the two horizontal $\tilde{Z}$ strings in Eq.~\eqref{numerator}. However, due to the $\mathbb{Z}_2$ Gauss law on every vertex tensor, the $\Tilde{Z}$ strings in the bra and ket layers cancel each other, and the matrices represented by the blue boxes become the identity matrix. 
 
 With the above fixed points, we can contract the tensor networks of the numerator and denominator in Eqs.~\eqref{numerator} and \eqref{denominator} from the left and right:
   \begin{equation}\label{contraction_1}
 \vcenter{\hbox{\includegraphics[width=6cm]{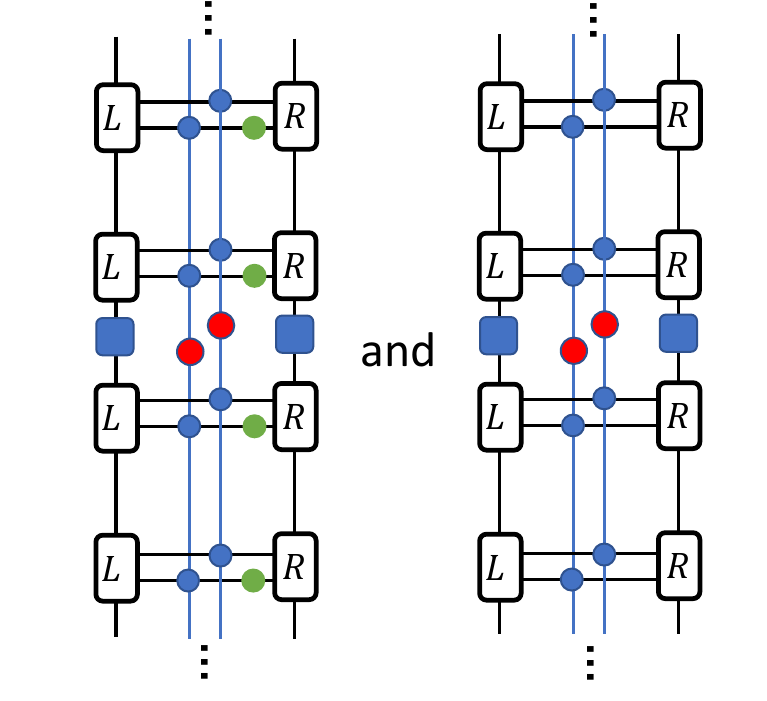}}}.
 \end{equation}
The above tensor networks can be further simplified using the relation $P_{\pm}\sigma_{L/R}=\sigma_{L/R} P_{\pm}$:
 \begin{equation}\label{contraction_2}
 \vcenter{\hbox{\includegraphics[width=6cm]{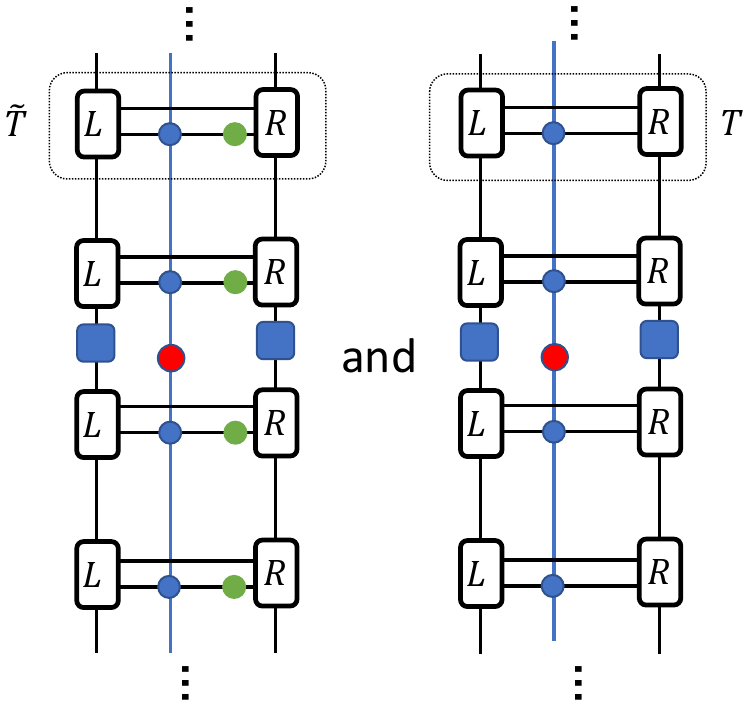}}}.
 \end{equation}
 The channel operators $\tilde{T}$ and $T$ can be defined from the above tensor networks, and it is easy to find their fixed points:
 \begin{eqnarray}
     \langle \tilde{U}_n| \tilde{T}&=&\tilde{t}\langle \tilde{U}_n|,\quad  \tilde{T}|\tilde{D}_n\rangle=\tilde{t}|\tilde{D}_n\rangle; \notag\\
     \langle U_n| T&=&t\langle U_n|,\quad\,\,  T|D_n\rangle=t|D_n\rangle.
 \end{eqnarray}
 Here $t,\tilde{t}\in \mathbb{R}$ are the dominant eigenvalues of the channel operators $T$ and $\tilde{T}$ respectively, and we specify the degenerate channel fixed points with a subscript $n$. Notice that the channel fixed points have to be biorthonormalized: $\langle U_{n}|D_{m}\rangle=\delta_{nm}$. Finally, by contracting the tensor networks using the channel fixed points from above and below, the MOP can be expressed as
 \begin{equation}
O_{\pmb{\alpha}}=\lim_{N\longrightarrow\infty}\left [\left(\frac{\tilde{t}}{t}\right)^N\frac{\tilde{F}_{\pmb{\alpha}}}{F_{\pmb{\alpha}}}\right]^{1/N}=\begin{cases}
0, \mbox{ if } \tilde{F}_{\pmb{\alpha}}/F_{\pmb{\alpha}}=0\\
t_v/t, \mbox{ if } \tilde{F}_{\pmb{\alpha}}/F_{\pmb{\alpha}}\neq0,
\end{cases}
 \end{equation}
 where
 \begin{equation}\label{E_tilde_and_E}
 \tilde{F}_{\pmb{\alpha}}=\sum_{n} \vcenter{\hbox{\includegraphics[width=1.8cm]{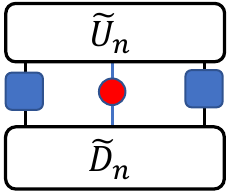}}},\quad F_{\pmb{\alpha}}=\sum_n\vcenter{\hbox{\includegraphics[width=1.8cm]{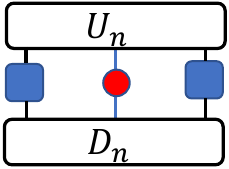}}}.
 \end{equation}
In the trivial phase, we find that $F_{\pmb{e}}$ and $F_{\pmb{f}}$ are zero, which is consistent with the fact that $\pmb{e}$ and $\pmb{f}$ are confined and the MES is no longer well-defined: $\langle\Psi_{\pmb{e}}|\Psi_{\pmb{e}}\rangle=\langle\Psi_{\pmb{f}}|\Psi_{\pmb{f}}\rangle=0$.
 
 \section{Degeneracy of entanglement spectrum and calculation of TEE using tensor networks}\label{TEE_TN}

 The key object for investigating entanglement properties of a quantum many-body wavefunction is the reduced density operator $\rho$ from bipartition. From Ref.~\cite{Cirac:2011}, it is known that the spectrum of a reduced density operator $\rho$ of a TNS is identical to the spectrum of $\sigma=\sigma_{L}^T\sigma_{R}$, where $\sigma_L$ and $\sigma_R$ are the fixed points of the transfer operator $\mathbb{T}$ of the TNS. The entanglement spectrum can be obtained by applying minus the logarithm to eigenvalues of $\sigma$. Moreover, considering the topological sectors, we have
 \begin{equation}
\sigma_{\pmb{1}}=\sigma_{\pmb{m}}=P_{+}\sigma,\quad
\sigma_{\pmb{e}}=\sigma_{\pmb{f}}=P_{-}\sigma,
 \end{equation}
 where $P_{\pm}$ is defined in Eq.~\eqref{projectors}. In the SET-TC phase, applying the $\mathbb{Z}_2^T$ symmetry on the TNS reveals the symmetry transformations on $\sigma_{\pmb{\alpha}}$:
 \begin{equation}
\tilde{U}^{\otimes N} \bar{\sigma}_{\pmb{\alpha}}\left(\tilde{U}^{-1}\right)^{\otimes N} =\sigma_{\pmb{\alpha}}, \quad
\tilde{Z}^{\otimes N}\sigma_{\pmb{\alpha}} \tilde{Z}^{\otimes N} =\sigma_{\pmb{\alpha}}.
\end{equation}
Since $\tilde{Z}^{\otimes N}P_{\pm}=\pm P_{\pm}$, we have
\begin{equation}
   \tilde{U}^{\otimes N}\tilde{\bar{U}}^{\otimes N}=\begin{cases}
       1,  \quad\pmb{\alpha}=\pmb{1},\pmb{m}\\
       -1,\,\, \pmb{\alpha}=\pmb{e},\pmb{f}
   \end{cases}. 
\end{equation}
Therefore, we can apply Kramers' theorem to $\sigma_{\pmb{e}}$ and $\sigma_{\pmb{f}}$, and derive that the entanglement spectra of the $\pmb{e}$ and $\pmb{f}$ sectors are even-fold degenerate in the SET-TC phase.

In the following, we show a method of directly calculating the TEE in the limit $N\rightarrow\infty$, which is similar to the MOP calculation. Since the transfer operator $\mathbb{T}$ is non-Hermitian, we calculate the second Renyi entropy using tensor networks. From Eq.~\eqref{n-Renyi entropy}, the second Renyi entropy is 
\begin{equation}\label{2_Renyi}
S^{[2]}_{\pmb{\alpha}}=2\log \Tr(\sigma_{\pmb{\alpha}})-\log\Tr(\sigma_{\pmb{\alpha}}^2), 
 \end{equation}
 where there is an extra term $2\log \Tr(\sigma_{\pmb{\alpha}})$ since  usually $\sigma_{\pmb{\alpha}}$ is not normalized in tensor-network calculations.
$\Tr(\sigma_{\pmb{\alpha}}^2)$ can be expressed in terms of a tensor network:
 \begin{equation}\label{rho_2_and_rho}
   \vcenter{\hbox{\includegraphics[width=4.5cm]{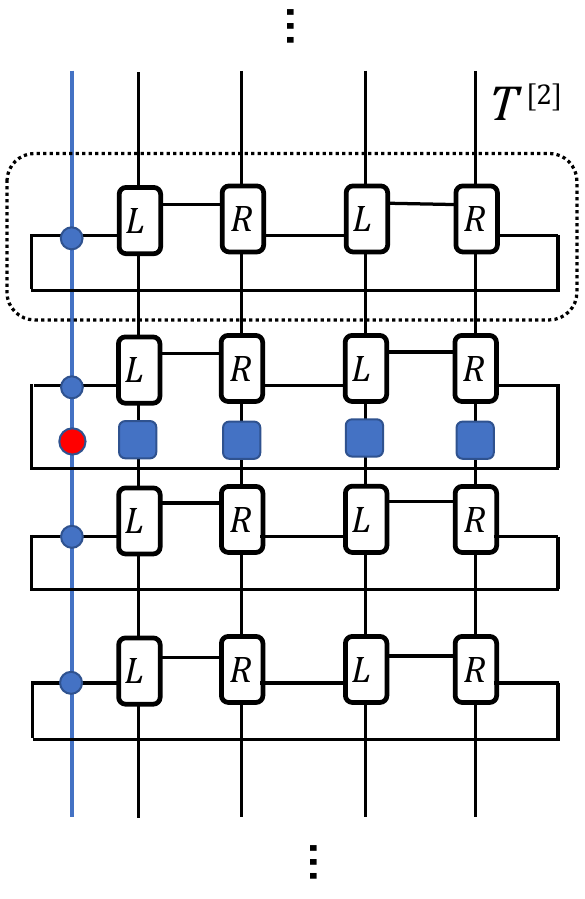}}}.
 \end{equation}
 The tensor network of $\Tr(\sigma_{\pmb{\alpha}})$ is the right hand side of Eq.~\eqref{contraction_2}. Defining another channel operator $T^{[2]}$, as shown in Eq.~\eqref{rho_2_and_rho}, its fixed points $U^{[2]}_n$, $D^{[2]}_n$ can be found
 \begin{equation}
     \langle U^{[2]}_n| T^{[2]}=t_{[2]}\langle U^{[2]}_n|,\quad  T^{[2]}|D^{[2]}_n\rangle=t_{[2]}|D^{[2]}_n\rangle,
 \end{equation}
 where the subscript $n$ specifies the degenerate fixed points and we impose the biorthonormality condition $\langle U^{[2]}_k|D^{[2]}_m\rangle=\delta_{km}$. We can contract the tensor networks of  $\Tr(\sigma_{\pmb{\alpha}}^2)$ and $\Tr(\sigma_{\pmb{\alpha}})$ using their channel fixed points
 \begin{equation}
\Tr(\sigma^2_{\pmb{\alpha}})=\lim_{N\rightarrow+\infty} t_{[2]}^N F^{[2]}_{\pmb{\alpha}},\quad      \Tr(\sigma_{\pmb{\alpha}})=\lim_{N\rightarrow+\infty} t^N F_{\pmb{\alpha}},
 \end{equation}
 where 
 \begin{equation}
     F_{\pmb{\alpha}}^{[2]}= \sum_n\vcenter{\hbox{\includegraphics[width=3cm]{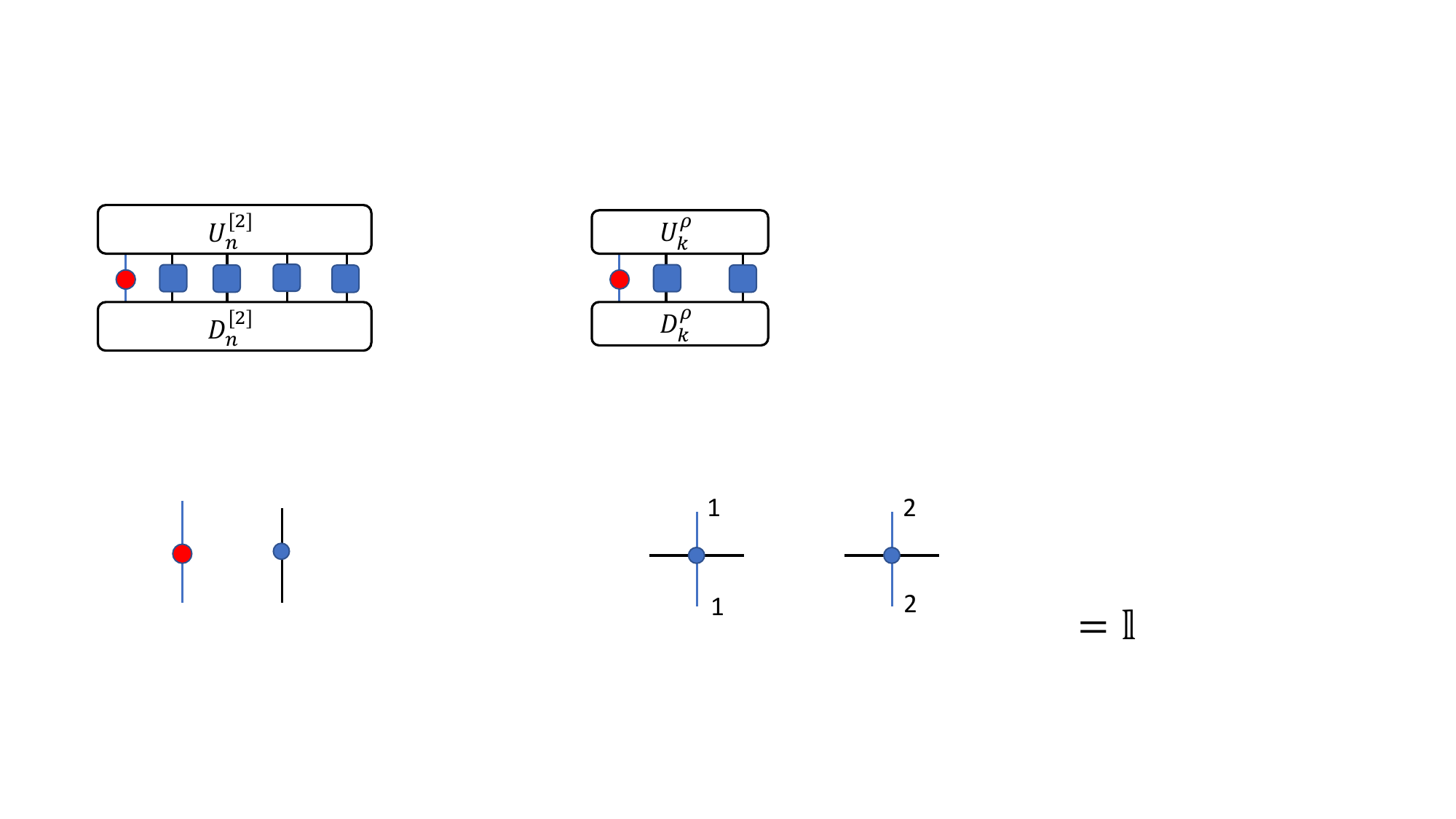}}}
 \end{equation}
 and $F_{\pmb{\alpha}}$ is defined in Eq.~\eqref{E_tilde_and_E}. Substituting these relations into Eq~\eqref{2_Renyi}, we obtain the second Renyi entanglement entropy in the limit $N \rightarrow\infty$
 \begin{equation}
     S^{[2]}_{\pmb{\alpha}}=\lim_{N\rightarrow\infty}N\log\frac{t^2}{t^{[2]}}-\log \frac{F^{[2]}_{\pmb{\alpha}}}{F_{\pmb{\alpha}}^2},
 \end{equation}
from which we can identify the TEE $\gamma=\log F^{[2]}_{\pmb{\alpha}}/F_{\pmb{\alpha}}^2$.

\end{document}